\documentclass[aps,prd,a4paper,twocolumn]{revtex4}
 
\usepackage{graphicx}
\usepackage{bm}
\usepackage{amsfonts}
\usepackage{amsmath}
\usepackage{color}
\usepackage{accents}
\usepackage{wasysym}

\usepackage[russian,ngerman,english]{babel}

\newcommand{\muas}[0]{\hbox{\rm $\mu$as}}

\newcommand{\ve}[1]{\mbox{\boldmath$#1$}}

\let\oldbibitem\bibitem
\renewcommand\bibitem[2][]{\oldbibitem{#2}}

\begin{document}

\title{Light deflection in the gravitational field of a solar system body with finite distance of source and observer 
for sub-micro-arcsecond astrometry} 

\author{Sven Zschocke}

\affiliation{Institute of Planetary Geodesy - Lohrmann Observatory, Technical University of Dresden, 
Helmholtzstrasse 10, D-01069 Dresden, Germany}

\begin{abstract}
The effect of deflection of a light signal that propagates through the gravitational field of a solar system body at rest is considered in the case that the source 
and the observer are at a finite distance from the body. The observer is assumed to be located somewhere nearby the Earth, for instance at Lagrange point $L_2$ of the 
Sun-Earth system, while the spatial position of the celestial light source is arbitrary. 
The gravitational fields in the exterior of the bodies are described by the full set of time-dependent mass-multipoles and spin-multipoles of these bodies. 
So the gravitating bodies can be of arbitrary shape, inner structure and rotational motion. The unit tangent vector of the light trajectory at the position 
of the observer is determined in the 1PN and 1.5PN approximation of the post-Newtonian (PN) scheme. A simplified expression for the unit tangent vector is obtained, where 
all terms are neglected that together contribute less than $10$ nano-arcseconds in light deflection for all astrometric configurations between source, body and observer. 
It is shown that in the case of an axi-symmetric body the unit tangent vector 
of the light ray as well as the light deflection are given in terms of Chebyshev polynomials. This fact allows for determining the effect of light deflection in the gravitational 
fields of the bodies up to any order of the mass-multipoles and spin-multipoles. It is shown that the total light deflection, that is the angle of light deflection 
where source and observer are located at infinite spatial distance from the body, represents an upper limit of the effect of light deflection. 
These investigations are aiming at astrometric measurements on the sub-micro-arcsecond level of accuracy.  
\end{abstract}

\pacs{95.10.Jk, 95.10.Ce, 95.30.Sf, 04.25.Nx, 04.80.Cc}

\maketitle 


\section{Introduction and Motivation}\label{Section0}

The primary aim of astrometric measurements concerns the determination of spatial positions of celestial objects, like 
stars, quasars, or solar system objects. The determination of these spatial positions is based on the observed direction 
of light signals that are emitted by these celestial light sources. This means that the primary aim of high precision 
astrometry is directly related to the measurement of the unit tangent vectors of these light trajectories at the position of the observer, 
which are on the scope of this investigation.  

Today's precision of astrometric measurements has achieved at the micro-arcsecond (\muas) scale of accuracy. In particular, the astrometry 
mission {\it Gaia} of the European Space Agency (ESA) has measured the parallax for approximately $2$ billion ($2 \times 10^9$) stars of our galaxy, 
mapping them in three dimensions with an accuracy of up to $10\,\muas$ \cite{Gaia1,Gaia2} for bright stellar sources. Meanwhile, there are 
several astrometry missions proposed to ESA, which are aiming at the sub-micro-arcsecond (sub-\muas) level of accuracy. The most promising candidates of 
such mission proposals are the astrometry missions {\it GaiaNIR} \cite{Gaia_NIR} and {\it Theia} \cite{Theia}. 
For instance, {\it GaiaNIR}, a successor mission of {\it Gaia}, aims to map about $50$ billion ($50 \times 10^9$) stars in near infrared. 
Also the astrometry missions {\it TOLIMAN} \cite{Toliman}, an Australian-American project, and {\it SHERA} \cite{Shera}, an American project, are mentioned, which are 
aiming at the sub-\muas{} level of accuracy. Their scientific goal is the detection of terrestrial exoplanets in the habitable zones of nearby stars 
by means of relative astrometry. 

An accuracy on the sub-\muas{} scale requires a highly precise modeling of light trajectories in the curved space-time of the solar system, in order to 
reduce astrometric measurements correctly and to trace the light rays back to the celestial light sources. But numerical integrations of the geodesic equation 
of light trajectories are unusable in view of the extensive calculations necessary for astrometry missions like {\it Gaia} \cite{Gaia1,Gaia2} or {\it GaiaNIR} \cite{Gaia_NIR}. 
Instead, analytical solutions for the light trajectories are very necessary in order to treat such a huge amount of astrometric observations. Additionally, analytical approaches 
are not only important in view of the era of big data reduction and exponentially growing data volumes of earth-grounded or space-based astrometry missions, but also because 
they allow one to identify all those terms in the analytical solutions, which have a certain impact on a given relativistic effect under consideration. 
Furthermore, analytical solutions allow for the validation of numerical solutions, as well as the investigation of all possible astrometric configurations. 
However, the higher the accuracy requirements are, the more complex the analytical expressions of light trajectories become.

Already in the case of the first few multipoles, like mass-quadrupole and spin-dipole, the mathematical expressions for the light trajectories become rather cumbersome. 
One reason for this complexity is laying in the fact, that the analytical solutions of light trajectory and, therefore, of the unit tangent vector of light trajectory, contain 
many terms that are not significant on the sub-micro-arcsecond level. In view of the huge amount of stellar objects for missions like {\it GaiaNIR}, it is an inevitable 
assignment to simplify these analytical solutions by taking account of those terms only that are relevant for a given accuracy. Afterwards one has to find compact expressions 
for these remaining terms of the analytical solutions, which allow for a highly effective reduction of observational data. 

The term {\it sub-micro-arcsecond level} refers to astrometric precisions in positional measurements of about $0.1\,\muas$ for bright celestial light sources. 
Such precisions require a relativistic model of light propagation which is about $10$ times more accurate than the end-of-mission accuracy. Accordingly, 
the given threshold of a sub-\muas{} model of light propagation is $0.01$ micro-arcsecond ($10$ nano-arcsecond). That means, in a theoretical model of light propagation on the 
{\it sub-micro-arcsecond level of precision} one may neglect all those terms that contribute in total less than $10$ nano-arcsecond to the angle of light deflection. 

In reality both the celestial light sources as well as the observer are located at finite distances from the massive solar system bodies. In particular, the observer can 
be located somewhere nearby the Earth, for instance at Lagrange point $L_2$ of the Sun-Earth system, as in the mission {\it Gaia} \cite{Gaia1,Gaia2} as well as 
in the missions {\it GaiaNIR} \cite{Gaia_NIR} and {\it Theia} \cite{Theia} proposed to ESA. The light sources are assumed to be located completely arbitrarily: 
they can be far stellar light sources as well as near solar system objects. That means, the spatial distances between the light sources and the observer 
can be arbitrarily large and arbitrarily small. In this investigation the solution of the mentioned problems is presented for such general astrometric configurations. 

The manuscript is organized as follows: 
In Section~\ref{Section1} the metric tensor of an isolated body in terms of the full set of its mass-multipoles and spin-multipoles, the geodesic equation of 
light propagation, and the light trajectory in the post-Newtonian scheme are given in the 1.5PN approximation of the Post-Newtonian (PN) scheme. 
The unit tangent vector of the light ray in the field of an arbitrary body at rest and at the observers position is presented in Section~\ref{Section2}. 
The fact that the individual terms of the unit tangent vector are of qualitatively different magnitude is considered in Section~\ref{Section_Magnitude}. 
The reduced unit tangent vector of a light ray is derived in Section~\ref{Simplified_1}, which is a step in between in order to arrive at the simplified 
tangent vector of the light trajectories. In Section~\ref{Section_Differential_Operations} we consider two 
differential operations that are necessary in order to get the simplified unit tangent vector. 
The simplified unit tangent vector of a light ray is derived in Section~\ref{Simplified_2}. 
The mass-multipoles and spin-multipoles of an axi-symmetric body are given in Section~\ref{Section3}. The simplified tangent vector of the light trajectory 
in the gravitational field of an axi-symmetric body is presented in Section~\ref{Simplified_3}.  
In Sections~\ref{Simplified_3} and \ref{Monopole_Quadrupole_Spin} the unit tangent vector in case of mass-monopole, spin-dipole,  
and mass-quadrupole are compared with results in the literature. The angle of light deflection, their upper limits, 
and numerical values are presented in Section~\ref{Section4}. Finally, a brief view on 2PN and 3PN terms and their magnitude is given in Section~\ref{Section_2PN_3PN}. 
A Summary and Outlook can be found in Section~\ref{Summary}. The notations, conventions, and some details of the calculations are relegated to a set of several appendixes.

\section{The light trajectory in the gravitational field of a body}\label{Section1} 

\subsection{Metric tensor}

For weak gravitational fields, the metric tensor can be separated as follows \cite{Einstein,Thorne,MTW,Poisson_Will,Carroll,Kopeikin_Efroimsky_Kaplan,Petrov_Kopeikin_Lompay_Tekin}, 
\begin{eqnarray}
        g_{\alpha\beta} &=& \eta_{\alpha\beta} + h_{\alpha\beta},
        \label{Linearized_Gravity_1}
\end{eqnarray}

\noindent
where $\eta_{\alpha\beta} = {\rm diag}(-1,+1,+1,+1)$ is the Minkowski metric, which implies that the flat background space-time is covered by 
Cartesian four-coordinates, $x^{\mu} = (x^0,x^1,x^2,x^3)$, where $x^0 = ct$ with coordinate time $t$, while $x^1,x^2,x^3$ are the spatial coordinates. The metric perturbations are 
assumed to be small, $|h_{\alpha\beta}| \ll 1$, and behave mathematically like symmetric tensorial fields (with respect to Lorentz transformations), which propagate in the flat background 
space-time \cite{MTW,Carroll,Petrov_Kopeikin_Lompay_Tekin}. In linearized gravity, the field equations of general relativity (GR) in harmonic gauge 
($\partial_{\nu} \overline{h}^{\mu\nu}=0$) read \cite{Einstein,Thorne,MTW,Poisson_Will,Carroll,Kopeikin_Efroimsky_Kaplan,Petrov_Kopeikin_Lompay_Tekin} 
\begin{eqnarray}
	\square\,\overline{h}_{\alpha\beta} &=& - \frac{16 \pi G}{c^4}\,T_{\alpha\beta}\,,
        \label{Linearized_Gravity_2}
\end{eqnarray}

\noindent
where $\overline{h}_{\alpha\beta} = h_{\alpha\beta} - \frac{1}{2}\,\eta_{\alpha\beta}\,h$ with $h = \eta^{\alpha\beta}\,h_{\alpha \beta}$, the flat d'Alembert operator is 
$\square = \eta^{\mu \nu}\partial_{\mu} \partial_{\nu}$ with the partial derivatives 
$\partial_{\mu} = \left(\frac{\partial}{\partial x^0},\frac{\partial}{\partial x^1},\frac{\partial}{\partial x^2},\frac{\partial}{\partial x^3}\right)$, and 
$T_{\alpha\beta}$ is the energy-momentum tensor of matter. In linearized gravity $h_{\alpha\beta} = \overline{h}_{\alpha\beta} - \frac{1}{2}\,\eta_{\alpha\beta}\,\overline{h}$ 
with $\overline{h} = \eta^{\alpha\beta}\,\overline{h}_{\alpha \beta}$. Hence, from $\overline{h}_{\alpha\beta}$ in (\ref{Linearized_Gravity_2}) one gets  
$h_{\alpha\beta}$ in (\ref{Linearized_Gravity_1}) up to terms of order ${\cal O}(G^2)$. 

If the source of matter is isolated (compact support of matter and no incoming radiation), then the integration of the linearized field equations (\ref{Linearized_Gravity_2}) leads to 
metric perturbations in the exterior of the body, which depend on a set of $10$ source-multipoles \cite{Thorne,Blanchet_Damour1,Multipole_Damour_2,2PN_Metric1}; for a detailed proof 
we refer to \cite{Zschocke_Theorem}. These multipoles are integrals over the energy-momentum tensor of the body. In the pioneering 
investigations \cite{Thorne,Blanchet_Damour1,Multipole_Damour_2,2PN_Metric1} it has been worked out, that the energy-momentum conservation of matter reduces the number of independent 
source-multipoles and leads to metric perturbations, which depend finally on a set of only $6$ source-multipoles: $I_L,J_L,W_L,X_L,Y_L,Z_L$. 

In case of weak gravitational fields and slow motions of matter, the metric perturbation in (\ref{Linearized_Gravity_1}) can further be series expanded in inverse powers of the speed  
of light, which is called the Post-Newtonian (PN) scheme \cite{Thorne,MTW,Poisson_Will,Kopeikin_Efroimsky_Kaplan,Petrov_Kopeikin_Lompay_Tekin,Comment1}. 
In the 1.5PN approximation this series expansion reads
\begin{eqnarray}
        h_{\alpha\beta}\left(t,\ve{x}\right) &=& h_{\alpha\beta}^{\left(2\right)}\left(t,\ve{x}\right) + h_{\alpha\beta}^{\left(3\right)}\left(t,\ve{x}\right) + {\cal O}(c^{-4}), 
        \label{PN_Expansion_1PN_15PN}
\end{eqnarray}

\noindent 
where the perturbations are of the order $h_{\alpha\beta}^{\left(2\right)} = {\cal O}(c^{-2})$ and $h_{\alpha\beta}^{\left(3\right)} = {\cal O}(c^{-3})$. These 
perturbations can be written as sum of individual perturbations caused by $N$ individual solar system bodies. That is why we can consider the influence of one body 
on light propagation. Finally one may sum over all these $N$ bodies. 

The harmonic gauge condition does not uniquely determine the metric tensor, but leaves freedom of a residual gauge. In the so-called canonical harmonic gauge, the metric 
perturbations (\ref{PN_Expansion_1PN_15PN}) depend finally on a set of only $2$ multipoles: mass-multipoles, $M_L = I_L + {\cal O}(c^{-5})$ \cite{Blanchet_Faye_Iyer_Sinha}, which 
account for shape and inner structure of the body, and spin-multipoles, $S_L = J_L + {\cal O}(c^{-5})$ \cite{Blanchet_Faye_Iyer_Sinha}, which account for rotational motions and 
inner currents of the body \cite{Thorne,Blanchet_Damour1,Multipole_Damour_2,2PN_Metric1}. The non-vanishing metric perturbations in canonical harmonic gauge  
are given by \cite{Thorne,Blanchet_Damour1,Multipole_Damour_2,Kopeikin_Efroimsky_Kaplan,Zschocke_2PM_Metric} 
\begin{eqnarray}
	h_{00}^{\left(2\right)}\left(t,\ve{x}\right) &=& \frac{2}{c^2} \sum\limits_{l=0}^{\infty} \frac{\left(-1\right)^l}{l!}\,\hat{M}_L\left(t\right)\,\hat{\partial}_L \frac{1}{r} \,,
        \label{Metric_00}
	\\
        h_{0i}^{\left(3\right)}\left(t,\ve{x}\right) &=& \frac{4}{c^3} \sum\limits_{l=1}^{\infty} \frac{\left(-1\right)^l\,l}{\left(l+1\right)!} \,
        \epsilon_{iab}\,\hat{S}_{b L-1}\left(t\right)\, \hat{\partial}_{a L-1} \frac{1}{r}\,,
        \label{Metric_0i}
        \end{eqnarray}

\noindent
and $h_{ij}^{\left(2\right)} = h_{00}^{\left(2\right)}\,\delta_{ij}$, where $r = \left|\ve{x}\right|$. The mass-multipoles and spin-multipoles are given by \cite{Multipole_Damour_2}
\begin{eqnarray}
	\hat{M}_L\left(t\right) &=& \underset{i_1 \dots i_l}{\rm STF}  \int d^3 x^{\prime} 
	\; x^{\prime}_L\;\frac{T^{00}(t,\ve{x}^{\prime})}{c^2} + {\cal O}\left(c^{-2}\right), 
\label{Mass_Multipoles}
\\
	\hat{S}_L\left(t\right) &=& \underset{i_1 \dots i_l}{\rm STF}\ \int d^3 x^{\prime} \;\epsilon_{ab i_l}\,x^{\prime}_{a L-1}
	\;\frac{T^{0b}(t,\ve{x}^{\prime})}{c} + {\cal O}\left(c^{-2}\right),
	\nonumber\\  
\label{Spin_Multipoles}
\end{eqnarray}

\noindent 
where the integration runs over the volume of the body and we recall $T^{00} = {\cal O}(c^2)$, $T^{0a} = {\cal O}(c^1)$, $T^{ab} = {\cal O}(c^0)$. 
The differential operator in (\ref{Metric_00}) and (\ref{Metric_0i}) is with respect to the global coordinate system and reads 
\begin{eqnarray}
	\hat{\partial}_L &=& \underset{i_1 \dots i_l}{\rm STF}\,\frac{\partial}{\partial x^{i_1}} \dots \frac{\partial}{\partial x^{i_l}}\;,  
        \label{Partial_Derivatives}
        \end{eqnarray}

\noindent
where $\underset{i_1 \dots i_l}{\rm STF}$ denotes an operation, which makes a Cartesian tensor symmetric and trace-free (STF) with respect to its 
spatial indices $L = i_1 \dots i_l$. Further details of STF operations are summarized in Appendix~\ref{Appendix_STF}. 

In Eqs.~(\ref{Metric_00}) - (\ref{Metric_0i}) the orthogonal three-vectors of the principal axes $\ve{e}_1,\ve{e}_2,\ve{e}_3$ of the body are arbitrarily 
oriented with respect to the global Cartesian coordinate system. The multipoles (\ref{Mass_Multipoles}) and (\ref{Spin_Multipoles}) are time-dependent, because the body is assumed to be 
rotating around an arbitrary axis (see also Figure~\ref{Diagram1}), hence its orientation continuously changes (precession), which inherently results in the time-dependence of its 
mass-multipoles and spin-multipoles. It is assumed that the origin of the spatial coordinates is located at the center-of-mass of that body. This circumstance implies that 
the mass-dipole vanishes: $\hat{M}_i = 0$. On this occasion it is mentioned that mass-monopole $M$ and spin-dipole $\hat{S}_i$ (angular momentum) are conserved 
\cite{Multipole_Damour_2,Thorne,Blanchet_Damour1,Kopeikin_Efroimsky_Kaplan}. Actually, these terms can change their values by emitting gravitational radiation, which is, however, an effect 
beyond 1.5PN approximation and extremely tiny. In what follows, the metric tensor (\ref{Linearized_Gravity_1}) with Eqs.~(\ref{Metric_00}) - (\ref{Metric_0i}), where the time-dependent multipoles 
are given by Eqs.~(\ref{Mass_Multipoles}) and (\ref{Spin_Multipoles}), will be called the metric of an {\it arbitrary body}.

\subsection{The geodesic equation}

In flat (Minkowskian) space-time, a light signal that is emitted at some initial time, $t_0$, into some three-direction, $\ve{\sigma}$, propagates along a straight trajectory,  
\begin{eqnarray}
        \ve{x}_{\rm N} &=& \ve{x}_0 + c \left(t - t_0\right) \ve{\sigma}\;. 
        \label{Unperturbed_Lightray_2}
\end{eqnarray}

\noindent
One may define the impact vector of this unperturbed light ray, 
\begin{eqnarray}
        \ve{d}_{\sigma} &=& \ve{\sigma} \times \left(\ve{x}_{\rm N} \times \ve{\sigma}\right) = \ve{\sigma} \times \left(\ve{x}_0 \times \ve{\sigma}\right).
        \label{impact_vector}
\end{eqnarray}

\noindent
This impact vector is a three-vector that points from the center-of-mass of the body toward the unperturbed light ray at their closest distance, as elucidated in Figure~\ref{Diagram1}. 
The absolute value of this impact vector is the impact parameter $d_{\sigma} = |\ve{d}_{\sigma}|$. In general relativity the propagation of light signals is governed by 
the geodesic equation, which in 1.5PN approximation, i.e. up to terms of order ${\cal O}(c^{-4})$, reads \cite{Brumberg1991}
\begin{eqnarray}
\frac{\ddot{x}^i \left(t\right)}{c^2} &=& \frac{\partial h_{00}^{(2)}}{\partial x^i}
- 2\,\frac{\partial h_{00}^{(2)}}{\partial x^j}\,\sigma^i \sigma^j - \frac{\partial h_{00}^{(2)}}{\partial x^0}\,\sigma^i 
\nonumber\\
&& - \frac{\partial h_{0i}^{(3)}}{\partial x^j}\,\sigma^j
+ \frac{\partial h_{0j}^{(3)}}{\partial x^i}\,\sigma^j
- \frac{\partial h_{0j}^{(3)}}{\partial x^k}\,\sigma^i \sigma^j \sigma^k.
\label{Geodesic_Equation_15PN}
\end{eqnarray}

\noindent 
The double-dot on the left-hand side in (\ref{Geodesic_Equation_15PN}) means twice of the total differentiation with respect to the coordinate time, and 
$h_{ij}^{\left(2\right)} = h_{00}^{\left(2\right)}\,\delta_{ij}$ has been taken into account.

\subsection{The light trajectory} 

The geodesic equation (\ref{Geodesic_Equation_15PN}) is a differential
equation of second order; thus a unique solution of (\ref{Geodesic_Equation_15PN}) requires two initial conditions (initial value problem) 
\cite{Brumberg1991,Kopeikin1997,KopeikinSchaefer1999_Gwinn_Eubanks,Klioner1991,KlionerKopeikin1992,Zschocke_1PN,Zschocke_15PN}: the unit-direction $\ve{\sigma}$ of 
the light ray at past infinity and the spatial position of light source $\ve{x}_0$ at the moment of emission of the light signal, 
\begin{eqnarray} 
        \ve{\sigma} &=& \frac{\dot{\ve{x}}\left(t\right)}{c}\,\bigg|_{t \rightarrow - \infty} \;,
        \label{vector_sigma}
        \\
        \ve{x}_0 \; &=& \; \ve{x}\left(t\right) \bigg|_{t = t_0}\;,
        \label{x_0}
\end{eqnarray}

\noindent
with $\ve{\sigma} \cdot \ve{\sigma} = 1$. The geodesic equation (\ref{Geodesic_Equation_15PN}) can be solved by iteration, and the solution of first and second integration reads formally 
\begin{eqnarray}
	\frac{\dot{\ve{x}}\left(t\right)}{c} &=& \ve{\sigma} 
	+ \sum\limits_{l=0}^{\infty} \frac{\Delta \dot{\ve{x}}^{M_L}_{\rm 1PN}\left(t\right)}{c} 
	+ \sum\limits_{l=1}^{\infty} \frac{\Delta \dot{\ve{x}}^{S_L}_{\rm 1.5PN}\left(t\right)}{c}\,, 
        \label{First_Integration_1PN}
        \\
        \ve{x}\left(t\right) &=& \ve{x}_{\rm N} + \sum\limits_{l=0}^{\infty} \Delta\ve{x}^{M_L}_{\rm 1PN}\left(t, t_0\right) 
	+ \sum\limits_{l=1}^{\infty} \Delta\ve{x}^{S_L}_{\rm 1.5PN}\left(t, t_0\right), 
        \nonumber\\ 
        \label{Second_Integration_1PN}
\end{eqnarray}

\noindent 
up to terms of the order ${\cal O}(c^{-4})$.

In reality, the light source as well as the observer are located at finite distances $\ve{x}_0$ and $\ve{x}_1$ from the gravitating solar system body.
This fact implies to solve the geodesic equation in terms of these two boundary values (boundary value problem),  
\begin{eqnarray}
        \ve{x}_0 &=& \ve{x}\left(t\right)\,\,\bigg|_{t = t_0}\,, 
        \label{boundary0}
        \\
        \ve{x}_1 &=& \ve{x}\left(t\right)\,\,\bigg|_{t = t_1}\,. 
        \label{boundary1}
\end{eqnarray}

\begin{figure}[t]
\begin{center}
\includegraphics[scale=0.147]{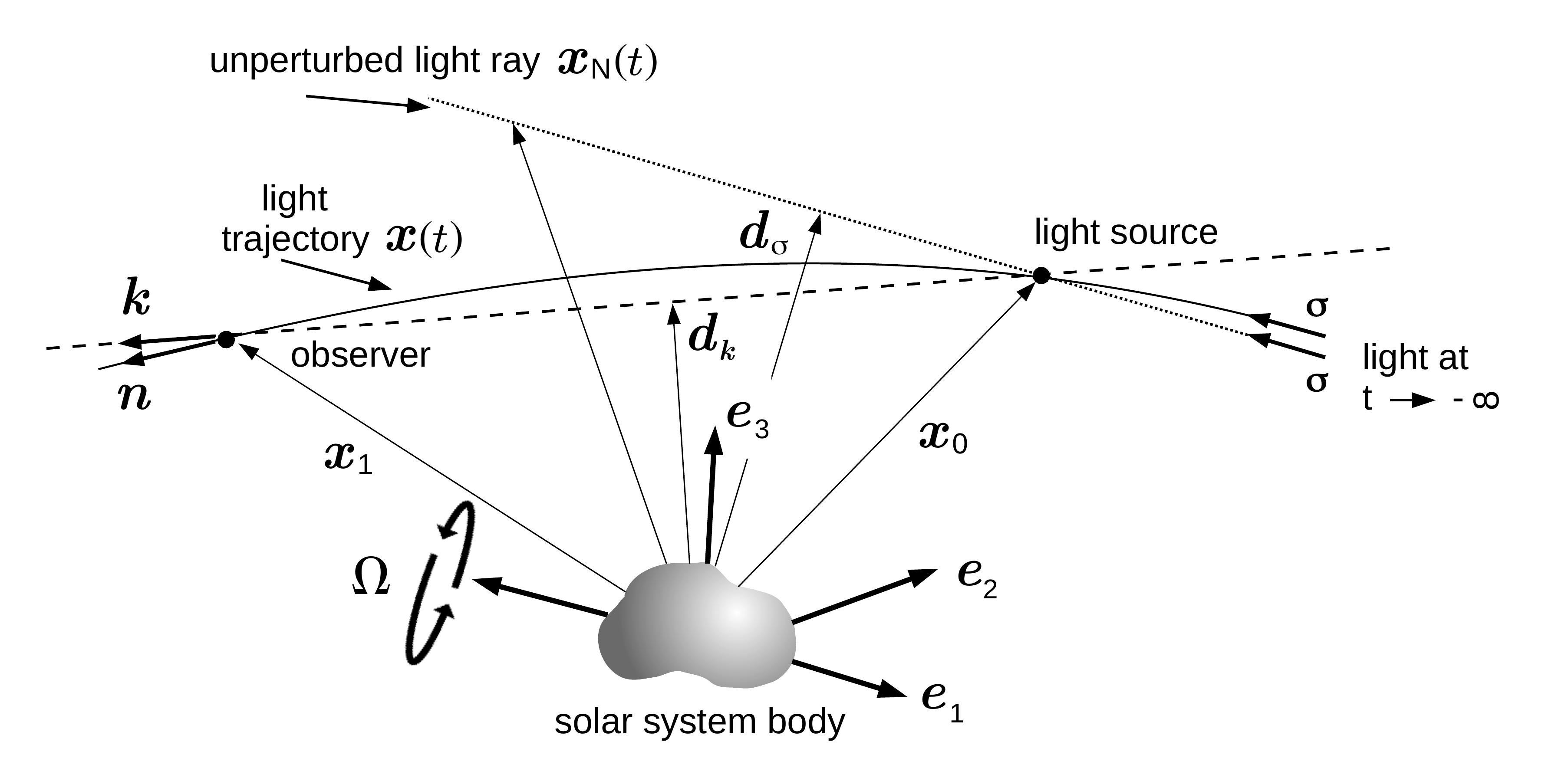}
\end{center}
\caption{
        A geometrical representation of the propagation of a light signal through the gravitational field of a body at rest of arbitrary shape, 
	inner structure and rotational motion around an arbitrarily oriented axis with angular velocity $\Omega$. 
	The orientation of the rotational axis is time-dependent and continuously changes due do precession of the body. 
        The origin of the spatial coordinates $\left(x^1,x^2,x^3\right)$ is assumed to be located at the center-of-mass of the body. 
        The orientation of the orthogonal principal axes $\ve{e}_1,\ve{e}_2,\ve{e}_3$ of the body with respect to the coordinate system is arbitrary. 
	They are co-rotating with the body. 
        The light signal is emitted by the light source at $\ve{x}_0$ and propagates along the exact light trajectory $\ve{x}\left(t\right)$.
        The unperturbed light ray $\ve{x}_{\rm N}\left(t\right)$ is given by Eq.~(\ref{Unperturbed_Lightray_2}) and propagates in the direction 
	of $\ve{\sigma}$ along a straight line through the position of the light source at $\ve{x}_0$. The impact vector $\ve{d}_{\sigma}$ is given 
	by Eq.~(\ref{impact_vector}) and the impact vector $\ve{d}_k$ is given by Eq.~(\ref{impact_vector_k}).}
\label{Diagram1}
\end{figure}

\noindent
These equations state, that the spatial positions of light signal at the time of emission, $t_0$, and at the time of reception, $t_1$, are in coincidence 
with the spatial position of the light source, $\ve{x}_0$, and observer, $\ve{x}_1$, respectively. In the boundary value problem one introduces the unit-vector 
\begin{eqnarray}
\ve{k} &=& \frac{\ve{x}_1 - \ve{x}_0}{ \left|\ve{x}_1 - \ve{x}_0\right|}\,, 
\label{vector_k}
\end{eqnarray}

\noindent
with $\ve{k} \cdot \ve{k} = 1$ and that is pointing from the light source toward the observer. The introduction of this three-vector in (\ref{vector_k}) implicates a new impact vector, 
\begin{eqnarray} 
        \ve{d}_k &=& \ve{k} \times \left(\ve{x}_0 \times \ve{k}\right) = \ve{k} \times \left(\ve{x}_1 \times \ve{k}\right), 
\label{impact_vector_k}
\end{eqnarray}
        
\noindent
which points from the center-of-mass of the body toward the closest point of the coordinate line between $\ve{x}_0$ and $\ve{x}_1$;
for a graphical elucidation see Figure~\ref{Diagram1}. The absolute value of this impact vector is the impact parameter $d_k = \left|\ve{d}_k\right|$. 
The transformation from $\ve{k}$ to $\ve{\sigma}$ in the 1.5PN approximation reads \cite{Klioner2003a}  
\begin{eqnarray}
	\ve{\sigma} = \ve{k} && - \frac{1}{R}\left[\ve{k}\times \bigg(\sum\limits_{l=0}^{\infty} \Delta\ve{x}^{M_L}_{\rm 1PN}\left(t_1,t_0\right)\times\ve{k}\bigg)\right]
	\nonumber\\ 
	&& - \frac{1}{R}\left[\ve{k}\times \bigg(\sum\limits_{l=1}^{\infty} \Delta\ve{x}^{S_L}_{\rm 1.5PN}\left(t_1,t_0\right)\times\ve{k}\bigg)\right].  
	\label{transformation_sigma_to_k}
\end{eqnarray}

\noindent 
To define and to determine the light deflection, the unit tangent vector at the moment of reception needs to be introduced 
\begin{eqnarray}
        \ve{n} &=& \frac{\dot{\ve{x}}\left(t_1\right)}{\left|\dot{\ve{x}}\left(t_1\right)\right|}\,.
        \label{vector_n}
\end{eqnarray}

\noindent
By inserting the transformation (\ref{transformation_sigma_to_k}) into (\ref{First_Integration_1PN}) one obtains 
for the unit tangent vector (\ref{vector_n}) the expression \cite{Klioner2003a}  
\begin{eqnarray}
	\ve{n} &=& \ve{k} + \sum\limits_{l=0}^{\infty} \Delta \ve{n}_{M_L} + \sum\limits_{l=1}^{\infty} \Delta \ve{n}_{S_L}
        \label{light_deflection}
\end{eqnarray}

\noindent
with
\begin{eqnarray}	
	\Delta \ve{n}_{M_L} &=& \ve{k} \left[\times \left(\frac{\Delta\dot{\ve{x}}^{M_L}_{\rm 1PN}\left(t_1\right)}{c}
	- \frac{\Delta\ve{x}^{M_L}_{\rm 1PN}\left(t_1,t_0\right)}{R}\right) \times \ve{k}\right], 
	\nonumber\\ 
        \label{light_deflection_M}
	\\
	\Delta \ve{n}_{S_L} &=& \ve{k} \left[\times \left(\frac{\Delta\dot{\ve{x}}^{S_L}_{\rm 1.5PN}\left(t_1\right)}{c}
        - \frac{\Delta\ve{x}^{S_L}_{\rm 1.5PN}\left(t_1,t_0\right)}{R}\right) \times \ve{k}\right],
        \nonumber\\
        \label{light_deflection_S}
\end{eqnarray}

\noindent
up to terms beyond 1.5PN approximation. 

In \cite{Kopeikin1997} advanced integration methods have been developed, which allow for integrating of the geodesic equations (\ref{Geodesic_Equation_15PN}). 
These integration methods will be considered below. Afterwards, the expressions of the light ray perturbations in (\ref{First_Integration_1PN}) 
and (\ref{Second_Integration_1PN}) will be presented in their explicit form by Eqs.~(\ref{First_Integration_5}) - (\ref{First_Integration_6}) and 
Eqs.~(\ref{Second_Integration_5}) - (\ref{Second_Integration_6}).

\subsection{The light trajectory in terms of new variables}

In the approach in \cite{Kopeikin1997} advanced integration methods were introduced, based on the new parameters,
\begin{eqnarray}
        c \tau &=& \ve{\sigma} \cdot \ve{x}_{\rm N}\;,
        \label{Parameter1}
        \\
	\hat{\xi}^i &=& P^i_j\,x_{\rm N}^j\;,  
        \label{Parameter2}
\end{eqnarray}

\noindent
which are independent of each other in the three-dimensional sub-space-time orthogonal to $\ve{\sigma}$. 
Then, the unperturbed light ray is given by Eq.~(\ref{Unperturbed_Lightray_2}) and the operator 
\begin{eqnarray}
        P^{ij} &=& \delta^{ij} - \sigma^i \sigma^j
        \label{Projection_Operator}
\end{eqnarray}

\noindent
is a projection operator onto the plane perpendicular to vector $\ve{\sigma}$. By inserting the unperturbed light ray (\ref{Unperturbed_Lightray_2}) and the projector 
(\ref{Projection_Operator}) into (\ref{Parameter2}), one may identify the auxiliary variable $\ve{\hat{\xi}}$ as impact vector $\ve{d}_{\sigma}$ of the unperturbed light ray defined 
by Eq.~(\ref{impact_vector}). 

The unperturbed light ray $\ve{x}_{\rm N}$ and its absolute value $r_{\rm N} = \left|\ve{x}_{\rm N}\right|$ can be parametrized in terms of these new variables and take the form
\begin{eqnarray}
        \ve{x}_{\rm N} &=& \ve{\hat{\xi}} + c \tau\,\ve{\sigma}\;,
        \label{Parameter3}
        \\
        r_{\rm N} &=& \sqrt{\hat{\xi}^2 + c^2 \tau^2}\;.
        \label{Parameter4}
\end{eqnarray}

\noindent 
The three-vector $\ve{\hat{\xi}}$ is laying in the plane perpendicular to $\ve{\sigma}$ and, therefore, only two of its three components are independent. 
This fact leads to the following partial derivatives (cf. Eq.~(23) in \cite{Kopeikin1997}, see also Eq.~(11.2.12) in \cite{Book_Soffel_Han})
\begin{eqnarray}
        \frac{\partial \hat{\xi}^i}{\partial \hat{\xi}^j} &=& P^i_j = P^{ij} = P_{ij}\;.
\label{Differentiation_5}
\end{eqnarray}

\noindent
Then, the spatial derivatives in (\ref{Metric_00}), when transformed in terms of these new variables, are given by (cf. Eq.~(20) in \cite{Kopeikin1997})
\begin{eqnarray}
        \frac{\partial}{\partial x^i} &=& \frac{\partial}{\partial \hat{\xi}^i} + \sigma_i\,\frac{\partial}{\partial c \tau}\;.  
        \label{spatial_derivative_1}
\end{eqnarray}

\noindent 
In practical calculations it is often more appropriate to consider the three spatial components of three-vector $\ve{\hat{\xi}}$ as independent of each other 
(cf. text above Eq.~(31) in \cite{KopeikinSchaefer1999_Gwinn_Eubanks}). This three-vector is denoted by $\ve{\xi}$. The parameters $c\tau$ and $\ve{\xi}$ are 
independent of each other in the four-dimensional space-time. Using $\ve{\xi}$ one gets the familiar result (see also Eq.~(11.2.13) in \cite{Book_Soffel_Han}) 
\begin{eqnarray}
        \frac{\partial \xi^i}{\partial \xi^j} &=& \delta^i_j = \delta^{ij} = \delta_{ij}\;, 
\label{Differentiation_10}
\end{eqnarray}

\noindent
with subsequent projection into the two-dimensional plane perpendicular to three-vector $\ve{\sigma}$, 
\begin{eqnarray}
        P^k_j\,\frac{\partial \xi^i}{\partial \xi^k} &=& P^i_j = P^{ij} = P_{ij}\;. 
\label{Differentiation_15}
\end{eqnarray}

\noindent 
That means, for the spatial derivatives, when transformed into derivatives expressed in terms of these new variables, one obtains 
\begin{eqnarray}
        \frac{\partial}{\partial x^i} &=& P_i^j\,\frac{\partial}{\partial \xi^j} + \sigma_i\,\frac{\partial}{\partial c \tau}\;. 
        \label{spatial_derivative_2}
\end{eqnarray}

\noindent
This relation coincides with Eq.~(33) in \cite{KopeikinSchaefer1999_Gwinn_Eubanks} in case of time-independent functions. It is noticed that (\ref{spatial_derivative_2}) in 
combination with (\ref{Differentiation_10}) is identical with (\ref{spatial_derivative_1}) in combination with (\ref{Differentiation_5}). Here, we will prefer this procedure 
in (\ref{spatial_derivative_2}), that means we will consider the spatial components of $\ve{\xi}$ as three independent components with a subsequent projection onto the 
two-dimensional plane perpendicular to the three-vector $\ve{\sigma}$. Then, using (\ref{spatial_derivative_2}) and the binomial theorem, 
\begin{eqnarray}
	\left(a + b\right)^l &=& \sum\limits_{p=0}^l {l \choose p} a^{l-p}\,b^{p}\;,  
        \label{binomial_theorem_1}
\end{eqnarray}

\noindent
where 
\begin{eqnarray}
        {l \choose p} &=& \frac{l!}{p! \left(l - p\right)!}
        \label{binomial_coefficients}
\end{eqnarray}

\noindent
are the binomial coefficients, one finds for these $l$ partial derivatives in (\ref{Partial_Derivatives}), when expressed in terms of these new variables, 
the following expression, 
\begin{eqnarray} 
	\widehat{\partial}^{\tau}_{L} &=& \underset{i_1 \dots i_l}{\rm STF} \sum\limits_{p=0}^{l} {l \choose p} \, 
\sigma_{i_1}\,\dots\,\sigma_{i_p}\,P_{i_{p+1}}^{j_{p+1}}\, \dots \,P_{i_l}^{j_l}  
\nonumber\\ 
&& \times \frac{\partial}{\partial \xi^{j_{p+1}}}\, \dots \, 
\frac{\partial}{\partial \xi^{j_{l}}}\,\left(\frac{\partial}{\partial c\tau}\right)^p \;,  
\label{Transformation_Derivative}
\end{eqnarray}

\noindent 
where the upper index $\tau$ in $\widehat{\partial}^{\tau}_{L}$ refers to the differentiation with respect to the parameter $\tau$ on the right-hand 
side of (\ref{Transformation_Derivative}). 
The wide hat in $\widehat{\partial}_L$ indicates the STF operation with respect to the indices $L = i_1 \dots i_l$. In addition, the wide hat also indicates the 
difference between the differential operator of Eq.~(\ref{Transformation_Derivative}) from the differential operator of Eq.~(\ref{Partial_Derivatives}) \cite{Comment3}.  

\noindent 
The solution of the first integration (\ref{First_Integration_1PN}) and the second integration (\ref{Second_Integration_1PN}) in terms of these new variables reads 
\begin{eqnarray}
        \frac{\dot{\ve{x}}\left(\tau\right)}{c} &=& \ve{\sigma} 
        + \sum\limits_{l=0}^{\infty} \frac{\Delta \dot{\ve{x}}^{M_L}_{\rm 1PN}\left(\tau\right)}{c} 
        + \sum\limits_{l=1}^{\infty} \frac{\Delta \dot{\ve{x}}^{S_L}_{\rm 1.5PN}\left(\tau\right)}{c}\,, 
        \label{First_Integration_1PN_New}
        \\
        \ve{x}\left(\tau\right) &=& \ve{x}_{\rm N} 
	+ \sum\limits_{l=0}^{\infty} \left[\Delta \ve{x}^{M_L}_{\rm 1PN}\left(\tau\right) - \Delta \ve{x}^{M_L}_{\rm 1PN}\left(\tau_0\right)\right] 
        \nonumber\\	
	&& \hspace{0.52cm} + \sum\limits_{l=1}^{\infty} \left[\Delta\ve{x}^{S_L}_{\rm 1.5PN}\left(\tau\right) - \Delta\ve{x}^{S_L}_{\rm 1.5PN}\left(\tau_0\right)\right], 
        \label{Second_Integration_1PN_New}
\end{eqnarray}

\noindent
up to terms of the order ${\cal O}(c^{-4})$, and where the unperturbed light ray $\ve{x}_{\rm N}$ in (\ref{Second_Integration_1PN_New}) is given by Eq.~(\ref{Parameter3}). 

The light ray perturbations of the first integration (\ref{First_Integration_1PN_New}) and the second integration (\ref{Second_Integration_1PN_New}) of the geodesic equation have been 
obtained in \cite{Kopeikin1997} for the case of time-independent multipoles of a body at rest. The advanced integration techniques, originally introduced in \cite{Kopeikin1997}, have 
later been further developed for the case of light propagation in the gravitational fields of a body at rest with the full set of time-dependent multipoles 
\cite{KopeikinSchaefer1999_Gwinn_Eubanks,KopeikinKorobkovPolnarev2006,KopeikinKorobkov2005} in the first post-Minkowskian (weak-field) approximation. 

Here, we are interested in the post-Newtonian (weak-field slow-motion) scheme. The case of light propagation in the gravitational field of slowly moving bodies with the full set of time-dependent 
multipoles has been solved in the 1PN and 1.5PN approximation in our investigations \cite{Zschocke_1PN} and \cite{Zschocke_15PN}, respectively. The first integration of geodesic equation was presented 
by Eqs.~(110) and (111) in \cite{Zschocke_15PN} for the mass-multipoles and by Eq.~(113) in \cite{Zschocke_15PN} for the spin-multipoles. The second integration of geodesic equation was presented by 
Eqs.~(118) and (119) in \cite{Zschocke_15PN} for the mass-multipoles and by Eq.~(121) in \cite{Zschocke_15PN} for the spin-multipoles. If one considers a body at rest, that means the orbital velocity 
of the body vanishes, $\ve{v}=0$ (but the body can still be in slow rotational motions) and if one neglects terms that contain the first time-derivative of the mass-multipoles, $\dot{M}_L$ \cite{Comment2},  
then these solutions in \cite{Zschocke_15PN} transform into the following expressions for the light ray perturbations: 
\begin{eqnarray}
	\frac{\Delta \dot{x}^{i\,M_L}_{\rm 1PN}\left(\tau\right)}{c} &=& 
	- \frac{2G\hat{M}_{L}\left(t\right)}{c^2}\,\frac{\left(-1\right)^l}{l!}\,\widehat{\partial}^{\tau}_{L}\,\frac{\sigma^i}{r_{\rm N}}
        \nonumber\\
        && \hspace{-0.5cm} - \frac{2G\hat{M}_{L}\left(t\right)}{c^2}\,\frac{\left(-1\right)^l}{l!}\, 
	\widehat{\partial}^{\tau}_{L}\,\frac{\hat{\xi}^i}{r_{\rm N} - c \tau}\,\frac{1}{r_{\rm N}}\,,  
\label{First_Integration_5}
\\
\nonumber\\
	\frac{\Delta \dot{x}^{i\,S_L}_{\rm 1.5PN}\left(\tau\right)}{c} &=& - \frac{4G\hat{S}_{bL-1}\left(t\right)}{c^3} 
        \epsilon_{iab} \frac{\left(-1\right)^l\,l}{\left(l+1\right)!}\,  
	\widehat{\partial}^{\tau}_{aL-1} \frac{1}{r_{\rm N}}
        \nonumber\\
	&& \hspace{-2.25cm} -\,\frac{4G\hat{S}_{bL-1}\left(t\right)}{c^3} 
	\epsilon_{abc}\,\sigma^c\,\frac{\left(-1\right)^l\,l}{\left(l+1\right)!}\,  
	\widehat{\partial}^{\tau}_{aL-1} \frac{\hat{\xi}^i}{r_{\rm N} - c \tau} \frac{1}{r_{\rm N}}\,,
\label{First_Integration_6}
\end{eqnarray}

\noindent
for the first integration (\ref{First_Integration_1PN_New}), and 
\begin{eqnarray}
           \Delta x^{i\,M_L}_{\rm 1PN}\left(\tau\right) &=& \frac{2G\hat{M}_{L}\left(t\right)}{c^2}\,{\sigma}^i\,
           \frac{\left(-1\right)^l}{l!}\,
           \widehat{\partial}^{\tau}_{L}\,\ln \left(r_{\rm N} - c \tau\right)
\nonumber\\
\nonumber\\
          && - \frac{2G\hat{M}_{L}\left(t\right)}{c^2}\,\frac{\left(-1\right)^l}{l!}\,
	  \widehat{\partial}^{\tau}_{L}\,\frac{\hat{\xi}^i}{r_{\rm N} - c \tau}\,,  
\label{Second_Integration_5}
\\
\nonumber\\
        \Delta x^{i\,S_L}_{\rm 1.5PN}\left(\tau\right) &=& \frac{4G\hat{S}_{bL-1}\left(t\right)}{c^3} 
        \epsilon_{iab} \frac{\left(-1\right)^l\,l}{\left(l+1\right)!}\,  
	\widehat{\partial}^{\tau}_{aL-1} \ln \left(r_{\rm N} - c \tau\right) 
        \nonumber\\
        && \hspace{-1.25cm} -\,\frac{4G\hat{S}_{bL-1}\left(t\right)}{c^3} 
        \epsilon_{abc}\,\sigma^c\,\frac{\left(-1\right)^l\,l}{\left(l+1\right)!}\,  
	\widehat{\partial}^{\tau}_{aL-1} \frac{\hat{\xi}^i}{r_{\rm N} - c \tau}\,, 
\label{Second_Integration_6}
\end{eqnarray}

\noindent 
for the second integration (\ref{Second_Integration_1PN_New}). 
In case of time-independent multipoles these solutions in (\ref{First_Integration_5}) - (\ref{Second_Integration_6}) agree with \cite{Kopeikin1997}. 

The Eqs.~\ref{First_Integration_5}) - (\ref{Second_Integration_6}) are given in terms of auxiliary variables as defined by Eqs.~(\ref{Parameter1}) and (\ref{Parameter2}). 
After performing all differentiations in these expressions with respect to the auxiliary variables, one obtains the solutions of the first and second integration of geodesic equation 
of Eqs.~(\ref{First_Integration_1PN}) and (\ref{Second_Integration_1PN}) just by replacing $c\tau$ by $\ve{\sigma} \cdot \ve{x}_{\rm N}$ and $P^i_j\,\xi^j$ by $d^i_{\sigma}$. 
Only in the arguments of the multipoles, $\hat{M}_{L}\left(t\right)$ and $\hat{S}_{L}\left(t\right)$, have we kept the standard time-variable $t$, because the differential 
operators $\widehat{\partial}^{\tau}_{L}$ do not act on these multipoles.

\section{The unit tangent vector}\label{Section2}

The unit tangent vector of a light ray at the moment of reception at the spatial position of the observer in terms of these new variables is given by 
\begin{eqnarray}
	\ve{n} &=& \frac{\dot{\ve{x}}\left(\tau_1\right)}{\left|\dot{\ve{x}}\left(\tau_1\right)\right|}\,. 
        \label{vector_n_new}
\end{eqnarray}

\noindent
From Eq.~(\ref{light_deflection}) with the expressions in (\ref{light_deflection_M}) and (\ref{light_deflection_S}) follows the unit tangent vector   
\begin{eqnarray}
	\ve{n} &=& \ve{k} + \sum\limits_{l=0}^{\infty}\Delta \ve{n}_{M_L} + \sum\limits_{l=1}^{\infty}\Delta \ve{n}_{S_L}\,,
        \label{light_deflection_new}
\end{eqnarray}

\noindent 
with $\ve{n} \cdot \ve{n} = 1$ and where 
\begin{eqnarray} 
	\Delta \ve{n}_{M_L} &=& \ve{k} \times \left[\left(\frac{\Delta\dot{\ve{x}}^{M_L}_{\rm 1PN}\left(\tau_1\right)}{c}
        - \frac{\Delta\ve{x}^{M_L}_{\rm 1PN}\left(\tau_1,\tau_0\right)}{R}\right) \times \ve{k}\right], 
        \nonumber\\ 
        \label{light_deflection_M_new}
	\\ 
	\Delta \ve{n}_{S_L} &=& \ve{k} \times \left[\left(\frac{\Delta\dot{\ve{x}}^{S_L}_{\rm 1.5PN}\left(\tau_1\right)}{c}
        - \frac{\Delta\ve{x}^{S_L}_{\rm 1.5PN}\left(\tau_1,\tau_0\right)}{R}\right) \times \ve{k}\right],
        \nonumber\\
        \label{light_deflection_S_new}
\end{eqnarray}

\noindent
up to terms beyond 1.5PN approximation. The distance between source and observer, $R$, is given by 
\begin{eqnarray} 
	R &=& |\ve{x}_1 - \ve{x}_0| = \ve{k} \cdot \ve{x}_1 - \ve{k} \cdot \ve{x}_0 
	\nonumber\\
	&=& \sqrt{r_1^2 + r_0^2 - 2\,r_0\,r_1\,\cos \delta(\ve{x}_0,\ve{x}_1)} \;,  
	\label{Spatial_Distance_1}
\end{eqnarray} 

\noindent
where $\delta(\ve{x}_0,\ve{x}_1)$ is the angle between the three-vectors $\ve{x}_0$ and $\ve{x}_1$, while $r_0 = |\ve{x}_0|$ and $r_1 = |\ve{x}_1|$ 
are the distances between body and source and body and observer, respectively. 
The perturbations in (\ref{light_deflection_M_new}) and (\ref{light_deflection_S_new}) are given by Eqs.~(\ref{First_Integration_5}) - (\ref{Second_Integration_6}), 
where the differential operator is given by Eq.~(\ref{Transformation_Derivative}). In view of relation (\ref{transformation_sigma_to_k}) we have 
\begin{eqnarray}
	\ve{\sigma} &=& \ve{k} + {\cal O} \left(c^{-2}\right),
	\label{sigma_k_1}
	\\
	\ve{\sigma} \cdot \ve{k} &=& 1 + {\cal O} \left(c^{-4}\right), 
	\label{sigma_k_2}
\end{eqnarray}

\noindent
where the three-vector $\ve{\sigma}$ is defined by Eq.~(\ref{vector_sigma}) and the three-vector $\ve{k}$ is defined by Eq.~(\ref{vector_k}). Therefore, in the 
perturbations (\ref{First_Integration_5}) - (\ref{Second_Integration_6}) one may replace three-vector $\ve{\sigma}$ by three-vector $\ve{k}$ in line with the 
1.5PN approximation. In particular, $\ve{\sigma}$ is replaced by $\ve{k}$ in the projector (\ref{Projection_Operator}), in the new parameters (\ref{Parameter1}) 
and (\ref{Parameter2}), and in the distance $r_{\rm N}$ in (\ref{Parameter4}). Furthermore, we will specify these quantities at the spatial positions of source 
and observer, $\ve{x}_0$ and $\ve{x}_1$, as well as at the moment of emission and reception of the light signal, $t_0$ and $t_1$ by using 
\begin{eqnarray}
	\ve{x}_0 &=& \ve{x}_{\rm N}\left(t_0\right),
        \label{xN_x0} 
        \\
	\ve{x}_1 &=& \ve{x}_{\rm N}\left(t_1\right) + {\cal O} \left(c^{-2}\right),   
        \label{xN_x1}
\end{eqnarray}

\noindent
where the unperturbed light ray is given by Eq.~(\ref{Unperturbed_Lightray_2}) and in terms of new variables by Eq.~(\ref{Parameter3}). 
In this way we get the following expressions, which will from now on be used in the subsequent considerations: 
\begin{eqnarray}
        P^{ij} &=& \delta^{ij} - k^i k^j\,,
        \label{Projection_Operator_k}
        \\
	c \tau_0 &=& \ve{k} \cdot \ve{x}_0\,,
        \label{Parameter0_k}
        \\
	c \tau_1 &=& \ve{k} \cdot \ve{x}_1\,,
        \label{Parameter1_k}
        \\
	\hat{\xi}^i &=& P^i_j\,x_0^j = P^i_j\,x_1^j\,,  
        \label{Parameter2_k}
	\\
	r_0 &=& \sqrt{\hat{\xi}^2 + c^2 \tau_0^2}\,,
	\label{r_0}
	\\
	r_1 &=& \sqrt{\hat{\xi}^2 + c^2 \tau_1^2}\,, 
        \label{r_1}
\end{eqnarray}

\noindent
where the projector (\ref{Projection_Operator_k}) is an operator that projects a three-vector onto the plane perpendicular to vector $\ve{k}$ and (\ref{r_0}) and (\ref{r_1}) are 
the spatial distances between the center-of-mass of the body and the position of source and observer, respectively. In view of relation (\ref{sigma_k_1}), these expressions in 
Eqs.~(\ref{Projection_Operator_k}) - (\ref{r_1}) differ from the previously defined quantities of Eqs.~(\ref{Parameter1}) - (\ref{Parameter4}) by terms of the order ${\cal O}(c^{-2})$. 
In favor of a simpler notation, we are not introducing a new label for the projector of Eq.~(\ref{Projection_Operator_k}). By these replacements and for these specific values, the 
differential operator in (\ref{Transformation_Derivative}) becomes (for the specific values $\tau_0$ and $\tau_1$) 
\begin{eqnarray} 
	\widehat{\partial}^{\tau_0}_{L} &=& \widehat{\partial}_{L} + \widehat{\partial}^{\tau_0,p \neq 0}_{L}\,, 
\label{Differential_Operator_k_0}
\\
        \widehat{\partial}^{\tau_1}_{L} &=& \widehat{\partial}_{L} + \widehat{\partial}^{\tau_1,p \neq 0}_{L}\,,
\label{Differential_Operator_k_1}
\end{eqnarray}

\noindent 
where the differential operator term with $p=0$ reads  
\begin{eqnarray} 
	\widehat{\partial}_{L} &=& \underset{i_1 \dots i_l}{\rm STF} P_{i_{1}}^{j_{1}}\, \dots \,P_{i_l}^{j_l} 
	\frac{\partial}{\partial \xi^{j_1}}\,\dots\,\frac{\partial}{\partial \xi^{j_{l}}}\,,
\label{Differential_Operator_p_0}
\end{eqnarray}
 
\noindent
and the differential operator terms with $p \neq 0$ are  
\begin{eqnarray}
 \widehat{\partial}^{\tau_0,p \neq 0}_{L} &=& \underset{i_1 \dots i_l}{\rm STF} \sum\limits_{p=1}^{l} {l \choose p}\,
k_{i_1}\,\dots\,k_{i_p}\,P_{i_{p+1}}^{j_{p+1}}\, \dots \,P_{i_l}^{j_l}
\nonumber\\
&& \times \frac{\partial}{\partial \xi^{j_{p+1}}}\, \dots \,
\frac{\partial}{\partial \xi^{j_{l}}}\,\left(\frac{\partial}{\partial c\tau_0}\right)^p \,,
\label{Differential_Operator_tau_0_p}
\\
 \widehat{\partial}^{\tau_1,p \neq 0}_{L} &=& \underset{i_1 \dots i_l}{\rm STF} \sum\limits_{p=1}^{l} {l \choose p}\,
k_{i_1}\,\dots\,k_{i_p}\,P_{i_{p+1}}^{j_{p+1}}\, \dots \,P_{i_l}^{j_l}
\nonumber\\
&& \times \frac{\partial}{\partial \xi^{j_{p+1}}}\, \dots \,
\frac{\partial}{\partial \xi^{j_{l}}}\,\left(\frac{\partial}{\partial c\tau_1}\right)^p \,.
\label{Differential_Operator_tau_1_p}
\end{eqnarray}

\noindent 
Similarly, because of relation (\ref{sigma_k_1}), these differential operators of Eqs.~(\ref{Differential_Operator_k_0}) and (\ref{Differential_Operator_k_1}) 
differ from the previously defined differential operator of Eq.~(\ref{Transformation_Derivative}) by terms of the order ${\cal O}(c^{-2})$, which is in line 
with the 1.5PN approximation. 

The unit tangent vector of the light trajectory at the observers position is given by Eq.~(\ref{light_deflection_new}) with the mass-multipole and spin-multipole terms 
of Eqs.~(\ref{light_deflection_M_new}) and (\ref{light_deflection_S_new}). 
The mass-multipole term in the first lines of Eqs.~(\ref{First_Integration_5}) and (\ref{Second_Integration_5}) do not contribute to this unit tangent vector, 
as one can see by inserting these terms into Eq.~(\ref{light_deflection_M_new}) and taking account of Eq.~(\ref{sigma_k_1}). Accordingly, by inserting (\ref{First_Integration_5}) 
and (\ref{Second_Integration_5}) into (\ref{light_deflection_M_new}) one obtains three mass-multipole terms, 
\begin{eqnarray} 
	\Delta \ve{n}_{M_L} &=& \Delta\ve{n}^1_{M_L} + \Delta \ve{n}^2_{M_L} + \Delta \ve{n}^3_{M_L}\,,  
	\label{delta_n_M}
\end{eqnarray}

\noindent
with the spatial components ($i=1,2,3$) 
\begin{eqnarray}
	\Delta n^{1\,i}_{M_L} &=& - \frac{2 G \hat{M}_{L}\left(t_1\right)}{c^2} \frac{\left(-1\right)^l}{l!}\, 
	\widehat{\partial}^{\tau_1}_{L}\,\frac{\hat{\xi}^i}{r_1 - c \tau_1} \frac{1}{r_1}\,,
	\label{delta_k1_M}
	\\
	\Delta n^{2\,i}_{M_L} &=& + \frac{2 G \hat{M}_{L}\left(t_1\right)}{c^2} \frac{\left(-1\right)^l}{l!}\,\frac{1}{R}\,
	\widehat{\partial}^{\tau_1}_{L}\,\frac{\hat{\xi}^i}{r_1 - c \tau_1}\,,
        \label{delta_k2_M}
	\\
	\Delta n^{3\,i}_{M_L} &=& - \frac{2 G \hat{M}_{L}\left(t_0\right)}{c^2} \frac{\left(-1\right)^l}{l!}\,\frac{1}{R}\, 
	\widehat{\partial}^{\tau_0}_{L}\,\frac{\hat{\xi}^i}{r_0 - c \tau_0}\,,
        \label{delta_k3_M}
\end{eqnarray}

\noindent 
where the differential operators $\widehat{\partial}^{\tau_0}_{L}$ and $\widehat{\partial}_{L}^{\tau_1}$ are given by Eqs.~(\ref{Differential_Operator_k_0}) 
and (\ref{Differential_Operator_k_1}). The double cross product in (\ref{light_deflection_M_new}) is identical with Eqs.~(\ref{delta_k1_M}) - (\ref{delta_k3_M}), 
because the vectorial expressions in the parentheses are orthogonal to three-vector $\ve{k}$. 

Similarly, by inserting (\ref{First_Integration_6}) and (\ref{Second_Integration_6}) into (\ref{light_deflection_S_new}) one obtains six spin-multipole terms, 
\begin{eqnarray} 
	\Delta \ve{n}_{S_L} &=& \sum\limits_{k=1}^{6} \Delta\ve{n}^k_{S_L}\,, 
        \label{delta_n_S}
\end{eqnarray}

\noindent
where the spatial components ($i=1,2,3$) of the first three terms read  
\begin{eqnarray}
	\Delta n^{1\,i}_{S_L} &=& - \frac{4 G \hat{S}_{bL-1}\left(t_1\right)}{c^3} 
        \epsilon_{abc}\,k^c\,\frac{\left(-1\right)^l\,l}{\left(l+1\right)!}\,  
	\widehat{\partial}^{\tau_1}_{aL-1} \frac{\hat{\xi}^i}{r_1 - c \tau_1} \frac{1}{r_1}\,,
	\nonumber\\ 
        \label{delta_k1_S}
        \\
	\Delta n^{2\,i}_{S_L} &=& + \frac{4 G \hat{S}_{bL-1}\left(t_1\right)}{c^3} 
	\epsilon_{abc}\,k^c\,\frac{\left(-1\right)^l\,l}{\left(l+1\right)!}\,\frac{1}{R}\,
	\widehat{\partial}^{\tau_1}_{aL-1} \frac{\hat{\xi}^i}{r_1 - c \tau_1}\,, 
	\nonumber\\ 
        \label{delta_k2_S}
        \\
	\Delta n^{3\,i}_{S_L} &=& - \frac{4 G \hat{S}_{bL-1}\left(t_0\right)}{c^3} 
        \epsilon_{abc}\,k^c\,\frac{\left(-1\right)^l\,l}{\left(l+1\right)!}\,\frac{1}{R}\,
	\widehat{\partial}^{\tau_0}_{aL-1} \frac{\hat{\xi}^i}{r_0 - c \tau_0}\,, 
	\nonumber\\ 
        \label{delta_k3_S}
\end{eqnarray}

\noindent 
while the spatial components ($i=1,2,3$) of the last three terms are given by 
\begin{eqnarray}
        \Delta n^{4\,i}_{S_L} &=& - \frac{4 G \hat{S}_{bL-1}\left(t_1\right)}{c^3}\, 
        \widehat{\epsilon}_{iab}\,\frac{\left(-1\right)^l\,l}{\left(l+1\right)!}\,  
        \left(\widehat{\partial}^{\tau_1}_{aL-1}\,\frac{1}{r_1}\right),
        \label{delta_k4_S}
        \\
        \Delta n^{5\,i}_{S_L} &=& - \frac{4 G \hat{S}_{bL-1}\left(t_1\right)}{c^3}\, 
        \widehat{\epsilon}_{iab}\,\frac{\left(-1\right)^l\,l}{\left(l+1\right)!}\,\frac{1}{R}\,
        \widehat{\partial}^{\tau_1}_{aL-1} \ln \left(r_1 - c \tau_1\right), 
        \nonumber\\ 
        \label{delta_k5_S}
        \\
        \Delta n^{6\,i}_{S_L} &=& + \frac{4 G \hat{S}_{bL-1}\left(t_0\right)}{c^3}\, 
        \widehat{\epsilon}_{iab}\,\frac{\left(-1\right)^l\,l}{\left(l+1\right)!}\,\frac{1}{R}\,
        \widehat{\partial}^{\tau_0}_{aL-1} \ln \left(r_0 - c \tau_0\right), 
        \nonumber\\ 
        \label{delta_k6_S}
\end{eqnarray}

\noindent
where we use the abbreviation $\widehat{\epsilon}_{iab} = \epsilon_{iab} - k_i\,\epsilon_{jab}\,k^j$, originating from the double cross product
in (\ref{light_deflection_S_new}). The differential operators were given by Eqs.~(\ref{Differential_Operator_k_0}) and (\ref{Differential_Operator_k_1}). 
These Eqs.~(\ref{delta_k1_M}) - (\ref{delta_k3_M}) and (\ref{delta_k1_S}) - (\ref{delta_k6_S}) are valid for an arbitrary body.

\section{On the magnitude of terms}\label{Section_Magnitude} 

The unit tangent vector of a light trajectory in the gravitational field of an arbitrary body is given by Eq.~(\ref{light_deflection_new}),
with the mass-multipole perturbations (\ref{light_deflection_M_new}) and spin-multipole perturbations (\ref{light_deflection_S_new}).
These perturbations are given in their explicit form by Eqs.~(\ref{delta_n_M}) - (\ref{delta_k3_M}) and Eqs.~(\ref{delta_n_S}) - (\ref{delta_k6_S}).
In the next Sections it will be demonstrated that these individual terms of the tangent vector of light trajectory are of qualitatively different magnitudes. 
In particular, for the mass-multipole terms of Eqs.~(\ref{delta_k1_M}) - (\ref{delta_k3_M}) we may distinguish three kinds of magnitudes,  
\begin{eqnarray}
	&& \epsilon^{M_L}_1 = A^{M_L}\,\frac{GM}{c^2}\,\frac{1}{d_k}\,J_l\,\left(\frac{P}{d_k}\right)^l \sim \muas\,,
	\label{magnitude_M_1}
	\\
	&& \epsilon^{M_L}_2 = B^{M_L}\,\frac{GM}{c^2}\,\frac{1}{r_1}\,J_l\,\left(\frac{P}{d_k}\right)^l \sim {\rm nas}\,,
        \label{magnitude_M_2}
	\\
	&& \epsilon^{M_L}_3 = C^{M_L}\,\frac{GM}{c^2}\,\frac{1}{r_1}\,J_l\,\left(\frac{P}{r_1}\right)^l \sim {\rm pas}\,,
        \label{magnitude_M_3}
\end{eqnarray}

\noindent
where $\muas$, ${\rm nas}$, and ${\rm pas}$ are micro- nano-, and pico-arcseconds; see also Appendix~\ref{Appendix0}. 
Similarly, for the spin-multipole terms of Eqs.~(\ref{delta_k1_S}) - (\ref{delta_k6_S}) we may also distinguish between three kind of magnitudes, 
\begin{eqnarray}
	&& \epsilon^{S_L}_1 = A^{S_L}\,\frac{GM}{c^3}\,\Omega\,J_{l-1}\,\left(\frac{P}{d_k}\right)^{l+1} \sim \muas \,,
        \label{magnitude_S_1}
        \\
	&& \epsilon^{S_L}_2 = B^{S_L}\,\frac{GM}{c^3}\,\Omega\,J_{l-1}\,\left(\frac{P}{r_1}\right) \left(\frac{P}{d_k}\right)^l \sim {\rm nas}\,,
        \label{magnitude_S_2}
        \\
	&& \epsilon^{S_L}_3 = C^{S_L}\,\frac{GM}{c^3}\,\Omega\,J_{l-1}\,\left(\frac{P}{r_1}\right)^{l+1} \sim {\rm pas}\,.
        \label{magnitude_S_3}
\end{eqnarray}

\noindent 
The coefficients $A^{M_L}, B^{M_L}, C^{M_L}$ and $A^{S_L}, B^{S_L}, C^{S_L}$ will be determined in the next Sections. The parameters  
$M$, $P$, $J_l$, and $\Omega$ are the mass, equatorial radius, zonal harmonic coefficient, and angular velocity of the solar system body, 
while $d_k$ is the impact parameter of the unperturbed light ray with respect to the body and $r_1$ is the distance between body and observer; 
a graphical elucidation is shown in Figure~\ref{Diagram1}. Because the impact of spin-multipoles of order $l$ on light deflection is a about $100$ 
times smaller than the impact of mass-multipoles of order $l+1$ on light deflection, relation (\ref{magnitude_S_2}) means actually less than $1\,{\rm nas}$. 

The terms of the unit tangent vector of the light ray, which are of the order (\ref{magnitude_M_1}) and (\ref{magnitude_S_1}), are the relevant terms for 
astrometry on the sub-\muas{} level. They are given below by Eqs.~(\ref{hat_M_L}) - (\ref{Simplified_Tangent_Vector_6}) and their upper limits are presented by 
Eqs.~(\ref{upper_limit_M_L}) - (\ref{upper_limit_S_1}). 

However, the unit tangent vector of the light ray contains also many terms that are of the order of Eqs.(\ref{magnitude_M_2}) - (\ref{magnitude_M_3}) and  
Eqs.~(\ref{magnitude_S_2}) - (\ref{magnitude_S_3}), which are suppressed by the ratio of equatorial radius of the body and the distance between body and observer. 
For instance, this ratio is about $5 \times 10^{-3}$ in case of the Sun and about $10^{-4}$ in case of Jupiter. 
Because of these small prefactors, the numerical magnitude of these terms to the effect of light deflection is at most a few nano-arcseconds or even less.  
Especially, for astrometry on the sub-micro-arcsecond level one may neglect all those terms that contribute only a few nano-arcseconds (nas) or a few pico-arcseconds (pas) 
to the angle of light deflection. Thus, by keeping only terms of the magnitude $\epsilon^{M_L}_1$ and $\epsilon^{S_L}_1$ and neglecting all terms of the magnitudes 
$\epsilon^{M_L}_2,\epsilon^{M_L}_3$ and $\epsilon^{S_L}_2,\epsilon^{S_L}_3$ one arrives at a considerably simplified unit tangent vector of the light ray. 
This simplified tangent vector is important for highly effective data reduction of future astrometric measurements on the sub-micro-arcsecond level of accuracy and 
will be derived in the subsequent Sections~\ref{Simplified_1} and \ref{Simplified_2}.

\section{The reduced tangent vector}\label{Simplified_1}

The unit tangent vector of the light ray in 1.5PN approximation is given by Eq.~(\ref{light_deflection_new}) with the mass-multipole terms given by Eq.~(\ref{delta_n_M}) with the 
terms in (\ref{delta_k1_M}) - (\ref{delta_k3_M}) and the spin-multipole terms of the unit tangent vector are given by Eq.~(\ref{delta_n_S}) with the terms 
in (\ref{delta_k1_S}) - (\ref{delta_k6_S}). As asserted in the previous Section, these expressions are much too complicated for an effective treatment of astrometric data reduction, 
because they contain many terms that are below the given threshold of $0.01\,\muas$ ($10$ nas) of sub-micro-arcsecond astrometry. To identify these tiny terms we separate our 
considerations into two Sections: in this Section~\ref{Simplified_1} we will arrive at the reduced tangent vector, which is a step in between toward the simplified tangent vector 
which is finally obtained in Section~\ref{Simplified_2}. The mass-multipole and spin-multipole terms are considered separately in the following two Subsections.

\subsection{The mass-multipole terms}

The mass-multipole terms of the unit tangent vector are given by Eq.~(\ref{delta_n_M}) with the terms in (\ref{delta_k1_M}) - (\ref{delta_k3_M}). 
According to (\ref{Differential_Operator_k_0}) and (\ref{Differential_Operator_k_1}), we may separate these perturbations into terms with parameter $p=0$
and with parameter $p \neq 0$. Then, the mass-multipole perturbations read
\begin{eqnarray} 
        \Delta \ve{n}_{M_L} &=& \Delta\ve{n}^{1,p=0}_{M_L} + \Delta \ve{n}^{2,p=0}_{M_L} + \Delta \ve{n}^{3,p=0}_{M_L}
        \nonumber\\
        &+& \Delta\ve{n}^{1,p \neq 0}_{M_L} + \Delta \ve{n}^{2,p \neq 0}_{M_L} + \Delta \ve{n}^{3,p \neq 0}_{M_L}\,.
        \label{delta_n_M_Simplified}
\end{eqnarray}

\noindent 
The terms in the first line of (\ref{delta_n_M_Simplified}) read  
\begin{eqnarray}
        \Delta \ve{n}^{1,p=0}_{M_L} &=& - \frac{2 G \hat{M}_{L}\left(t_1\right)}{c^2} \frac{\left(-1\right)^l}{l!}\, 
        \widehat{\partial}_{L}\,\frac{\ve{\hat{\xi}}}{r_1 - c \tau_1} \frac{1}{r_1},
        \label{delta_k1_M_p_0}
        \\
        \Delta \ve{n}^{2,p=0}_{M_L} &=& + \frac{2 G \hat{M}_{L}\left(t_1\right)}{c^2} \frac{\left(-1\right)^l}{l!}\,\frac{1}{R}\,
        \widehat{\partial}_{L}\,\frac{\ve{\hat{\xi}}}{r_1 - c \tau_1},
        \label{delta_k2_M_p_0}
        \\
        \Delta \ve{n}^{3,p=0}_{M_L} &=& - \frac{2 G \hat{M}_{L}\left(t_0\right)}{c^2} \frac{\left(-1\right)^l}{l!}\,\frac{1}{R}\, 
        \widehat{\partial}_{L}\,\frac{\ve{\hat{\xi}}}{r_0 - c \tau_0},
        \label{delta_k3_M_p_0}
\end{eqnarray}

\noindent
where $l\ge 0$ and the differential operator is given by Eq.~(\ref{Differential_Operator_p_0}). 
The terms in the second line of (\ref{delta_n_M_Simplified}) read 
\begin{eqnarray}
        \Delta \ve{n}^{1,p \neq 0}_{M_L} &=& - \frac{2 G \hat{M}_{L}\left(t_1\right)}{c^2} \frac{\left(-1\right)^l}{l!}\, 
        \widehat{\partial}^{\tau_1, p \neq 0}_{L} \frac{\ve{\hat{\xi}}}{r_1 - c \tau_1} \frac{1}{r_1},
        \label{delta_k1_M_p_neq_0}
        \\
        \Delta \ve{n}^{2,p \neq 0}_{M_L} &=& + \frac{2 G \hat{M}_{L}\left(t_1\right)}{c^2} \frac{\left(-1\right)^l}{l!} \frac{1}{R}\,
        \widehat{\partial}^{\tau_1, p \neq 0}_{L} \frac{\ve{\hat{\xi}}}{r_1 - c \tau_1},
        \label{delta_k2_M_p_neq_0}
        \\
        \Delta \ve{n}^{3,p \neq 0}_{M_L} &=& - \frac{2 G \hat{M}_{L}\left(t_0\right)}{c^2} \frac{\left(-1\right)^l}{l!} \frac{1}{R}\, 
        \widehat{\partial}^{\tau_0, p \neq 0}_{L} \frac{\ve{\hat{\xi}}}{r_0 - c \tau_0}, 
        \label{delta_k3_M_p_neq_0}
\end{eqnarray}

\noindent 
where $l \ge 2$ and the differential operators are given by Eqs.~(\ref{Differential_Operator_tau_0_p}) and (\ref{Differential_Operator_tau_1_p}). 
In Appendixes~\ref{Proof1} - \ref{Proof2} it is shown that the upper limit of the term of Eq.~(\ref{delta_k1_M_p_neq_0}) and the upper limit 
of the sum of Eqs.~(\ref{delta_k2_M_p_neq_0}) and (\ref{delta_k3_M_p_neq_0}) to the angle of light deflection is given by 
\begin{eqnarray} 
        \left|\delta\left(\ve{k},\Delta \ve{n}^{1,p\neq 0}_{M_L}\right)\right| &=& C_1^{M_L}\frac{G M}{c^2 r_1} \left|J_l\right| \left(\frac{P}{r_1}\right)^l,
        \label{scaling_delta_k1_M_p_neq_0} 
        \\
        \left|\delta\left(\ve{k},\Delta\ve{n}^{2+3,p\neq 0}_{M_L}\right)\right| &\le& B_1^{M_L}\,\frac{G M}{c^2 r_1} \left|J_{l}\right| 
        \left(\frac{P}{d_k}\right)^l .  
        \label{scaling_delta_k2_k3_M_p_neq_0}
\end{eqnarray}

\noindent 
In order to get these upper limits (\ref{scaling_delta_k1_M_p_neq_0}) - (\ref{scaling_delta_k2_k3_M_p_neq_0}) we have assume the body to be an axi-symmetric body. 
The coefficients of these equations have been determined in the Appendixes~\ref{Proof1} and \ref{Proof2}. They are given in Table~\ref{Table_C_1_M} of Appendix~\ref{Proof1} 
and in Table~\ref{Table_B_2_M} of Appendix~\ref{Proof2}. 

The term (\ref{scaling_delta_k1_M_p_neq_0}) is of the order of Eq.~(\ref{magnitude_M_3}) and contributes much less than $1\,{\rm nas}$, 
\begin{eqnarray} 
	\left|\delta\left(\ve{k},\Delta \ve{n}^{1,p\neq 0}_{M_L}\right)\right| &\ll& 1\,{\rm nas}\,, 
	\label{scaling_delta_k1_M_p_neq_0_magnitude}
\end{eqnarray}

\noindent 
while the term (\ref{scaling_delta_k2_k3_M_p_neq_0}) is of the order of Eq.~(\ref{magnitude_M_2}).  
The terms of Eqs.~(\ref{delta_k2_M_p_neq_0}) and (\ref{delta_k3_M_p_neq_0}) are much larger than the term of Eq.~(\ref{delta_k1_M_p_neq_0}). The reason is that the 
distance $R$ between source and observer appears in the denominator of these terms. This distance can in principle be arbitrarily small. Therefore, one has to 
estimate the sum of Eqs.~(\ref{delta_k2_M_p_neq_0}) and (\ref{delta_k3_M_p_neq_0}), which is a finite quantity, but considerably larger than (\ref{delta_k1_M_p_neq_0}). 
Nevertheless, according to  (\ref{scaling_delta_k2_k3_M_p_neq_0}) these terms contribute at most a few ${\rm nas}$. 

Numerical values of these terms of Eq.~(\ref{scaling_delta_k2_k3_M_p_neq_0}) 
are presented in Table~\ref{Table_delta_3_4_5} for the case of grazing rays at Jupiter and Saturn. For all other solar system bodies the magnitude of these terms is even less. 
Such tiny effects are negligible for astrometry missions like {\it GaiaNIR} \cite{Gaia_NIR} or {\it Theia} \cite{Theia}, aiming at astrometry on the sub-\muas{} scale of accuracy, 
that means aiming at an accuracy of about $0.1\,\muas$. Even for future astrometry missions aiming at an accuracy of about $0.01\,\muas$ ($10$ nas) these terms remain negligible. 
\begin{table}[t]
        \caption{The contribution of the terms of Eq.~(\ref{scaling_delta_k2_k3_M_p_neq_0}) to the angle of light deflection 
	in case of a grazing ray at Jupiter ($\jupiter$) and Saturn ($\saturn$). For all other solar system bodies these terms are even smaller. All values are given in nano-arcseconds (nas). 
	A blank entry means less than $1\,{\rm nas}$.}
\begin{tabular}{| c | c | c |} 
\hline
	&&\\[-12pt]
	$l$ &\hbox to 30mm{\hfill $|\delta_{\jupiter}(\ve{k},\Delta \ve{n}^{2+3, p\neq 0}_{M_L})|$ \hfill} &\hbox to 30mm{\hfill $|\delta_{\saturn}(\ve{k},\Delta \ve{n}^{2+3, p\neq 0}_{M_L})|$ \hfill} \\[3pt]
\hline
	&&\\[-12pt]
	$2$ & $20.3$ & $6.7$ \\[3pt]
	$4$ & $6.7$  & $3.2$ \\[3pt]
	$6$ & $1.7$  & $1.3$ \\[3pt]
	$8$ & $0.5$ & $0.6$ \\[3pt]
	$10$& $0.2$ & $0.5$ \\[3pt]
\hline
\end{tabular}
\label{Table_delta_3_4_5}
\end{table}

\subsection{The spin-multipole terms}

The spin-multipole terms of the unit tangent vector are given by Eq.~(\ref{delta_n_S}) with the terms (\ref{delta_k1_S}) - (\ref{delta_k6_S}). 
Before we proceed further, it is useful to consider the magnitude of the terms of Eqs.~(\ref{delta_k4_S}) - (\ref{delta_k6_S}) at this stage, because these terms
turn out to be negligibly small. The upper limit of the term of Eq.~(\ref{delta_k4_S}) and the sum of Eqs.~(\ref{delta_k5_S}) and (\ref{delta_k6_S}) to the angle of light 
deflection are given by 
\begin{eqnarray}
	&& \hspace{-0.5cm}\left|\delta\left(\ve{k},\Delta \ve{n}^{4}_{S_L}\right)\right| \le C_1^{S_L} \frac{G M}{c^3}\,\Omega\,\left|J_{l-1}\right| \left(\frac{P}{r_1}\right)^{l+1},
        \label{scale_delta_k4_S}
        \\
	&& \hspace{-0.5cm} \left|\delta\left(\ve{k},(\Delta\ve{n}^{5+6}_{S_L}\right)\right| \le B_1^{S_L} \frac{G M}{c^3}\,\Omega\,\left|J_{l-1}\right| 
	\left(\frac{P}{r_1}\right) \left(\frac{P}{d_k}\right)^l. 
        \label{scale_delta_k5_k6_S}
\end{eqnarray}
 
\noindent
These terms of Eqs.~(\ref{scale_delta_k4_S}) and (\ref{scale_delta_k5_k6_S}) are terms of the order (\ref{magnitude_S_3}) and (\ref{magnitude_S_2}), respectively. A proof 
of these relations is obtained by similar considerations as shown in Appendixes~\ref{Proof1} and \ref{Proof2} for the mass-multipole terms. But actually, only the very first 
few spin-multipoles are relevant on the sub-\muas-level: the spin-dipole term $l=1$ and the spin-hexapole term $l=3$. To get an idea about 
the magnitude of these terms we consider the case of spin-dipole, which is the most dominating term. Inserting the spin-dipole (\ref{S}) and the differential 
operators (\ref{Differential_Operator_k_0}) and (\ref{Differential_Operator_k_1}) into Eqs.~(\ref{delta_k4_S}) - (\ref{delta_k6_S}) yields the coefficients for the case of 
spin-dipole: $C_1^{S_1} = 2\,\kappa^2$ and $B_1^{S_1} = 2\,\kappa^2$; for a proof see Appendix~\ref{Appendix_Spin}. 
In order to obtain these upper limits (\ref{scale_delta_k4_S}) - (\ref{scale_delta_k5_k6_S}) we assume the body to be an axi-symmetric body in uniform rotation. 
By inserting the parameters of Table~\ref{Table1} into (\ref{scale_delta_k4_S}) and (\ref{scale_delta_k5_k6_S}), 
one obtains for grazing light rays at any solar system body, that these upper limits contribute much less than $1\,{\rm nas}$ to the angle of light deflection. It is clear that the 
contribution of higher spin-multipoles is less than the spin-dipole term. Thus, the contribution of these terms of Eqs.~(\ref{delta_k4_S}) - (\ref{delta_k6_S}) to the angle of 
light deflection is much less than $1\,{\rm nas}$ to any spin-multipole order, 
\begin{eqnarray}
	\left|\delta\left(\ve{k},\Delta \ve{n}^{4}_{S_L}\right)\right| &\ll& 1\,{\rm nas}\,,
        \label{numerical_magnitude_Spin_Multipole_1}
        \\
        \left|\delta\left(\ve{k},(\Delta\ve{n}^{5+6}_{S_L}\right)\right| &\ll& 1\,{\rm nas} \,. 
	\label{numerical_magnitude_Spin_Multipole_2}
\end{eqnarray}

\noindent 
Therefore, these terms of Eqs.~(\ref{delta_k4_S}) - (\ref{delta_k6_S}) can be neglected for astrometric measurements on the sub-micro-arcsecond level. 
Accordingly, only the following three spin-multipole terms are considered,
\begin{eqnarray} 
        \Delta \ve{n}_{S_L} &=& \Delta\ve{n}^1_{S_L} + \Delta\ve{n}^2_{S_L} + \Delta\ve{n}^3_{S_L}\,, 
        \label{delta_n_S_new}
\end{eqnarray}

\noindent
with the expressions of Eqs.~(\ref{delta_k1_S}) - (\ref{delta_k3_S}). 
According to (\ref{Differential_Operator_k_0}) and (\ref{Differential_Operator_k_1}), we may separate these perturbations into terms with parameter $p=0$
and with parameter $p \neq 0$. Then, the spin-multipole perturbations read
\begin{eqnarray} 
        \Delta \ve{n}_{S_L} &=& \Delta\ve{n}^{1,p=0}_{S_L} + \Delta\ve{n}^{2,p=0}_{S_L} + \Delta\ve{n}^{3,p=0}_{S_L}
        \nonumber\\
        &+& \Delta\ve{n}^{1,p \neq 0}_{S_L} + \Delta\ve{n}^{2,p \neq 0}_{S_L} + \Delta\ve{n}^{3,p \neq 0}_{S_L}. 
        \label{delta_n_S_Simplified}
\end{eqnarray}

\noindent
The terms in the first line of (\ref{delta_n_S_Simplified}) read 
\begin{eqnarray}
        \Delta \ve{n}^{1,p=0}_{S_L} &=& - \frac{4 G \hat{S}_{bL-1}\!\left(t_1\right)}{c^3} 
        \epsilon_{abc} k^c \frac{\left(-1\right)^l l}{\left(l+1\right)!}   
        \widehat{\partial}_{aL-1} \frac{\ve{\hat{\xi}}}{r_1 - c \tau_1} \frac{1}{r_1},
        \nonumber\\ 
        \label{delta_k1_S_p_0}
        \\
        \Delta \ve{n}^{2,p=0}_{S_L} &=& + \frac{4 G \hat{S}_{bL-1}\!\left(t_1\right)}{c^3} 
        \epsilon_{abc} k^c\,\frac{\left(-1\right)^l l}{\left(l+1\right)!} \frac{1}{R} 
        \widehat{\partial}_{aL-1} \frac{\ve{\hat{\xi}}}{r_1 - c \tau_1}, 
        \nonumber\\ 
        \label{delta_k2_S_p_0}
        \\
        \Delta \ve{n}^{3,p=0}_{S_L} &=& - \frac{4 G \hat{S}_{bL-1}\!\left(t_0\right)}{c^3} 
        \epsilon_{abc} k^c \frac{\left(-1\right)^l l}{\left(l+1\right)!}\,\frac{1}{R} 
        \widehat{\partial}_{aL-1} \frac{\ve{\hat{\xi}}}{r_0 - c \tau_0}, 
        \nonumber\\ 
        \label{delta_k3_S_p_0}
\end{eqnarray}

\noindent
where $l \ge 1$ and the differential operator is given by Eq.~(\ref{Differential_Operator_p_0}).
The terms in the second line of (\ref{delta_n_S_Simplified}) read
\begin{eqnarray}
        \Delta \ve{n}^{1,p \neq 0}_{S_L} &=& - \frac{4 G \hat{S}_{bL-1}\!\left(t_1\right)}{c^3} 
        \epsilon_{abc} k^c \frac{\left(-1\right)^l l}{\left(l+1\right)!} 
        \widehat{\partial}^{\tau_1, p \neq 0}_{aL-1} \frac{\ve{\hat{\xi}}}{r_1 - c \tau_1} \frac{1}{r_1},
        \nonumber\\ 
        \label{delta_k1_S_p_neq_0}
        \\
        \Delta \ve{n}^{2,p \neq 0}_{S_L} &=& + \frac{4 G \hat{S}_{bL-1}\!\left(t_1\right)}{c^3} 
        \epsilon_{abc} k^c \frac{\left(-1\right)^l l}{\left(l+1\right)!} \frac{1}{R}
        \widehat{\partial}^{\tau_1, p \neq 0}_{aL-1} \frac{\ve{\hat{\xi}}}{r_1 - c \tau_1}, 
        \nonumber\\ 
        \label{delta_k2_S_p_neq_0}
        \\
        \Delta \ve{n}^{3,p \neq 0}_{S_L} &=& - \frac{4 G \hat{S}_{bL-1}\!\left(t_0\right)}{c^3} 
	\epsilon_{abc} k^c \frac{\left(-1\right)^l l}{\left(l+1\right)!} \frac{1}{R}
        \widehat{\partial}^{\tau_0, p \neq 0}_{aL-1} \frac{\ve{\hat{\xi}}}{r_0 - c \tau_0}, 
        \nonumber\\ 
        \label{delta_k3_S_p_neq_0}
\end{eqnarray}

\noindent
where $l \ge 1$ and the differential operators are given by Eqs.~(\ref{Differential_Operator_tau_0_p}) and (\ref{Differential_Operator_tau_1_p}).

For the upper limit of the term of Eq.~(\ref{delta_k1_S_p_neq_0}) and for the upper limit of the sum of the terms of Eqs.~(\ref{delta_k2_S_p_neq_0}) 
and (\ref{delta_k3_S_p_neq_0}) one obtains   
\begin{eqnarray}
	&& \left|\delta\left(\ve{k},\Delta \ve{n}^{1, p\neq 0}_{S_L}\right)\right| \le C_2^{S_L}\,\frac{G M}{c^3}\,\Omega\,\left|J_{l-1}\right| \left(\frac{P}{r_1}\right)^{l+1},
        \label{scaling_delta_k1_S_p_neq_0}
        \\
	&& \left|\delta\left(\ve{k},\Delta\ve{n}^{2+3,p \neq 0}_{S_L}\right)\right| \le B_2^{S_L}\,\frac{G M}{c^3}\,\Omega\,\left|J_{l-1}\right| 
	\left(\frac{P}{r_1}\right) \left(\frac{P}{d_k}\right)^{l}. 
	\nonumber\\ 
        \label{scaling_delta_k2_k3_S_p_neq_0}
\end{eqnarray}
 
\noindent
The proof of these relations is similar to the considerations of Appendixes~\ref{Proof1} and \ref{Proof2} for the mass-multipole terms. But, as mentioned above, only the 
very first few spin-multipoles are relevant on the sub-\muas-level: the spin-dipole term $l=1$ and the spin-hexapole term $l=3$. In particular, for the spin-dipole one 
obtains $C_2^{S_1}=0$ and $B_2^{S_1}=0$, which can be deduced by very similar steps as presented in Appendix~\ref{Appendix_Spin}. 
From these relations (\ref{scaling_delta_k1_S_p_neq_0}) and (\ref{scaling_delta_k2_k3_S_p_neq_0}) it is obvious that the 
contributions of these terms to the angle of light deflection is much less than $1\,{\rm nas}$ in higher orders of the spin-multipoles, 
\begin{eqnarray}
        \left|\delta\left(\ve{k},\Delta \ve{n}^{1, p\neq 0}_{S_L}\right)\right| &\ll& 1\,{\rm nas}\,,
        \label{numerical_magnitude_Spin_dipole_3}
        \\
\left|\delta\left(\ve{k},\Delta\ve{n}^{2+3,p \neq 0}_{S_L}\right)\right| &\ll& 1\,{\rm nas} \,,
        \label{numerical_magnitude_Spin_dipole_4}
\end{eqnarray}

\noindent
while in case of spin-dipole these terms vanish. Therefore, these terms (\ref{delta_k1_S_p_neq_0}) - (\ref{delta_k3_S_p_neq_0}) can be neglected in astrometric 
measurements on the sub-micro-arcsecond level. 

These facts allow us to simplify the unit tangent vector (\ref{light_deflection_new}) of a light ray considerably by neglecting the terms of 
Eqs.~(\ref{delta_k1_M_p_neq_0}) - (\ref{delta_k3_M_p_neq_0}) as well as of Eqs.~(\ref{delta_k4_S}) - (\ref{delta_k6_S}) and of 
Eqs.~(\ref{delta_k1_S_p_neq_0}) - (\ref{delta_k3_S_p_neq_0}). In other words, for sub-\muas{} astrometry it is sufficient to take into 
account only the terms of Eqs.~(\ref{delta_k1_M_p_0}) - (\ref{delta_k3_M_p_0}) and of Eqs.~(\ref{delta_k1_S_p_0}) - (\ref{delta_k3_S_p_0}).

\subsection{The reduced tangent vector}

As discussed in the previous Subsections, for astrometry on the sub-\muas{} level one needs to account only the terms in the first line of Eq.~(\ref{delta_n_M_Simplified}) and 
in the first line of Eq.~(\ref{delta_n_S_Simplified}). Then one arrives at the reduced unit tangent vector of the light ray at the position of the observer, given by 
\begin{eqnarray}
	\widetilde{\ve{n}} &=& \ve{k} + \sum\limits_{l=0}^{\infty}\Delta\widetilde{\ve{n}}_{M_L} + \sum\limits_{l=1}^{\infty}\Delta\widetilde{\ve{n}}_{S_L}\,. 
        \label{tangent_vector_reduced}
\end{eqnarray}

\noindent 
The individual terms of the reduced unit tangent vector of Eq.~(\ref{tangent_vector_reduced}) are given by  
\begin{eqnarray} 
	\Delta \widetilde{\ve{n}}_{M_L} &=& \Delta\ve{n}^{1,p=0}_{M_L} + \Delta \ve{n}^{2,p=0}_{M_L} + \Delta \ve{n}^{3,p=0}_{M_L}\,,
        \label{delta_n_M_Simplified_Final}
	\\ 
	\Delta \widetilde{\ve{n}}_{S_L} &=& \Delta\ve{n}^{1,p=0}_{S_L} + \Delta\ve{n}^{2,p=0}_{S_L} + \Delta\ve{n}^{3,p=0}_{S_L}\,,
        \label{delta_n_S_Simplified_Final}
\end{eqnarray}

\noindent 
with the mass-multipole terms of Eqs.~(\ref{delta_k1_M_p_0}) - (\ref{delta_k3_M_p_0}) and spin-multipole terms of Eqs.~(\ref{delta_k1_S_p_0}) - (\ref{delta_k3_S_p_0}). In the expression 
of the reduced unit tangent vector (\ref{tangent_vector_reduced}) all those terms have been neglected which contribute at most a few nano-arcseconds to the angle of light deflection. 
The reduced unit tangent vector (\ref{tangent_vector_reduced}) carries a wide tilde in order to distinguish it from the unit tangent vector of Eq.~(\ref{light_deflection_new}) which 
contains the complete set of terms in the 1.5PN approximation. In Section~\ref{Simplified_2} the reduced tangent vector (\ref{tangent_vector_reduced}) will further be simplified. 
For that we need two differential operations that are considered in the subsequent Section.

\section{Two differential operations}\label{Section_Differential_Operations}

By considering the mass-multipole terms of Eqs.~(\ref{delta_n_M_Simplified_Final}) with (\ref{delta_k1_M_p_0}) - (\ref{delta_k3_M_p_0}) and the spin-multipole 
terms (\ref{delta_n_S_Simplified_Final}) with (\ref{delta_k1_S_p_0}) - (\ref{delta_k3_S_p_0}) one encounters two typical differential operations: 
\begin{eqnarray}
	\ve{U}_L\left(c\tau,r\right)  &=& \widehat{\partial}_{L}\,\frac{\ve{\hat{\xi}}}{r - c \tau} \frac{1}{r} \,,
        \label{Differential_Operator_1}
	\\
	\ve{V}_L\left(c\tau,r\right)  &=& \widehat{\partial}_{L}\,\frac{\ve{\hat{\xi}}}{r - c \tau} \,,  
        \label{Differential_Operator_2}
\end{eqnarray}

\noindent 
with $r = \sqrt{\hat{\xi}^2 + c^2 \tau^2}$. 
For these differential operations we are using the following two relations, which are shown in Appendix~\ref{Appendix_Relation_1_and_2}, 
\begin{eqnarray}
	\ve{U}_L\left(c\tau,r\right)  &=& \ve{U}^1_L\left(c\tau,r\right) + \ve{U}^2_L\left(c\tau,r\right),  
        \label{Relation_n_1}
\end{eqnarray}

\noindent 
with 
\begin{eqnarray}
	\ve{U}^1_L\left(c\tau,r\right) &=& \left(-1\right)^{l+1} \underset{i_1 \dots i_l}{\rm STF}\,\frac{\partial}{\partial \ve{\hat{\xi}}} 
        \Bigg[\sum\limits_{n=0}^{[l/2]} G_n^l \,P_{i_1 i_2}\,\dots\,P_{i_{2n-1} i_{2n}}
        \nonumber\\
        && \times \left(1 + \frac{c\tau}{r}\right)
        \frac{\hat{\xi}_{i_{2n+1}} \dots \hat{\xi}_{i_l}}{\left(\hat{\xi}\right)^{2l-2n}}\Bigg],
        \label{T_1}
	\\
	\ve{U}^2_L\left(c\tau,r\right) &=& \left(-1\right)^{l+1} \underset{i_1 \dots i_l}{\rm STF}\,\frac{\partial}{\partial \ve{\hat{\xi}}}
        \Bigg[\sum\limits_{n=0}^{[l/2]} G_n^l \,P_{i_1 i_2}\,\dots\,P_{i_{2n-1} i_{2n}}
        \nonumber\\
	&& \times \accentset{\ast}{F}^l_n\left(c\tau,r\right) \frac{\hat{\xi}_{i_{2n+1}} \dots \hat{\xi}_{i_l}}{\left(\hat{\xi}\right)^{2l-2n}}\Bigg],
        \label{T_2}
\end{eqnarray}

\noindent 
and 
\begin{eqnarray}
	\ve{V}_L\left(c\tau,r\right) &=& \ve{V}^1_L\left(c\tau,r\right) + \ve{V}^2_L\left(c\tau,r\right),
        \label{Relation_n_2}
\end{eqnarray}

\noindent
with
\begin{eqnarray}
	\ve{V}^1_L\left(c\tau,r\right) &=& \left(-1\right)^{l+1} \underset{i_1 \dots i_l}{\rm STF}\,\frac{\partial}{\partial \ve{\hat{\xi}}}
        \Bigg[\sum\limits_{n=0}^{[l/2]} G_n^l \,P_{i_1 i_2}\,\dots\,P_{i_{2n-1} i_{2n}}  
        \nonumber\\
	&& \hspace{-1.0cm} \times \; c\tau \left(1 + \frac{c\tau}{r}\right) \frac{\hat{\xi}_{i_{2n+1}} \dots \hat{\xi}_{i_l}}{\left(\hat{\xi}\right)^{2l-2n}}\Bigg],  
	\label{S_1} 
	\\
	\ve{V}^2_L\left(c\tau,r\right) &=& \left(-1\right)^{l+1} \underset{i_1 \dots i_l}{\rm STF}\,\frac{\partial}{\partial \ve{\hat{\xi}}}
        \Bigg[\sum\limits_{n=0}^{[l/2]} G_n^l \,P_{i_1 i_2}\,\dots\,P_{i_{2n-1} i_{2n}}
        \nonumber\\
	&& \hspace{-1.0cm} \times \left(c\tau\,\accentset{\ast}{F}^l_n\left(c\tau,r\right) + r W^l_n \left(\frac{\hat{\xi}}{r}\right)^{2l-2n}\right) 
        \frac{\hat{\xi}_{i_{2n+1}} \dots \hat{\xi}_{i_l}}{\left(\hat{\xi}\right)^{2l-2n}} \Bigg]. 
        \nonumber\\
	\label{S_2} 
\end{eqnarray}

\noindent
The dimensionless scalar function $\accentset{\ast}{F}^l_n\left(c\tau,r\right)$ is defined by Eq.~(\ref{Function_F_Star}). Let us notice here that this 
scalar function exists only for $l\ge 2$ and vanishes for $l=1$. 
Below it is shown that these terms of Eqs.~(\ref{T_2}) and (\ref{S_2}) can be neglected on the sub-micro-arcsecond level. 
In Eqs.~(\ref{T_1}) - (\ref{T_2}) and Eqs.~(\ref{S_1}) - (\ref{S_2}) we have used the abbreviation \cite{Kopeikin1997}
\begin{eqnarray}
	\frac{\partial}{\partial \hat{\xi}^i} = P^{ij}\,\frac{\partial}{\partial \xi^j}  
        \label{Abbreviation}
\end{eqnarray}

\noindent
for the spatial components ($i=1,2,3$) of the operation $\partial/\partial \ve{\hat{\xi}}$, where the projector is defined by Eq.~(\ref{Projection_Operator_k}). 
In these relations (\ref{Relation_n_1}) and (\ref{Relation_n_2}) the coefficients 
\begin{eqnarray}
	&& \hspace{-1.0cm}  G^l_n = \left(-1\right)^n\,2^{l-2n-1}\,\frac{l!}{n!}\,\frac{\left(l-n-1\right)!}{\left(l-2n\right)!}\,,
        \label{Coefficients_G_l_n}
	\\
	&& \hspace{-1.0cm} W^l_n = \frac{\left(2l - 2n - 3\right)!!}{\left(2l - 2n - 2\right)!!} = \frac{1}{2^{2l-2n-2}} {2l-2n-2 \choose l-n-1}\,,
        \label{Coefficients_W_l_n}
\end{eqnarray}

\noindent
have been introduced. The coefficient (\ref{Coefficients_G_l_n}) is identical with the coefficient given by Eq.~(55) in \cite{Zschocke_Total_Light_Deflection_15PN}. 

A further comment is in order here. Because the STF operation of any tensor vanishes if it contains a Kronecker symbol 
(cf. Eqs~(\ref{STF_comment_1}) - (\ref{STF_comment_2b}) in Appendix~\ref{Appendix_STF}), one may replace the projectors (\ref{Projection_Operator_k}) in
these relations (\ref{Relation_n_1}) and (\ref{Relation_n_2}) by
\begin{eqnarray}
        P_{i_{p+1} i_{p+2}} \,\dots\, P_{i_{p+2n-1} i_{p+2n}} \rightarrow \left(-1\right)^n k_{i_{p+1}}\,\dots\,k_{i_{p+2n}} .
        \nonumber\\ 
        \label{Replacement}
\end{eqnarray}

\noindent 
For later purposes we give here a slightly more general version of that replacement, while in these relations (\ref{Relation_n_1}) and (\ref{Relation_n_2}) one only
needs the case with $p=0$. In case of the spin-multipoles (\ref{delta_k1_S_p_0}) - (\ref{delta_k3_S_p_0}) one would seemingly encounter $\widehat{\partial}_{aL-1}$
instead of $\widehat{\partial}_{L}$ of Eqs.~(\ref{Differential_Operator_1}) and (\ref{Differential_Operator_2}). But by changing the dummy index from $a$ to $i_l$
one arrives at the very same expressions as given by Eqs.~(\ref{Differential_Operator_1}) and (\ref{Differential_Operator_2}). That means, one may apply the same
replacement (\ref{Replacement}) also in case of spin-multipole terms.

\section{The simplified tangent vector}\label{Simplified_2}

The reduced tangent vector of the light ray is given by Eq.~(\ref{tangent_vector_reduced}) with (\ref{delta_n_M_Simplified_Final}) and (\ref{delta_n_S_Simplified_Final}), 
where the individual mass-multipole terms are given by Eqs.~(\ref{delta_k1_M_p_0}) - (\ref{delta_k3_M_p_0}) and the individual spin-multipole terms are given by
Eqs.~(\ref{delta_k1_S_p_0}) - (\ref{delta_k3_S_p_0}). The reduced tangent vector is a step in between to arrive at the simplified tangent vector, 
which is the primary aim of this investigation and which is obtained in this Section.

\subsection{Two further simplifications}

The expressions (\ref{Relation_n_1}) and (\ref{Relation_n_2}) have to be implemented into Eqs.~(\ref{delta_k1_M_p_0}) - (\ref{delta_k3_M_p_0}) and into 
Eqs.~(\ref{delta_k1_S_p_0}) - (\ref{delta_k3_S_p_0}). In doing so one obtains expressions, which are still much too cumbersome for an effective treatment of astrometric data. 
In particular, they do contain many terms that contribute less than a few ${\rm nas}$ to the effect of light deflection. Therefore, two further simplifications are performed:

\subsubsection{First simplification} 

The term (\ref{T_2}) is neglected for both the mass-multipole and spin-multipole terms. 
In order to demonstrate that (\ref{T_2}) can be neglected on the sub-\muas{} level, we consider the mass-multipole terms. That means, we insert (\ref{T_2}) into  
Eq.~(\ref{delta_k1_M_p_0}) and obtain 
\begin{eqnarray}
	\ve{U}_2^{M_L}\left(c\tau_1,r_1\right) &=& - \frac{2 G \hat{M}_L\left(t_1\right)}{c^2} \frac{(-1)^l}{l!}\,\ve{U}^2_L\left(c\tau_1,r_1\right),\;\; 
        \label{Proof_T2_1}
\end{eqnarray}

\noindent 
where $\ve{U}_2^{M_L}$ is a three-vector in space. The tensorial indices of multi-index $L$ of $\ve{U}^2_L$ and of the STF mass-multipoles $\hat{M}_L$ are contracted with each other. 
In Appendix~\ref{Appendix_Neglection_1} its contribution to the angle of light deflection (\ref{light_deflection_M_L}) is determined and given by (for $l \ge 2$) 
\begin{eqnarray}
	|\delta(\ve{k},\ve{U}_2^{M_L})| &\le& C_2^{M_L}\,\frac{G M}{c^2 r_1} \left(\frac{P}{r_1}\right) \left|J_l\right| \left(\frac{P}{d_k}\right)^{l-1}. 
	\label{Proof_T2_35}
\end{eqnarray}
 
\noindent 
The coefficients $C_2^{M_L}$ are presented in Table~\ref{Table_A_l} of Appendix~\ref{Appendix_Neglection_1}. This term (\ref{Proof_T2_35}) is virtually of 
the order (\ref{magnitude_M_3}) and contributes less much than $1\,{\rm nas}$ to the effect of light deflection at any body of the solar system, 
\begin{eqnarray}
	|\delta(\ve{k},\ve{U}_2^{M_L})| &\ll& 1\,{\rm nas}\,. 
        \label{Proof_T2_35_magnitude}
\end{eqnarray}

\noindent 
The same statement is valid for the spin-multipole terms, that means we insert (\ref{T_2}) into
Eq.~(\ref{delta_k1_S_p_0}) and obtain
\begin{eqnarray}
        \ve{U}_2^{S_L}\left(c\tau_1,r_1\right) &=& - \frac{4 G \hat{S}_{bL-1}\left(t_1\right)}{c^3} \epsilon_{abc}\,k^c 
	\nonumber\\ 
	&& \times \frac{\left(-1\right)^l\,l}{\left(l+1\right)!}\,\ve{U}^2_{aL-1}\left(c\tau_1,r_1\right), 
        \label{Proof_T2_1_Spin}
\end{eqnarray}

\noindent 
where $\ve{U}_2^{S_L}$ is a three-vector in space. In order to determine the contribution of (\ref{Proof_T2_1_Spin}) to the angle of light deflection, we  
approximate the body as an axi-symmetric body in uniform rotation. Then, by implementing the spin-multipoles (\ref{S_L}) into (\ref{Proof_T2_1_Spin}), one 
finds that the contribution of (\ref{Proof_T2_1_Spin}) to the angle of light deflection for $l>1$ is given by  
\begin{eqnarray}
	&& \hspace{-0.75cm} |\delta(\ve{k},\ve{U}_2^{S_L})| \le C_3^{S_L}\,\frac{G M}{c^3} \Omega \left|J_{l-1}\right| \left(\frac{P}{r_1}\right)^2 \left(\frac{P}{d_k}\right)^{l-1}.
        \label{Proof_T2_35_Spin}
\end{eqnarray}

\noindent 
For the spin-dipole $l=1$ one obtains $C_3^{S_1} = 0$; see also the corresponding remark below Eq.~(\ref{S_2}). 
The proof of the step from (\ref{Proof_T2_1_Spin}) to (\ref{Proof_T2_35_Spin}) goes similar as the approach of Appendix~\ref{Appendix_Neglection_2} for the mass-multipoles. 
From these relations it is clear that these terms contribute much less than $1\,{\rm nas}$ to the angle of light deflection in higher orders of the spin-multipoles. 
By inserting the parameters of Table~\ref{Table1} into (\ref{Proof_T2_35_Spin}), one obtains for grazing light rays at any solar system body 
\begin{eqnarray} 
        \left|\delta\left(\ve{k},\ve{U}_2^{S_L}\right)\right| &\ll& 1\,{\rm nas}\,.
        \label{numerical_magnitude_Spin_Multipole_5}
\end{eqnarray}

\noindent 
Therefore, these terms (\ref{Proof_T2_1_Spin}) can be neglected for astrometry on the sub-micro-arcsecond level of accuracy.

\subsubsection{Second simplification} 

The term (\ref{S_2}) is neglected for both the mass-multipole and spin-multipole terms. 
In order to demonstrate that (\ref{S_2}) can be neglected on the sub-\muas{} level, we consider the mass-multipole terms. That means, we insert (\ref{S_2}) into
Eqs.~(\ref{delta_k2_M_p_0}) and (\ref{delta_k3_M_p_0}) and obtain
\begin{eqnarray}
	\ve{V}_2^{M_L} &=& \frac{2 G \hat{M}_L\left(t_1\right)}{c^2} \frac{(-1)^l}{l!} \frac{1}{R} \left(\ve{V}^2_L\left(c\tau_1,r_1\right) - \ve{V}^2_L\left(c\tau_0,r_0\right)\right),
	\nonumber\\ 
        \label{Proof_S2_1}
\end{eqnarray}

\noindent
where $\ve{V}_2^{M_L}$ is a three-vector in space. A series expansion yields $\hat{M}_L\left(t_0\right) = \hat{M}_L\left(t_1\right)$ up to time derivatives of the mass-multipoles  
which are, however, neglected \cite{Comment2}. The tensorial indices of multi-index $L$ of $\ve{V}^2_L$ and of the STF mass-multipoles $\hat{M}_L$ are contracted with each other.
\begin{table}[t]
        \caption{The contribution of the terms of Eq.~(\ref{Proof_S2_35}) to the angle of light deflection
	in case of a grazing ray at at Jupiter ($\jupiter$) and Saturn ($\saturn$). For all other solar system bodies these terms are even smaller. All values are given in nano-arcseconds (nas). 
	A blank entry means less than $1\,{\rm nas}$.}
\begin{tabular}{| c | c | c |} 
\hline
	&&\\[-12pt]
	$l$ &\hbox to 32mm{\hfill $|\delta_{\jupiter}(\ve{k},\ve{V}_2^{M_L})|$ \hfill} &\hbox to 32mm{\hfill $|\delta_{\saturn}(\ve{k},\ve{V}_2^{M_L})|$ \hfill} \\[3pt]
\hline
	&&\\[-12pt]
	$2$ & $5.0$ & $1.7$ \\[3pt]
        $4$ & $1.4$ & $0.6$ \\[3pt]
        $6$ & $0.3$ & $0.2$ \\[3pt]
        $8$ & $0.1$ & $0.1$ \\[3pt]
        $10$ & $-$ & $-$ \\[3pt]
\hline
\end{tabular}
\label{Table_T2_35_S2_35}
\end{table}

\noindent 
In Appendix~\ref{Appendix_Neglection_2} its contribution to the angle of light deflection (\ref{light_deflection_M_L}) is determined and given by
\begin{eqnarray}
	 |\delta(\ve{k},\ve{V}_2^{M_L})| &\le& \left|B_2^{M_L}\right|\,\frac{G M}{c^2 r_1} \left|J_l\right| \left(\frac{P}{d_k}\right)^{l}.
        \label{Proof_S2_35}
\end{eqnarray}

\noindent
The absolute value of the coefficients $B_2^{M_L}$ are presented in Table~\ref{Table_B_l} of Appendix~\ref{Appendix_Neglection_2}. This term is of the order (\ref{magnitude_M_2}). 
Numerical values of this term are presented in Table~\ref{Table_T2_35_S2_35}. 
It is noticed here, that the term of Eq.~(\ref{Proof_S2_1}) is much larger than the term of Eq.~(\ref{Proof_T2_1}). 
The reason is that the distance $R$ between source and observer appears in the denominator of Eq.~(\ref{Proof_S2_1}). 
The distance $R$ can in principle be arbitrarily small. A similar observation has been made above; see text below Eq.~(\ref{scaling_delta_k1_M_p_neq_0_magnitude}). 
Nevertheless, the term of Eq.~(\ref{Proof_S2_1}) is finite and, according to (\ref{Proof_S2_35}), contributes at most a few ${\rm nas}$.

The same statement is valid for the spin-multipole terms, that means we insert (\ref{S_2}) into Eqs.~(\ref{delta_k2_S_p_0}) and (\ref{delta_k3_S_p_0}) and obtain
\begin{eqnarray}
        \ve{V}_2^{S_L} &=&  \frac{4 G \hat{S}_{bL-1}\left(t_1\right)}{c^3} 
        \epsilon_{abc}\,k^c\,\frac{\left(-1\right)^l\,l}{\left(l+1\right)!}\,\frac{1}{R} 
	\nonumber\\
	&& \times \left(\ve{V}^2_{aL-1}\left(c\tau_1,r_1\right) - \ve{V}^2_{aL-1}\left(c\tau_0,r_0\right)\right),
        \label{Proof_S2_1_Spin}
\end{eqnarray}

\noindent
where $\ve{V}_2^{S_L}$ is a three-vector in space. A series expansion yields $\hat{S}_{bL-1}\left(t_0\right) = \hat{S}_{bL-1}\left(t_1\right)$ up to time derivatives of the 
spin-multipoles which are, however, neglected \cite{Comment2}. 
For the determination of the contribution of (\ref{Proof_S2_1_Spin}) to the angle of light deflection, 
we approximate the body as an axi-symmetric body in uniform rotation. Then, by implementing the spin-multipoles (\ref{S_L}) into (\ref{Proof_S2_1_Spin}), 
one finds that the contribution of (\ref{Proof_S2_1_Spin}) to the angle of light deflection is given by 
\begin{eqnarray}
	|\delta(\ve{k},\ve{V}_2^{S_L})| &\le& B_3^{S_L}\,\frac{G M}{c^3} \Omega \left|J_{l-1}\right| \left(\frac{P}{r_1}\right) \left(\frac{P}{d_k}\right)^{l}.
        \label{Proof_S2_35_Spin}
\end{eqnarray}

\noindent 
The proof of the step from (\ref{Proof_S2_1_Spin}) to (\ref{Proof_S2_35_Spin}) proceeds in the same way as the approach presented in Appendix~\ref{Appendix_Neglection_2} 
for the mass-multipoles. As mentioned in the text below Eq.~(\ref{scale_delta_k5_k6_S}) only the spin-dipole term $l=1$ and the spin-hexapole term $l=3$ are relevant 
on the sub-\muas{} level. In order to get an idea about the magnitude of these terms in (\ref{Proof_S2_35_Spin}) we consider the case of spin-dipole, which is the most 
dominating term. By using very similar steps as done in Appendix~\ref{Appendix_Spin}, one obtains $B_3^{S_1} = 2\,\kappa^2$. By inserting the parameters of 
Table~\ref{Table1} into (\ref{Proof_S2_35_Spin}) one finds that this term contributes much less than $1\,{\rm nas}$ to the angle of light deflection in case of spin-dipole. 
The coefficients $B_3^{S_L}$ of Eq.~(\ref{Proof_S2_35_Spin}) are not given here explicitly. But from that relation it is obvious that the contributions of higher spin-multipoles 
to the angle of light deflection is much less than $1\,{\rm nas}$ to any order of the spin-multipole index, 
\begin{eqnarray} 
        \left|\delta\left(\ve{k},\ve{V}_2^{S_L}\right)\right| &\ll& 1\,{\rm nas}\,,
        \label{numerical_magnitude_Spin_dipole_6}
\end{eqnarray}

\noindent
for grazing light rays at any solar system body. Therefore, these terms (\ref{Proof_S2_1_Spin}) can be neglected for astrometry on the sub-micro-arcsecond level of accuracy.

\subsection{The simplified tangent vector}

The reduced tangent vector was given by Eq.~(\ref{tangent_vector_reduced}) with the individual terms of Eqs.~(\ref{delta_n_M_Simplified_Final}) and (\ref{delta_n_S_Simplified_Final}), 
where the mass-multipole terms are given by Eqs.~(\ref{delta_k1_M_p_0}) - (\ref{delta_k3_M_p_0}) and the spin-multipole terms by Eqs.~(\ref{delta_k1_S_p_0}) - (\ref{delta_k3_S_p_0}). 
By taking into account these two simplifications of the previous Subsection, one arrives from the reduced tangent vector at the simplified tangent vector 
of a light ray. This simplified unit tangent vector is written in the form 
\begin{eqnarray}
	\widehat{\ve{n}} &=& \ve{k} + \sum\limits_{l=0}^{\infty} \Delta \widehat{\ve{n}}_{M_L} + \sum\limits_{l=1}^{\infty} \Delta \widehat{\ve{n}}_{S_L}.   
        \label{unit_tangent_vector_simplified}
\end{eqnarray}

\noindent 
The unit tangent vector (\ref{unit_tangent_vector_simplified}) and the perturbations are denoted by a {\it hat}, in order to distinguish them from the unit tangent vector 
of Eq.~(\ref{light_deflection_new}) and its perturbations. These mass-multipole and spin-multipole terms in (\ref{unit_tangent_vector_simplified}) are given in the next 
two Subsections. This simplified tangent vector (\ref{unit_tangent_vector_simplified}) of a light ray in the gravitational field of an arbitrary solar system body 
contains all those terms that contribute at least $0.01\,\muas$ ($10$ nas) to the angle of light deflection. It will be shown below in Section~\ref{Section4} that only the first few 
mass-multipoles with multipole index $0 \le l \le 8$ and the first two spin-multipoles with multipole index $1 \le l \le 3$ are relevant on the sub-\muas{} level of 
accuracy.

\subsubsection{The mass-multipole terms}

\noindent 
The mass-multipole terms in (\ref{unit_tangent_vector_simplified}) read  
\begin{eqnarray}
	\Delta \widehat{\ve{n}}_{M_L} &=& \Delta \widehat{\ve{n}}^{1}_{M_L} + \Delta \widehat{\ve{n}}^{2}_{M_L} + \Delta \widehat{\ve{n}}^{3}_{M_L},
         \label{hat_M_L}
\end{eqnarray}

\noindent
with the individual terms 
\begin{eqnarray}
	\Delta \widehat{\ve{n}}^{1}_{M_L} &=& + \frac{2 G \hat{M}_{L}\left(t_1\right)}{c^2} \frac{1}{l!} \left(1 + \frac{\ve{k} \cdot \ve{x}_1}{r_1}\right) 
        \nonumber\\
        && \hspace{-1.25cm} \times \frac{\partial}{\partial\ve{\hat{\xi}}} 
        \sum\limits_{n=0}^{[l/2]} \left(-1\right)^n G^l_n  
	k_{i_1}\,\dots\,k_{i_{2n}}\; \frac{\hat{\xi}_{i_{2n+1}}\,\dots\,\hat{\xi}_{i_{l}}}{(\hat{\xi})^{2l-2n}} , 
        \label{Simplified_Tangent_Vector_1}
        \\
        \nonumber\\
	\Delta \widehat{\ve{n}}^{2}_{M_L} &=& 
	- \frac{2 G \hat{M}_{L}\left(t_1\right)}{c^2} \frac{1}{l!}\,\frac{\ve{k} \cdot \ve{x}_1}{R} \left(1 + \frac{\ve{k} \cdot \ve{x}_1}{r_1}\right) 
        \nonumber\\
        && \hspace{-1.25cm} \times \frac{\partial}{\partial\ve{\hat{\xi}}} 
        \sum\limits_{n=0}^{[l/2]} \left(-1\right)^n G^l_n 
	k_{i_1}\,\dots\,k_{i_{2n}}\; \frac{\hat{\xi}_{i_{2n+1}}\,\dots\,\hat{\xi}_{i_{l}}}{(\hat{\xi})^{2l-2n}} ,
        \label{Simplified_Tangent_Vector_2}
        \\ 
        \nonumber\\
	\Delta \widehat{\ve{n}}^{3}_{M_L} &=& 
	+ \frac{2 G \hat{M}_{L}\left(t_0\right)}{c^2} \frac{1}{l!}\,\frac{\ve{k} \cdot \ve{x}_0}{R} \left(1 + \frac{\ve{k} \cdot \ve{x}_0}{r_0}\right) 
        \nonumber\\
        && \hspace{-1.25cm} \times \frac{\partial}{\partial\ve{\hat{\xi}}} 
        \sum\limits_{n=0}^{[l/2]} \left(-1\right)^n G^l_n 
	k_{i_1}\,\dots\,k_{i_{2n}}\; \frac{\hat{\xi}_{i_{2n+1}}\,\dots\,\hat{\xi}_{i_{l}}}{(\hat{\xi})^{2l-2n}},
        \label{Simplified_Tangent_Vector_3}
\end{eqnarray}

\noindent 
where the ${\rm STF}_{i_1 \dots i_l}$ operation has been omitted, because the mass-multipoles $\hat{M}_{L}$ are STF tensors with respect to indices $L$, hence we may apply 
relation (\ref{STF_comment_3}). 

\subsubsection{The spin-multipole terms}

\noindent 
The spin-multipole terms in (\ref{unit_tangent_vector_simplified}) read  
\begin{eqnarray}
        \Delta \widehat{\ve{n}}_{S_L} &=& \Delta \widehat{\ve{n}}^{1}_{S_L} + \Delta \widehat{\ve{n}}^{2}_{S_L} + \Delta \widehat{\ve{n}}^{3}_{S_L},
        \label{hat_S_L}
\end{eqnarray}

\noindent 
with the individual terms 
\begin{eqnarray}
	\Delta \widehat{\ve{n}}^{1}_{S_L} &=& + \frac{4 G \hat{S}_{bL-1}\left(t_1\right)}{c^3}\,\epsilon_{i_l bc}\,k^c 
        \frac{l}{\left(l+1\right)!} \left(1 + \frac{\ve{k} \cdot \ve{x}_1}{r_1}\right) 
        \nonumber\\
        && \hspace{-1.0cm} \times 
        \frac{\partial}{\partial\ve{\hat{\xi}}} \sum\limits_{n=0}^{[l/2]} \left(-1\right)^n G^l_n 
	\;\underset{i_1 \dots i_l}{\rm STF}\,k_{i_1}\,\dots\,k_{i_{2n}}\; \frac{\hat{\xi}_{i_{2n+1}}\,\dots\,\hat{\xi}_{i_{l}}}{(\hat{\xi})^{2l-2n}} \,,
        \label{Simplified_Tangent_Vector_4}
        \\
        \nonumber\\
	\Delta \widehat{\ve{n}}^{2}_{S_L} &=& - \frac{4 G \hat{S}_{bL-1}\left(t_1\right)}{c^3}\,\epsilon_{i_l bc}\,k^c 
	\frac{l}{\left(l+1\right)!}\,\frac{\ve{k} \cdot \ve{x}_1}{R} \left(1 + \frac{\ve{k} \cdot \ve{x}_1}{r_1}\right) 
        \nonumber\\
        && \hspace{-1.0cm} \times 
        \frac{\partial}{\partial\ve{\hat{\xi}}} \sum\limits_{n=0}^{[l/2]} \left(-1\right)^n G^l_n 
	\;\underset{i_1 \dots i_l}{\rm STF}\,k_{i_1}\,\dots\,k_{i_{2n}}\; \frac{\hat{\xi}_{i_{2n+1}}\,\dots\,\hat{\xi}_{i_{l}}}{(\hat{\xi})^{2l-2n}} \,,
        \label{Simplified_Tangent_Vector_5}
        \\
        \nonumber\\
	\Delta \widehat{\ve{n}}^{3}_{S_L} &=& + \frac{4 G \hat{S}_{bL-1}\left(t_0\right)}{c^3}\,\epsilon_{i_l bc}\,k^c 
        \frac{l}{\left(l+1\right)!}\,\frac{\ve{k} \cdot \ve{x}_0}{R} \left(1 + \frac{\ve{k} \cdot \ve{x}_0}{r_0}\right) 
        \nonumber\\
        && \hspace{-1.0cm} \times 
        \frac{\partial}{\partial\ve{\hat{\xi}}} \sum\limits_{n=0}^{[l/2]} \left(-1\right)^n G^l_n 
	\;\underset{i_1 \dots i_l}{\rm STF}\,k_{i_1}\,\dots\,k_{i_{2n}}\; \frac{\hat{\xi}_{i_{2n+1}}\,\dots\,\hat{\xi}_{i_{l}}}{(\hat{\xi})^{2l-2n}} \,,  
        \label{Simplified_Tangent_Vector_6}
\end{eqnarray}
  
\noindent
where the ${\rm STF}_{i_1 \dots i_l}$ operation needs to be taken into account explicitly, because the spin-multipoles $\hat{S}_{bL-1}$ are only STF with respect to indices $bL-1$ and 
not with respect to indices $L$. A further comment is made about the time-argument of the multipoles in Eqs.~(\ref{Simplified_Tangent_Vector_1}) - (\ref{Simplified_Tangent_Vector_3})
and in Eqs.~(\ref{Simplified_Tangent_Vector_4}) - (\ref{Simplified_Tangent_Vector_6}). In view of the slow
rotational motions of the solar system bodies it can certainly be assumed that the effect of light deflection is not very sensitive to the time-dependence of 
the rotational axis caused by the precession of the bodies, even on the sub-\muas{} scale of accuracy. 

In Eqs.~(\ref{Simplified_Tangent_Vector_1}) - (\ref{Simplified_Tangent_Vector_3}) and Eqs.~(\ref{Simplified_Tangent_Vector_4}) - (\ref{Simplified_Tangent_Vector_6}) the 
new variables $c\tau_0$ and $c\tau_1$ have been replaced by $\ve{k} \cdot \ve{x}_0$ and $\ve{k} \cdot \ve{x}_1$, respectively, in line with the statements below 
Eq.~(\ref{Second_Integration_6}) and by taking account of relations (\ref{xN_x0}) and (\ref{xN_x1}).

There are two factors $(-1)^n$ in these expressions: one factor is contained in the coefficients $G^l_n$ and one factor appears in each of these equations. Below it becomes clear, why 
it is more appropriate not to cancel these terms against each other, but to keep them explicitly as is, namely in the step from (\ref{Function_F_l_M}) to (\ref{Function_F_l_M_final1}) 
and in the step from (\ref{Function_F_l_S}) to (\ref{Function_F_l_S_final1}).

\subsection{The complete sum of all neglected terms}\label{Total_Sum_Neglected_Terms} 

As noticed in the introductory Section, the term {\it sub-micro-arcsecond astrometry} refers to an astrometric precision of $0.1$ micro-arcseconds in positional measurements. 
A relativistic model of light propagation should be $10$ times more accurate than this accuracy. Thus, in a theoretical model of light propagation on the sub-\muas{} level of 
precision one may neglect terms that contribute less than $10$ nano-arcsecond to the angle of light deflection. The effect of light deflection is an integral effect of
all multipoles of the concrete massive body under consideration. Accordingly, the simplified tangent vector (\ref{unit_tangent_vector_simplified}) should differ from 
the tangent vector (\ref{light_deflection_new}), which is exact in the 1.5PN approximation, only by terms whose total sum over all multipoles 
contributes less than $10$ nano-arcsecond to the angle of light deflection.  

The neglected spin-multipole terms were given by Eqs.~(\ref{delta_k4_S}) - (\ref{delta_k6_S}) and by Eqs.~(\ref{delta_k1_S_p_neq_0}) - (\ref{delta_k3_S_p_neq_0}) 
as well as by Eqs.~(\ref{Proof_T2_1_Spin}) and (\ref{Proof_S2_1_Spin}). All these neglected terms contribute much less than $1\,{\rm nas}$ to the angle of light 
deflection, as indicated by Eqs.~(\ref{numerical_magnitude_Spin_Multipole_1}), (\ref{numerical_magnitude_Spin_Multipole_2}), (\ref{numerical_magnitude_Spin_dipole_3}), 
(\ref{numerical_magnitude_Spin_dipole_4}), (\ref{numerical_magnitude_Spin_Multipole_5}), and (\ref{numerical_magnitude_Spin_dipole_6}). Therefore, these neglected 
spin-multipole terms need not to be further discussed here. 

The neglected mass-multipole terms were given by Eqs.~(\ref{delta_k1_M_p_neq_0}) - (\ref{delta_k3_M_p_neq_0}) as well as Eqs.~(\ref{Proof_T2_1}) and (\ref{Proof_S2_1}). 
The numerical magnitude of the terms of Eqs.~(\ref{delta_k1_M_p_neq_0}) and (\ref{Proof_T2_1}) are much less than $1\,{\rm nas}$, as indicated by 
Eqs.~(\ref{scaling_delta_k1_M_p_neq_0_magnitude}) and (\ref{Proof_T2_35_magnitude}), and need also not to be further discussed here. 

So we are left with the neglected terms that were given by Eqs.~(\ref{delta_k2_M_p_neq_0}) - (\ref{delta_k3_M_p_neq_0}) as well as by Eq.~(\ref{Proof_S2_1}).
The numerical magnitude of these terms are given by Table~\ref{Table_delta_3_4_5} and Table~\ref{Table_T2_35_S2_35} for grazing light rays at Jupiter and Saturn, which is 
the impact vector $d_k$ equals the radius $P$ of the body. These numerical values sum up to a total amount of 
\begin{eqnarray}
	\sum\limits_{l=2}^{10} \left[|\delta_{\jupiter}(\ve{k},\Delta \ve{n}^{2+3, p\neq 0}_{M_L})| + |\delta_{\jupiter}(\ve{k},\ve{V}_2^{M_L})|\right] 
	&\le& 36.2\,{\rm nas}\,, 
	\nonumber\\ 
	\label{Total_Sum_1_Jupiter}
	\\
	\sum\limits_{l=2}^{10} \left[|\delta_{\saturn}(\ve{k},\Delta \ve{n}^{2+3, p\neq 0}_{M_L})| + |\delta_{\saturn}(\ve{k},\ve{V}_2^{M_L})|\right] 
	&\le& 14.9\,{\rm nas}\,, 
	\nonumber\\ 
        \label{Total_Sum_1_Saturn}
\end{eqnarray}

\noindent
for Jupiter (\ref{Total_Sum_1_Jupiter}) and Saturn (\ref{Total_Sum_1_Saturn}). For all other solar system bodies the complete sum of neglected terms over all multipoles would 
be smaller than the given threshold of $10\,{\rm nas}$ for all astrometric configurations including grazing light rays. 

Let us discuss the total sum of neglected terms for grazing rays at Jupiter and Saturn as given by Eqs.~(\ref{Total_Sum_1_Jupiter}) and (\ref{Total_Sum_1_Saturn}), respectively. 
The total sum of all neglected terms for grazing ray at Saturn is $14.9\,{\rm nas}$, which is very near the given threshold of $10\,{\rm nas}$. That means, these terms can 
certainly be neglected for sub-\muas{} astrometry. On the other side, in case of Jupiter, the neglected terms sum up in total to $36.2\,{\rm nas}$, which is slightly above the 
goal accuracy of $10\,{\rm nas}$. However, in case the impact vector would be a little bit larger than the radius of Jupiter, namely $d_k \ge 1.7 \times P$, then we get in total only 
\begin{eqnarray}
	\sum\limits_{l=2}^{10} \left[|\delta_{\jupiter}(\ve{k},\Delta \ve{n}^{2+3, p\neq 0}_{M_L})| + |\delta_{\jupiter}(\ve{k},\ve{V}_2^{M_L})|\right] 
	&\le& 9.8\,{\rm nas}\,, 
	\nonumber\\ 
        \label{Total_Sum_2}
\end{eqnarray}

\noindent 
which is below the given threshold of $10\,{\rm nas}$. It is clear that for real astrometric measurements the neglect of these terms of Eqs.~(\ref{Total_Sum_1_Jupiter}) and (\ref{Total_Sum_1_Saturn}), 
which are of complicated analytical structure, is well-justified for sub-\muas{} astrometry. 

Nevertheless, a further comment is made about the total sum of neglected terms (\ref{Total_Sum_1_Jupiter}) in case of grazing rays at Jupiter. According to the numerical values presented by 
Tables~\ref{Table_delta_3_4_5} and \ref{Table_T2_35_S2_35}, the dominant contribution of neglected terms originates from the mass-quadrupole, which in case of grazing rays at 
Jupiter amounts in total to $25.3\,{\rm nas}$ \cite{Comment4}. For such extreme configurations in the very vicinity of Jupiter one may replace the simplified mass-quadrupole 
term (i.e. the term with $l=2$ in Eqs.~(\ref{Simplified_Tangent_Vector_1}) - (\ref{Simplified_Tangent_Vector_3})) 
by the exact mass-quadrupole term of the unit tangent vector in 1PN approximation (i.e. the term with $l=2$ of Eqs.~(\ref{delta_k1_M}) - (\ref{delta_k3_M})). 
That quadrupole term of the tangent vector, which is exact in the 1PN approximation, has also been given in its explicit form by Eq.~(14) in \cite{Zschocke6} or 
by Eq.~(47) in \cite{Zschocke7}. In this way one would arrive at a still considerably simplified unit tangent vector, which differs from the exact unit tangent vector (\ref{light_deflection_new}) 
only by terms, which in total contribute less than $11\,{\rm nas}$ for all multipoles with $l \ge 0$ and for all possible astrometric configurations, including the case of grazing light rays. 
The same procedure can, of course, also be applied  in case of grazing ray at Saturn. 

In summary of this Section: the simplified unit tangent vector (\ref{unit_tangent_vector_simplified}) differs from the unit tangent vector (\ref{light_deflection_new}), which is exact in 
the 1.5PN approximation, by a series of neglected terms that are of cumbersome structure, but which contribute in total less than $10\,{\rm nas}$ for all solar system bodies and all 
possible astrometric configurations. Only in the very vicinity of Jupiter, where the light rays of bright sources have an impact parameter of smaller than $1.7$ times the radius of 
Jupiter, one might have to implement the full mass-quadrupole term in the 1PN approximation, in explicit form given in \cite{Zschocke6,Zschocke7}.

\section{Multipoles of axi-symmetric body}\label{Section3}

In each of the expressions of Eqs.~(\ref{Simplified_Tangent_Vector_1}) - (\ref{Simplified_Tangent_Vector_3}) there appears the mass-multipole, $\hat{M}_L$, for an arbitrarily 
shaped body, as given by Eq.~(\ref{Mass_Multipoles}), while in each of the expressions of Eqs.~(\ref{Simplified_Tangent_Vector_4}) - (\ref{Simplified_Tangent_Vector_6}) there 
appears the spin-multipole, $\hat{S}_L$, for an arbitrarily shaped body, as given by Eq.~(\ref{Spin_Multipoles}). In order to determine the magnitude of the individual terms 
in these equations, one needs a model for these solar system bodies to determine these multipoles. To get an idea about the magnitude of these individual terms, 
one may approximate the Sun and the giant planets by a rigid axi-symmetric body with radial dependent mass distribution, having the shape
\begin{eqnarray}
        \frac{\left(x^1\right)^2}{A^2} + \frac{\left(x^2\right)^2}{B^2} + \frac{\left(x^3\right)^2}{C^2} &=& 1\;,
        \label{axisymmetry}
\end{eqnarray}
 
\noindent
where $A = B \neq C$ are the principal axes of the body. 
\begin{figure}[t]
\begin{center}
\includegraphics[scale=0.147]{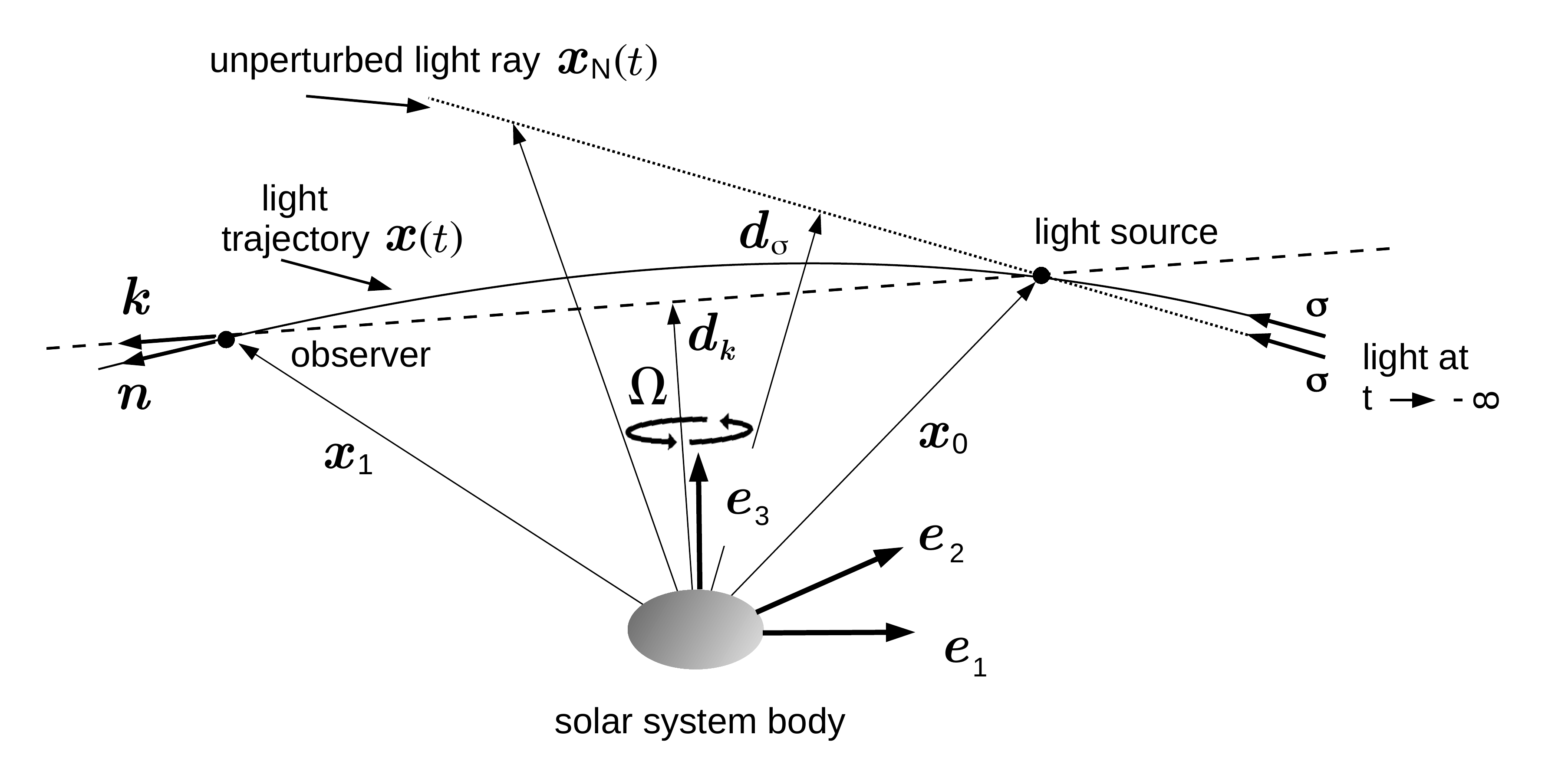}
\end{center}
\caption{
        A geometrical representation of the propagation of a light signal through the gravitational field of an axi-symmetric body at rest. The origin of the spatial 
	Cartesian coordinates $\left(x^1,x^2,x^3\right)$ is assumed to be located at the center-of-mass of the body. The unit vectors $\ve{e}_1,\ve{e}_2,\ve{e}_3$ 
	point in the spatial directions of the principal axes of the massive body, which are orthogonal to each other and co-rotating with the body. The body is in uniform rotational motion around 
	its symmetry axis $\ve{e}_3$ with angular velocity $\Omega$. The rotational axis of the massive body is aligned with the $x^3$-axis of the coordinate system, 
	$\ve{e}_3 = \left(0, 0, 1\right)$. The light signal is emitted by the light source at $\ve{x}_0$ and propagates along the exact light trajectory $\ve{x}\left(t\right)$. 
        The unperturbed light ray $\ve{x}_{\rm N}\left(t\right)$ is given by Eq.~(\ref{Unperturbed_Lightray_2}) and propagates in the direction of $\ve{\sigma}$ 
	along a straight line through the position of the light source at $\ve{x}_0$. The impact vector $\ve{d}_{\sigma}$ is given by Eq.~(\ref{impact_vector})
        and the impact vector $\ve{d}_k$ is given by Eq.~(\ref{impact_vector_k}).}
\label{Diagram2}
\end{figure}

\noindent 
The body is assumed to be in uniform rotational motion with angular velocity $\Omega$ around it symmetry axis $\ve{e}_3$; see Figure~\ref{Diagram2}. This assumption together 
with the axi-symmetry (\ref{axisymmetry}) of the body allows us to assume that the multipoles are time-independent. 

If the coordinate system is chosen such that this symmetry axis $\ve{e}_3$ of the massive body is aligned with the $x^3$-axis of the coordinate system, $\ve{e}_3 = \left(0, 0, 1\right)$, then 
the mass-multipoles (\ref{Mass_Multipoles}) and spin-multipoles (\ref{Spin_Multipoles}) for such a body are given by \cite{Thorne,Poisson_Will,Zschocke_Total_Light_Deflection_15PN,Zschocke_Time_Delay_2PN} 
\begin{eqnarray}
        \hat{M}_0 &=& - M \left(P\right)^0\,J_0\;, 
        \label{M}
        \\ 
        \hat{M}_L &=& - M \left(P\right)^l\,J_l\,\delta^{\;3}_{<{i_1}}\,\dots\,\delta^3_{{i_l}>} \;,
        \label{M_L}
        \\ 
        \hat{S}_{a} &=& - M\,\Omega\left(P\right)^2\,J_0\,\kappa^2\,\delta^3_{a}\;, 
        \label{S}
        \\
        \hat{S}_L &=& - M\,\Omega \left(P\right)^{l+1} \, J_{l-1}\,\frac{l+1}{l+4}\,\delta^{\;3}_{<{i_1}} \; \dots \;\delta^3_{{i_l}>} \;, 
        \label{S_L}
\end{eqnarray}

\noindent
where (\ref{M_L}) is valid for a natural number of $l \ge 2$, while (\ref{S_L}) is valid for a natural number of $l \ge 3$, and $\delta^{3}_{a}$ is the ath component 
of the unit vector $\ve{e}_3$ of the principal axis of the body, which points in the $z$-direction of the coordinate system: $\delta^{3}_{a} = (0,0,1)$. 
Here, $M$ is the Newtonian mass of the body, $P$ its equatorial radius of the body, and $\Omega$ is the angular velocity of the rotating body. 
The parameter $\kappa^2$ in (\ref{S}) is defined by \cite{Ellipticity} (see also Eqs.~(B60) - (B62) in \cite{Zschocke_Time_Delay_2PN})
\begin{eqnarray}
         \kappa^2 &=& \frac{I}{M\,P^2}\;, 
        \label{kappa}
\end{eqnarray}

\noindent
where $I$ is the moment of inertia of the real solar system body under consideration, which is related to the body's angular momentum via $|\ve{S}| = I\,\Omega$. For a spherically
symmetric body with uniform density $\kappa^2 = 2/5$, while for real solar system bodies $\kappa^2 < 2/5$ because the mass densities are increasing toward the center of the massive
bodies. The values of $\kappa^2$ are given in Table~\ref{Table1} for the Sun and giant planets of the solar system bodies.

The parameter $J_l$ are the actual zonal harmonic coefficients of index $l$, defined by \cite{Teyssandier1,Poisson_Will} 
\begin{eqnarray}
        J_l &=& - \frac{1}{M\,\left(P\right)^l} \int d^3 x\,r^l\,\frac{T^{00} + T^{kk}}{c^2}\,P_l\left(\cos \theta\right),
        \label{zonal_harmonic_coefficients}
\end{eqnarray}

\noindent
where $P_l$ are the Legendre polynomials (\ref{Legendre_Polynomials_2}), while the angle $\theta$ is the
colatitude. The zonal harmonic coefficients (\ref{zonal_harmonic_coefficients}) are defined for an axi-symmetric body \cite{Teyssandier1,Poisson_Will}. 
They are model-dependent in the sense that they depend on assumptions made for the
mass distribution in the interior of the solar system bodies. Therefore, it is preferable to use actual zonal harmonics which are deduced from measurements
of the gravitational fields of the Sun and giant planets. Their numerical values are given in Table~\ref{Table1} for the Sun and
the giant planets of the solar system.

These expressions in (\ref{M_L}) and (\ref{S_L}) have been derived in Appendix~B in \cite{Zschocke_Time_Delay_2PN} for an axi-symmetric body. The zonal harmonic coefficient of the 
mass-monopole term is $J_0 = - 1$, as it follows from (\ref{zonal_harmonic_coefficients}). The mass-dipole term vanishes in case the origin of coordinate system is located at the 
center-of-mass of the body: $J_1 = 0$ \cite{Poisson_Will,Kopeikin_Efroimsky_Kaplan,Thorne} and will, therefore, not be considered in what follows. 
\begin{table*}[t]
\centering
\caption{Numerical parameter for mass $M$, equatorial radius $P$, actual zonal harmonic coefficients $J_l$, angular velocity $\Omega = 2\pi/T$ (with rotational period $T$),
dimensionless moment of inertia $\kappa^2$ of the Sun and the giant planets of the solar system. The values for $G M/c^2$ and $P$ are taken from \cite{Ellipticity}.
The value for $J_l$ of the Sun are taken from \cite{J_n_Sun}. The values $J_l$ with $l=2,4,6$ of Jupiter and Saturn are taken
from \cite{Book_Zonal_Harmonics}, while $J_l$ with $l=8,10$ of Jupiter and Saturn are taken from \cite{Zonal_Harmonics_Jupiter} and \cite{Zonal_Harmonics_Saturn}, respectively.
The values $J_l$ with $l=2,4,6$ of Uranus and Neptune are taken from \cite{Zonal_Harmonics_Uranus_Neptune}, while $J_8$ of Uranus and Neptune is taken
from \cite{Zonal_Harmonics_Uranus_Neptune_J8}. The angular velocities $\Omega$ are taken from NASA planetary fact sheets. The values for the dimensionless moment of
inertia $\kappa^2$ are taken from \cite{Ellipticity}. The minimal distance between body and observer, $r^{\rm min}_1$, is computed under the assumption that the observer
is located at Lagrange point $L_2$, i.e. $1.5 \times 10^9\,{\rm m}$ from the Earth's orbit. A blank entry means the values are not known.}
\begin{tabular}{| c | c | c | c | c | c|}
\hline
&&&&&\\[-12pt]
Parameter
&\hbox to 24mm{\hfill Sun \hfill}
&\hbox to 24mm{\hfill Jupiter \hfill}
&\hbox to 24mm{\hfill Saturn \hfill}
&\hbox to 24mm{\hfill Uranus \hfill}
&\hbox to 24mm{\hfill Neptune \hfill}\\[3pt]
\hline
&&&&&\\[-12pt]
$GM/c^2\,[{\rm m}]$ & $1476.8$ & $1.410$ & $0.422$ & $0.064$  & $0.076$ \\[3pt]
$P\,[{\rm m}]$ & $696 \times 10^6$ & $71.49 \times 10^6$ & $60.27 \times 10^6$ & $25.56 \times 10^6$  & $24.76 \times 10^6$ \\[3pt]
$r^{\rm min}_1\,[{\rm m}]$ & $0.147 \times 10^{12}$ & $0.59 \times 10^{12}$ & $1.20 \times 10^{12}$ & $2.57 \times 10^{12}$  & $4.35 \times 10^{12}$ \\[3pt]
$J_2$ & $+ 2.21 \times 10^{-7}$ & $+ 14.696 \times 10^{-3}$ & $+ 16.291 \times 10^{-3}$ & $+ 3.341 \times 10^{-3}$  & $+ 3.408 \times 10^{-3}$ \\[3pt]
$J_4$ & $- 4.46 \times 10^{-9} $ & $ - 0.587 \times 10^{-3}$ & $ - 0.936 \times 10^{-3}$ & $ - 0.031 \times 10^{-3}$  & $ - 0.031 \times 10^{-3}$ \\[3pt]
$J_6$ & $- 2.80 \times 10^{-10} $ & $+ 0.034 \times 10^{-3}$ & $+ 0.086 \times 10^{-3}$ & $+ 0.444 \times 10^{-6}$  & $+ 0.433 \times 10^{-6}$ \\[3pt]
$J_8$ & $+ 1.49 \times 10^{-11} $ & $ - 2.5 \times 10^{-6}$ & $ - 10.0 \times 10^{-6}$ & $ - 0.008 \times 10^{-6}$  & $ - 0.007 \times 10^{-6}$ \\[3pt]
$J_{10}$ & $ - $ & $+ 0.21 \times 10^{-6}$ & $+ 2.0 \times 10^{-6}$ & $ - $  & $ - $ \\[3pt]
$\Omega\,[{\rm sec}^{-1}]$ & $2.865 \times 10^{-6}$ & $1.758 \times 10^{-4}$ & $1.638 \times 10^{-4}$ & $1.012 \times 10^{-4}$  & $1.083 \times 10^{-4}$ \\[3pt]
$\kappa^2$ & $0.059$ & $0.254$ & $0.210$ & $0.225$  & $0.240$ \\[3pt]
\hline
\end{tabular}
\label{Table1}
\end{table*}

\noindent 
The STF tensor ${\rm STF}_{i_1 \dots i_l}\;\delta^{3}_{i_1} \dots \delta^{3}_{i_l}$ in relations (\ref{M_L}) and (\ref{S_L}) denotes products of Kronecker symbols
which are symmetric and traceless with respect to their indices $i_1 \dots i_l$. They are given by the
formula \cite{Blanchet_Damour1,Zschocke_Total_Light_Deflection_15PN,Zschocke_Time_Delay_2PN,Poisson_Will}
\begin{eqnarray}
        \underset{i_1 \dots i_l}{\rm STF}\;\delta^3_{i_1} \, \dots \, \delta^3_{i_l} &=& \sum\limits_{s=0}^{[l/2]} H^l_s\,
        \delta_{\{ i_1 i_2}\,\dots\, \delta_{i_{2s - 1} i_{2s}} \,\delta^{3}_{i_{2s + 1}}\,\dots\, \delta^{3}_{i_l \}}
        \nonumber\\
\label{STF_Expansion}
\end{eqnarray}

\noindent
where the coefficients are given by
\begin{eqnarray}
        H^l_s &=& \left(-1\right)^s\,\frac{\left(2 l - 2 s - 1\right)!!}{\left(2 l - 1 \right)!!}\,.
\label{Coefficient_H_l_s}
\end{eqnarray}

\noindent
The total number $N$ of terms under the sum in (\ref{STF_Expansion}) for a given value of $l$ and $s$ is given by \cite{Poisson_Will}
\begin{eqnarray}
        N^l_s &=& \frac{l!}{(l-2s)!\,(2s)!!}\;.
\label{N_l_s}
\end{eqnarray}

\noindent 
The multipoles in (\ref{M}) - (\ref{S_L}) are valid for a coordinate system, $\left(x^1,x^2,x^3\right)$, where the symmetry axis $\ve{e}_3$ of the massive body is aligned 
with the $x^3$-axis. One may transform the multipoles into another coordinate system, $\left(x^{\prime\,1},x^{\prime\,2},x^{\prime\,3}\right)$, which has the same origin 
of spatial coordinates, but which is rotated to the previous one \cite{MTW,Kopeikin_Efroimsky_Kaplan,Poisson_Will,Arfken_Weber},
\begin{eqnarray}
        x^{a} &=& R^a_b\,x^{\prime\,b}\;,
        \label{Rotation_Coordinate_System}
\end{eqnarray}

\noindent
where the orthogonal matrix of rotation $R^a_b$ can be parametrized, for instance, by three Euler angles, and is given, for example, by Eq.~(3.94) in \cite{Arfken_Weber}.
Then, the multipoles in (\ref{M}) - (\ref{S_L}) in coordinate system $\{x^a\}$ and the multipoles in coordinate system $\{x^{\prime\,a}\}$ are related to each other
by the standard transformation of tensors in three-space \cite{MTW,Kopeikin_Efroimsky_Kaplan,Poisson_Will,Arfken_Weber},
\begin{eqnarray}
        \hat{M}_{i_1 \dots i_l} &=& \hat{M}^{\prime}_{j_1 \dots j_l}\,R^{j_1}_{i_1}\,\dots R^{j_l}_{i_l}\;, 
        \label{Rotation_Mass_Multipoles}
        \\
        \hat{S}_{i_1 \dots i_l} &=& \hat{S}^{\prime}_{j_1 \dots j_l}\,R^{j_1}_{i_1}\,\dots R^{j_l}_{i_l}\;.
        \label{Rotation_Spin_Multipoles}
\end{eqnarray}

\noindent
These relations (\ref{Rotation_Mass_Multipoles}) and (\ref{Rotation_Spin_Multipoles}) allow one to switch the tensorial components of mass-multipoles and spin-multipoles from 
one coordinate system to the other. Clearly, after this coordinate transformation (\ref{Rotation_Coordinate_System}) the body still rotates around its symmetry axis $\ve{e}_3$. 
For an explicit example in case of mass-quadrupole we refer to Eqs.~(48) - (53) in \cite{Klioner2003a}. In what follows, the metric tensor (\ref{Linearized_Gravity_1}) with 
Eqs.~(\ref{Metric_00}) - (\ref{Metric_0i}), but with the time-independent multipoles as given by Eqs.~(\ref{M}) - (\ref{S_L}),  
will be called the metric of an {\it axi-symmetric body}, which is, of course, time-independent because the body's rotation is around the symmetry axis $\ve{e}_3$ 
with uniform angular velocity $\Omega$.

\section{Simplified tangent vector in case of an axi-symmetric body}\label{Simplified_3}

According to Eq.~(\ref{unit_tangent_vector_simplified}), the simplified unit tangent vector of light trajectory is given by 
\begin{eqnarray}
        \widehat{\ve{n}} &=& \ve{k} + \sum\limits_{l=0}^{\infty} \Delta \widehat{\ve{n}}_{M_L} + \sum\limits_{l=1}^{\infty} \Delta \widehat{\ve{n}}_{S_L},
        \label{unit_tangent_vector_simplified_axisymmetric}
\end{eqnarray}

\noindent
with the mass-multipole terms of Eqs.~(\ref{Simplified_Tangent_Vector_1}) - (\ref{Simplified_Tangent_Vector_3}) 
and the spin-multipole terms in  Eqs.~(\ref{Simplified_Tangent_Vector_4}) - (\ref{Simplified_Tangent_Vector_6}). 
In order to find the unit tangent vector of the light ray in the gravitational field of an axi-symmetric body, as elucidated by Figure~\ref{Diagram2}, 
we have to insert the mass-multipoles (\ref{M}) - (\ref{M_L}) and the spin-multipoles (\ref{S}) - (\ref{S_L}) into these equations.

\subsection{The mass-multipole terms}

According to Eq.~(\ref{hat_M_L}) the mass-multipole terms in (\ref{unit_tangent_vector_simplified_axisymmetric}) are  
\begin{eqnarray}
        \Delta \widehat{\ve{n}}_{M_L} &=& \Delta \widehat{\ve{n}}^{1}_{M_L} + \Delta \widehat{\ve{n}}^{2}_{M_L} + \Delta \widehat{\ve{n}}^{3}_{M_L}, 
         \label{hat_M_L_axisymmetric}
\end{eqnarray}

\noindent 
with the individual terms of Eqs.~(\ref{Simplified_Tangent_Vector_1}) - (\ref{Simplified_Tangent_Vector_3}).

\subsubsection{The mass-monopole term: $l=0$}

By inserting the mass-monopole (\ref{M}) into these Eqs.~(\ref{Simplified_Tangent_Vector_1}) - (\ref{Simplified_Tangent_Vector_3}) one obtains for $l=0$,  
\begin{eqnarray}
	\Delta \widehat{\ve{n}}_{M_{l=0}} &=& - \frac{2 G M}{c^2}\,\frac{\ve{d_k}}{\left(d_k\right)^2}\,\frac{r_0 r_1 - \ve{x}_1 \cdot \ve{x}_1}{R\,r_1}, 
        \label{delta_n_Mass_Monopole} 
\end{eqnarray}

\noindent
where we have used $R = \ve{k} \cdot \ve{x}_1 - \ve{k} \cdot \ve{x}_0$ and the parameter $P^{ij} \xi_j$ has been replaced by the impact vector $d^i_k$. The same result can be 
obtained from the mass-monopole terms (\ref{delta_k1_M}) - (\ref{delta_k3_M}) which are exact in the 1PN approximation. That means, the mass-monopole terms of the simplified 
tangent vector (\ref{unit_tangent_vector_simplified_axisymmetric}) are identical with the mass-monopole terms of the tangent vector (\ref{light_deflection_new}), which is 
exact in the 1.5PN approximation.  
The monopole term (\ref{delta_n_Mass_Monopole}) coincides with the monopole term of Eq.~(24) in \cite{Article_Zschocke1} for Parameterized Post-Newtonian parameter $\gamma=1$.

\subsubsection{The mass-multipole terms: $l \ge 2$}

As mentioned above, the mass-dipole term vanishes, so we need to consider only higher mass-multipoles with $l \ge 2$. By inserting the mass-multipoles (\ref{M_L}) of an 
axi-symmetric body into these Eqs.~(\ref{Simplified_Tangent_Vector_1}) - (\ref{Simplified_Tangent_Vector_3}) one obtains 
for the higher mass-multipoles 
\begin{eqnarray}
	\Delta \widehat{\ve{n}}^{1}_{M_L} &=& - \frac{2 G M}{c^2} \frac{J_l}{l}  \left(1 + \frac{\ve{k} \cdot \ve{x}_1}{r_1}\right) 
	\frac{\partial}{\partial\ve{\hat{\xi}}} \left(\frac{P}{\hat{\xi}}\right)^l F^l_M\,,
        \label{Simplified_n1}
        \\
        \nonumber\\
	\Delta \widehat{\ve{n}}^{2}_{M_L} &=& + \frac{2 G M}{c^2} \frac{J_l}{l} 
	\frac{\ve{k} \cdot \ve{x}_1}{R} \left(1 + \frac{\ve{k} \cdot \ve{x}_1}{r_1}\right) 
	\frac{\partial}{\partial\ve{\hat{\xi}}} \left(\frac{P}{\hat{\xi}}\right)^l F^l_M\,, 
        \nonumber\\
        \label{Simplified_n2}
        \\ 
        \nonumber\\
	\Delta \widehat{\ve{n}}^{3}_{M_L} &=& - \frac{2 G M}{c^2} \frac{J_l}{l} 
	\frac{\ve{k} \cdot \ve{x}_0}{R} \left(1 + \frac{\ve{k} \cdot \ve{x}_0}{r_0}\right) 
	\frac{\partial}{\partial\ve{\hat{\xi}}} \left(\frac{P}{\hat{\xi}}\right)^l F^l_M\,, 
        \nonumber\\
        \label{Simplified_n3}
\end{eqnarray} 

\noindent
which is valid for $l \ge 2$. The scalar function reads 
\begin{eqnarray}
	F^l_M &=& \frac{1}{\left(l-1\right)!}\;\sum\limits_{n=0}^{[l/2]} G_n^l  
	\left(\underset{i_1 \dots i_l}{\rm STF}\;\delta^3_{i_1} \, \dots \, \delta^3_{i_l}\right) 
        \nonumber\\ 
        && \times\,\left(-1\right)^n k_{i_1} \, \dots \, k_{i_{2 n}}
	\frac{{\hat{\xi}}^{i_{2 n + 1}}\,\dots\,\hat{\xi}^{i_l}}{(\hat{\xi})^{l-2n}} \,,
        \label{Function_F_l_M}
\end{eqnarray}

\noindent
where the STF operation in the second line in (\ref{Function_F_l_M}) can be omitted in view of relation (\ref{STF_comment_3}). The structure of this function of Eq.~(\ref{Function_F_l_M}) 
is, with regard to relation (\ref{sigma_k_1}), identical with the function of Eq.~(D3) in our previous investigation \cite{Zschocke_Total_Light_Deflection_15PN}, where it has been shown 
that (\ref{Function_F_l_M}) can be written in the following form (cf. Eq.~(92) in \cite{Zschocke_Total_Light_Deflection_15PN}), 
\begin{eqnarray}
	F^l_M &=& \frac{1}{\left(l-1\right)!}\;\sum\limits_{n=0}^{[l/2]} G_n^l \left(1 - \left(\ve{k} \cdot \ve{e}_3\right)^2\right)^n  
	\left(\frac{\ve{d}_k \cdot \ve{e}_3}{d_k}\right)^{l-2n}, 
	\nonumber\\ 
        \label{Function_F_l_M_final1}
\end{eqnarray}

\noindent
where $\ve{\hat{\xi}}$ has been replaced by the impact vector $\ve{d}_k$. 
It was one of the primary results in our previous investigation \cite{Zschocke_Total_Light_Deflection_15PN}, that this function can be expressed in terms of 
Chebyshev polynomials of first kind (cf. Eq.~(95) in \cite{Zschocke_Total_Light_Deflection_15PN})  
\begin{eqnarray}
	F^l_M &=& \left(1 - \left(\ve{k} \cdot \ve{e}_3\right)^2\right)^{[l/2]}\; T_l\left(x\right), 
        \label{Function_F_l_M_final2}
\end{eqnarray}

\noindent 
where the argument of these Chebyshev polynomials is given by (cf. Eq.~(93) in \cite{Zschocke_Total_Light_Deflection_15PN}) 
\begin{eqnarray}
	x &=& \left(1 - \left(\ve{k} \cdot \ve{e}_3\right)^2\right)^{-1/2}\,\left(\frac{\ve{d}_k \cdot \ve{e}_3}{d_k}\right),  
	 \label{Argument_x}
\end{eqnarray}

\noindent 
which is a real number of the closed interval $x \in \left[-1,+1\right]$.

\subsection{The spin-multipole terms} 

According to Eq.~(\ref{hat_S_L}) the spin-multipole terms in (\ref{unit_tangent_vector_simplified_axisymmetric}) are 
\begin{eqnarray}
        \Delta \widehat{\ve{n}}_{S_L} &=& \Delta \widehat{\ve{n}}^{1}_{S_L} + \Delta \widehat{\ve{n}}^{2}_{S_L} + \Delta \widehat{\ve{n}}^{3}_{S_L},  
         \label{hat_S_L_axisymmetric}
\end{eqnarray}

\noindent 
with the individual terms of Eqs.~(\ref{Simplified_Tangent_Vector_4}) - (\ref{Simplified_Tangent_Vector_6}).

\subsubsection{The spin-dipole term: $l=1$} 

By inserting the spin-dipole (\ref{S}) into these Eqs.~(\ref{Simplified_Tangent_Vector_4} - (\ref{Simplified_Tangent_Vector_6}) one obtains for $l=1$,  
\begin{eqnarray}
	\Delta \widehat{\ve{n}}_{S_{l=1}} &=& - \frac{2 G M}{c^3} \,\Omega\,\kappa^2 \left(\frac{P}{d_k}\right)^2 \left(1 + \frac{\ve{k} \cdot \ve{x}_1}{r_1}\right) 
        \nonumber\\
        && \hspace{-0.0cm} \times \Bigg[2\,\frac{\left(\ve{k} \times \ve{d}_k\right) \cdot \ve{e}_3}{d_k}\,\frac{\ve{d}_k}{d_k} 
        + \left(\ve{k} \times \ve{e}_3\right)\Bigg]  
        \nonumber\\ 
        \nonumber\\ 
        && \hspace{-1.25cm} + \frac{2 G M}{c^3} \,\Omega\,\kappa^2 \left(\frac{P}{d_k}\right)^2 
	\left(1 + \frac{\left(\ve{k} \cdot \ve{x}_1\right)^2}{R\,r_1} - \frac{\left(\ve{k} \cdot \ve{x}_0\right)^2}{R\,r_0}\right)  
        \nonumber\\
        && \hspace{-0.0cm} \times \Bigg[2\,\frac{\left(\ve{k} \times \ve{d}_k\right) \cdot \ve{e}_3}{d_k}\,\frac{\ve{d}_k}{d_k}
        + \left(\ve{k} \times \ve{e}_3\right)\Bigg], 
        \label{delta_n_Spin_Dipole} 
\end{eqnarray}

\noindent
where we have used $R = \ve{k} \cdot \ve{x}_1 - \ve{k} \cdot \ve{x}_0$ and the parameter $\ve{\hat{\xi}}$ has been replaced by the impact vector $\ve{d}_k$. 
In the limit of spatial infinity of the source the second term of Eq.~(\ref{delta_n_Spin_Dipole}) vanishes. Then, in the limit of spatial infinity of the 
observer, the first term of Eq.~(\ref{delta_n_Spin_Dipole}) reduces to Eq.~(60) in \cite{Klioner1991}, 
Eq.~(72) in \cite{Kopeikin_Mashhoon}, and Eq.~(98) in \cite{Zschocke_Total_Light_Deflection_15PN}.

\subsubsection{The spin-multipole terms: $l \ge 3$} 

By inserting the spin-multipoles (\ref{S_L}) of an axi-symmetric body into these Eqs.~(\ref{Simplified_Tangent_Vector_4}) - (\ref{Simplified_Tangent_Vector_6}) one 
obtains for the higher spin-multipoles  
\begin{eqnarray}
	\Delta \widehat{\ve{n}}^{1}_{S_L} &=& - \frac{4 G M}{c^3}\,\Omega\,P\, 
	\frac{J_{l-1}}{l+4} \left(1 + \frac{\ve{k} \cdot \ve{x}_1}{r_1}\right) \frac{\partial}{\partial\ve{\hat{\xi}}} \left(\frac{P}{\hat{\xi}}\right)^l F^l_S\,, 
        \nonumber\\
        \label{Simplified_n4}
        \\
        \nonumber\\
	\Delta \widehat{\ve{n}}^{2}_{S_L} &=& + \frac{4 G M}{c^3}\,\Omega\,P\,\frac{J_{l-1}}{l+4}\,
	\frac{\ve{k} \cdot \ve{x}_1}{R} \left(1 + \frac{\ve{k} \cdot \ve{x}_1}{r_1}\right) 
        \nonumber\\
	&& \times\,\frac{\partial}{\partial\ve{\hat{\xi}}} \left(\frac{P}{\hat{\xi}}\right)^l F^l_S\,,
        \label{Simplified_n5}
        \\
        \nonumber\\
	\Delta \widehat{\ve{n}}^{3}_{S_L} &=& - \frac{4 G M}{c^3}\,\Omega\,P\,\frac{J_{l-1}}{l+4}\,
	\frac{\ve{k} \cdot \ve{x}_0}{R} \left(1 + \frac{\ve{k} \cdot \ve{x}_0}{r_0}\right) 
        \nonumber\\
	&& \times\,\frac{\partial}{\partial\ve{\hat{\xi}}} \left(\frac{P}{\hat{\xi}}\right)^l F^l_S\,,
        \label{Simplified_n6}
\end{eqnarray}

\noindent
which is valid for $l > 1$. The pseudo-scalar function reads 
\begin{eqnarray}
	&& \hspace{-1.0cm} F^l_S = \frac{1}{\left(l-1\right)!}\,\epsilon_{i_l bc}\,k^c 
	\left(\underset{b \dots i_{l-1}}{\rm STF}\;\delta^3_{b} \, \delta^3_{i_1} \, \dots \, \delta^3_{i_{l-1}}\right) 
        \nonumber\\ 
        && \hspace{-0.9cm} \times\,\underset{i_1 \dots i_l}{\rm STF}   
        \left[\sum\limits_{n=0}^{[l/2]} G_n^l\,\left(-1\right)^n k_{i_1}\,\dots\,k_{i_{2 n}} 
	\frac{\hat{\xi}^{i_{2 n +1 }}\,\dots\,\hat{\xi}^{i_{l}}}{(\hat{\xi})^{l - 2 n}} \right].  
        \label{Function_F_l_S}
\end{eqnarray}

\noindent 
The structure of this function of Eq.~(\ref{Function_F_l_S}) is, with regard to relation (\ref{sigma_k_1}), identical with the function of Eq.~(E3) in our previous 
investigation \cite{Zschocke_Total_Light_Deflection_15PN}, where it has been shown that (\ref{Function_F_l_S}) can be written in the following form (cf. Eq.~(100) 
in \cite{Zschocke_Total_Light_Deflection_15PN}),
\begin{eqnarray}
	F^l_S &=& \frac{1}{\left(l-1\right)!}\,\frac{\left(\ve{k} \times \ve{d}_k\right)\cdot \ve{e}_3}{d_k}
	\nonumber\\ 
	&& \hspace{-0.75cm} \times \sum\limits_{n=0}^{[l/2]} G_n^l \frac{l-2n}{l} \,\left(1 - \left(\ve{k} \cdot \ve{e}_3\right)^2\right)^{n}
	\left(\frac{\ve{d}_k \cdot \ve{e}_3}{d_k}\right)^{l-2n-1}. 
	\nonumber\\ 
        \label{Function_F_l_S_final1}
\end{eqnarray}

\noindent 
It was one of the primary results in our investigation, that this function can be expressed in terms of Chebyshev polynomials of second kind 
(cf. Eq.~(105) in \cite{Zschocke_Total_Light_Deflection_15PN}), 
\begin{eqnarray}
	F^l_S &=& \frac{\left(\ve{k} \times \ve{d}_k\right) \cdot \ve{e}_3}{d_k} \,\left(1 - \left(\ve{k} \cdot \ve{e}_3\right)^2\right)^{[l/2]}\,U_{l-1}\left(x\right),
	\nonumber\\ 
\end{eqnarray}

\noindent 
where the argument of these Chebyshev polynomials has been given by Eq.~(\ref{Argument_x}).

\section{Comparison with the literature}\label{Monopole_Quadrupole_Spin}

The mass-monopole and spin-dipole terms of the unit tangent vector of a light ray in the field of an axi-symmetric body, as presented by Eqs.~(\ref{delta_n_Mass_Monopole}) 
and (\ref{delta_n_Spin_Dipole}), have already been compared with the literature. 

Results for the unit tangent vector of light rays in case of finite distances of source and observer for higher multipoles beyond the mass-monopole and spin-dipole are very rare. 
So we are restricted here to consider only the next term of the multipole expansion, which is the mass-quadrupole term ($l=2$) of the unit tangent vector, with existing results 
in the literature. From Eqs.~(\ref{Simplified_n1}) - (\ref{Simplified_n3}) we get for $l=2$: 
\begin{eqnarray}
	\Delta \widehat{\ve{n}}_{M_{l=2}} &=& - \frac{GM}{c^2}\,J_2 \left(\frac{P}{d_k}\right)^2 {\cal A} 
	\nonumber\\
	&& \hspace{-1.5cm} \times \Bigg[ \frac{\ve{d}_k}{d_k} - \left(\ve{k} \cdot \ve{e}_3\right)^2 \frac{\ve{d}_k}{d_k} 
	- 4 \left(\frac{\ve{d}_k \cdot \ve{e}_3}{d_k}\right)^2 \frac{\ve{d}_k}{d_k} 
        \nonumber\\
	&& \hspace{-1.5cm} +\, 2 \left(\frac{\ve{d}_k \cdot \ve{e}_3}{d_k}\right) \ve{e}_3 
	- 2 \left(\ve{k} \cdot \ve{e}_3\right) \left(\frac{\ve{d}_k \cdot \ve{e}_3}{d_k}\right) \ve{k} \bigg] 
	\label{Comparison_Quadrupole_1}
\end{eqnarray}

\noindent
with the dimensionless scalar function 
\begin{eqnarray}
	{\cal A} &=& + \left(1 + \frac{\ve{k} \cdot \ve{x}_0}{r_0}\right) \frac{r_0 + \ve{k} \cdot \ve{x}_0}{R} 
	\nonumber\\ 
	&& - \left(1 + \frac{\ve{k} \cdot \ve{x}_1}{r_1}\right) \frac{r_1 + \ve{k} \cdot \ve{x}_1}{R}
	+ 2 \left(1 + \frac{\ve{k} \cdot \ve{x}_1}{r_1}\right),
	\nonumber\\ 
        \label{Comparison_Quadrupole_2}
\end{eqnarray}

\noindent
where the parameter $P^{ij} \xi_j$ has been replaced by $d^i_k$, in line with the statements below Eq.~(\ref{Second_Integration_6}) and 
taking account of relations (\ref{xN_x0}) and (\ref{xN_x1}). 
The scalar function of Eq.~(\ref{Comparison_Quadrupole_2}) is, up to a conventional constant $\left(d_k\right)^3$, identical to the scalar function of Eq.~(A.10) in \cite{Zschocke6} 
and to the scalar function of Eq.~(48) in \cite{Zschocke7}, up to a term of the order $\displaystyle \left(P/r_1\right)^2$ which can safely be neglected for sub-micro-arcsecond 
astrometry. That means, the expression in (\ref{Comparison_Quadrupole_1}) further simplifies the previous results in our articles \cite{Zschocke6,Zschocke7} for the mass-quadrupole. 
Similarly, the term in the square brackets of Eq.~(\ref{Comparison_Quadrupole_1}) is, up to a conventional constant $d_k $, identical to the term in the square brackets of Eq.~(57) 
in our article \cite{Zschocke7}.

\section{The angle of light deflection}\label{Section4}

\subsection{Light deflection at finite spatial distances}

The light deflection is defined as angle between three-vector $\ve{k}$ of Eq.~(\ref{vector_k}) and three-vector $\ve{n}$ of Eq.~(\ref{vector_n}), 
\begin{eqnarray}
        \delta\left(\ve{k},\ve{n}\right) &=& \arcsin \left|\ve{k} \times \ve{n}\right|.  
        \label{light_deflection_10}
\end{eqnarray}

\noindent
A graphical elucidation is presented by Figures~\ref{Diagram1} and \ref{Diagram2}. 
By inserting (\ref{light_deflection_new}) into (\ref{light_deflection_10}) the light deflection can be written in the following form,
\begin{eqnarray} 
        \delta\left(\ve{k},\ve{n}\right) &=& \left|\sum\limits_{l=0}^{\infty} \ve{k} \times \Delta\ve{n}_{M_L}  
	+ \sum\limits_{l=1}^{\infty} \ve{k} \times \Delta\ve{n}_{S_L}\right| + {\cal O}\left(c^{-4}\right), 
	\nonumber\\ 
        \label{light_deflection_20}
\end{eqnarray}

\noindent
where $\arcsin x = x + {\cal O}(x^3)$ for $x \ll 1$ has been used. This equation can also be compared with Eq.~(59) in \cite{Zschocke_Total_Light_Deflection_15PN},
which was only valid for the total light deflection, that means for infinite distances between source and observer from the solar system body, while (\ref{light_deflection_20}) 
is the angle of light deflection at finite distances between source and observer from the solar system body. The individual terms of Eq.~(\ref{light_deflection_20}) are given 
by Eqs.~(\ref{light_deflection_M_new}) and (\ref{light_deflection_S_new}).

The effect of the monopole term is by far the most dominant term of light deflection and several orders larger than the higher multipole terms in (\ref{light_deflection_20}).
Accordingly, it is appropriate to take the monopole term, given by Eq.~(\ref{delta_n_Mass_Monopole}), in front of the right-hand side of Eq.~(\ref{light_deflection_20}) and to perform 
a series expansion. In this way one obtains the following form for the light deflection, 
\begin{eqnarray}
        \delta\left(\ve{k},\ve{n}\right) &=& \sum\limits_{l=0}^{\infty} \delta\left(\ve{k},\Delta\ve{n}_{M_L}\right) 
	+ \sum\limits_{l=1}^{\infty} \delta\left(\ve{k},\Delta\ve{n}_{S_L}\right) + {\cal O}\left(c^{-4}\right),
	\nonumber\\ 
        \label{light_deflection_25}
\end{eqnarray}

\noindent
where the individual multipole terms are given by
\begin{eqnarray}
	\delta\left(\ve{k},\Delta\ve{n}_{M_L}\right) &=& - \Delta\ve{n}_{M_L} \cdot \frac{\ve{\hat{\xi}}}{\hat{\xi}}\,,  
        \label{light_deflection_M_L}
	\\
	\delta\left(\ve{k},\Delta\ve{n}_{S_L}\right) &=& - \Delta\ve{n}_{S_L} \cdot \frac{\ve{\hat{\xi}}}{\hat{\xi}}\,.
        \label{light_deflection_S_L}
\end{eqnarray}

\noindent
A similar step has been used in \cite{Klioner1991,Kopeikin1997,Zschocke_Total_Light_Deflection_15PN}. One may compare Eqs.~(\ref{light_deflection_M_L}) 
and (\ref{light_deflection_S_L}) with Eqs.~(61) and (62) in \cite{Zschocke_Total_Light_Deflection_15PN}, respectively, which were valid for the case of 
source and observer at infinite spatial distances from the body, while here we consider the case of source and observer at finite distances from the 
gravitating solar system body. 

Now we will use the mass-multipole and spin-multipole terms of the simplified unit tangent vector (\ref{unit_tangent_vector_simplified}) 
and obtain for the light deflection angle (\ref{light_deflection_25}), 
\begin{eqnarray}
	\delta\left(\ve{k},\widehat{\ve{n}}\right) &=& \sum\limits_{l=0}^{\infty} \delta\left(\ve{k},\Delta\widehat{\ve{n}}_{M_L}\right) 
	+ \sum\limits_{l=1}^{\infty} \delta\left(\ve{k},\Delta\widehat{\ve{n}}_{S_L}\right) + {\cal O}\left(c^{-4}\right),
        \nonumber\\ 
        \label{light_deflection_26}
\end{eqnarray}

\noindent
where the individual multipole terms are given by
\begin{eqnarray}
	&& \hspace{-0.85cm} 
	\delta\left(\ve{k},\Delta\widehat{\ve{n}}_{M_L}\right) = 
	- \left(\Delta\widehat{\ve{n}}_{M_L}^{1} + \Delta\widehat{\ve{n}}_{M_L}^{2} + \Delta\widehat{\ve{n}}_{M_L}^{3} \right) \cdot \frac{\ve{\hat{\xi}}}{\hat{\xi}}\,,
        \label{light_deflection_Final_M_L}
	\\
	&& \hspace{-0.75cm}
	\delta\left(\ve{k},\Delta\widehat{\ve{n}}_{S_L}\right) = 
	- \left(\Delta\widehat{\ve{n}}_{S_L}^{1} + \Delta\widehat{\ve{n}}_{S_L}^{2} + \Delta\widehat{\ve{n}}_{S_L}^{3} \right) \cdot \frac{\ve{\hat{\xi}}}{\hat{\xi}}\,,
        \label{light_deflection_Final_S_L}
\end{eqnarray}

\noindent
with the individual terms of Eqs.~(\ref{Simplified_n1}) - (\ref{Simplified_n3}) and Eqs.~(\ref{Simplified_n4}) - (\ref{Simplified_n6}). The difference between these equations 
for the light deflection angle is, that Eq.~(\ref{light_deflection_25}) contains all terms of the unit tangent vector, while Eq.~(\ref{light_deflection_26}) contains only those 
terms that are relevant for the given goal accuracy on the level of $0.01\,\muas$ ($10$ nas). 

In order to get the angle of Eqs.~(\ref{light_deflection_Final_M_L}) and (\ref{light_deflection_Final_S_L}) one would have to perform the 
differentiation $\partial/\partial \ve{\hat{\xi}}$ of Eqs.~(\ref{Simplified_n1}) - (\ref{Simplified_n3}) as well as Eqs.~(\ref{Simplified_n4}) - (\ref{Simplified_n6}). 
A compact form for these individual terms can be achieved by means of the following relations  
\begin{eqnarray}
	\frac{\ve{\hat{\xi}}}{\hat{\xi}} \cdot \left(\frac{\partial}{\partial \ve{\hat{\xi}}}\, f\left(\ve{\hat{\xi}}\right)\right) 
	&=& \frac{\partial}{\partial \hat{\xi}}\,f\left(\ve{\hat{\xi}}\right), 
        \label{Relation_A}
\end{eqnarray}

\noindent
which is valid for scalar functions which depend on the absolute value $\xi = |\ve{\xi}|$, and 
\begin{eqnarray}
	\frac{\ve{\hat{\xi}}}{\hat{\xi}} \cdot \left(\frac{\partial}{\partial \ve{\hat{\xi}}}\,\hat{\xi}_{i_n} \right) 
	&=& \frac{\partial}{\partial \hat{\xi}}\,\hat{\xi}_{i_n} = \hat{n}_{i_n}\,, 
        \label{Relation_B}
\end{eqnarray}

\noindent 
which is valid for vectorial components of $\ve{\hat{\xi}}$ and where $\hat{n}_{i_n} = \hat{\xi}_{i_n}/\hat{\xi}$ is the unit direction of the impact vector; 
we also note that 
\begin{eqnarray}
	\frac{\partial}{\partial \hat{\xi}}\,\hat{n}_{i_n} &=& \frac{\partial}{\partial \hat{\xi}}\,\frac{\hat{\xi}_{i_n}}{\hat{\xi}} = 0 
        \label{Relation_C}
\end{eqnarray}

\noindent 
which reflects the fact that the unit direction of a three-vector is independent of the spatial length of this three-vector. 
Here we refer to the abbreviation which has been introduced by Eq.~(\ref{Abbreviation}). 
Then, using relations (\ref{Relation_A}) and (\ref{Relation_B}), the calculation of these individual terms simplifies considerably 
and yield for the mass-multipole terms 
\begin{eqnarray}
	\Delta\widehat{\ve{n}}_{M_L}^{1}\cdot \frac{\ve{\hat{\xi}}}{\hat{\xi}} &=& - \frac{2 G M}{c^2}\,\frac{J_{l}}{l} 
	\left(1 + \frac{\ve{k} \cdot \ve{x}_1}{r_1}\right) 
        \nonumber\\ 
	&& \times\,\frac{\partial}{\partial \hat{\xi}} \left(\frac{P}{\hat{\xi}}\right)^l  \left|\ve{k} \times \ve{e}_3\right|^{l}\;T_l\left(x\right), 
        \label{light_deflection_angle_term_1}
	\\
	\nonumber\\
	\Delta\widehat{\ve{n}}_{M_L}^{2}\cdot \frac{\ve{\hat{\xi}}}{\hat{\xi}} &=& + \frac{2 G M}{c^2}\,\frac{J_{l}}{l}\,
        \frac{\ve{k} \cdot \ve{x}_1}{R} \left(1 + \frac{\ve{k} \cdot \ve{x}_1}{r_1}\right) 
        \nonumber\\
	&& \times\,\frac{\partial}{\partial \hat{\xi}} \left(\frac{P}{\hat{\xi}}\right)^l \left|\ve{k} \times \ve{e}_3\right|^{l}\;T_l\left(x\right), 
        \label{light_deflection_angle_term_2}
	\\
	\nonumber\\
	\Delta\widehat{\ve{n}}_{M_L}^{3}\cdot \frac{\ve{\hat{\xi}}}{\hat{\xi}} &=& - \frac{2 G M}{c^2}\,\frac{J_{l}}{l}\,
        \frac{\ve{k} \cdot \ve{x}_0}{R} \left(1 + \frac{\ve{k} \cdot \ve{x}_0}{r_0}\right) 
        \nonumber\\
	&& \times\,\frac{\partial}{\partial \hat{\xi}} \left(\frac{P}{\hat{\xi}}\right)^l 
	\left|\ve{k} \times \ve{e}_3\right|^{l}\;T_l\left(x\right), 
        \label{light_deflection_angle_term_3}
\end{eqnarray}

\noindent 
and for the spin-multipole terms 
\begin{eqnarray}
	\Delta\widehat{\ve{n}}_{S_L}^{1}\cdot \frac{\ve{\hat{\xi}}}{\hat{\xi}} &=& - \frac{4 G M}{c^3}\,\Omega\,P\, 
        \frac{J_{l-1}}{l+4} \left(1 + \frac{\ve{k} \cdot \ve{x}_1}{r_1}\right) 
        \nonumber\\ 
	&& \hspace{-1.5cm} \times\,\frac{\partial}{\partial \hat{\xi}} \left(\frac{P}{\hat{\xi}}\right)^l 
	\frac{\left(\ve{k} \times \ve{\hat{\xi}}\right)\cdot \ve{e}_3}{\hat{\xi}} \,\left|\ve{k} \times \ve{e}_3\right|^{l}\,U_{l-1}\left(x\right), 
        \label{light_deflection_angle_term_4}
        \\
	\nonumber\\
	\Delta\widehat{\ve{n}}_{S_L}^{2}\cdot \frac{\ve{\hat{\xi}}}{\hat{\xi}} &=& + \frac{4 G M}{c^3}\,\Omega\,P\,\frac{J_{l-1}}{l+4}\,
        \frac{\ve{k} \cdot \ve{x}_1}{R} \left(1 + \frac{\ve{k} \cdot \ve{x}_1}{r_1}\right) 
        \nonumber\\
	&& \hspace{-1.5cm} \times\,\frac{\partial}{\partial \hat{\xi}} \left(\frac{P}{\hat{\xi}}\right)^l 
	\frac{\left(\ve{k} \times \ve{\hat{\xi}}\right) \cdot \ve{e}_3}{\hat{\xi}} \,\left|\ve{k} \times \ve{e}_3\right|^{l}\,U_{l-1}\left(x\right), 
        \label{light_deflection_angle_term_5}
        \\
	\nonumber\\
	\Delta\widehat{\ve{n}}_{S_L}^{3}\cdot \frac{\ve{\hat{\xi}}}{\hat{\xi}} &=& - \frac{4 G M}{c^3}\,\Omega\,P\,\frac{J_{l-1}}{l+4}\,
        \frac{\ve{k} \cdot \ve{x}_0}{R} \left(1 + \frac{\ve{k} \cdot \ve{x}_0}{r_0}\right) 
        \nonumber\\
	&& \hspace{-1.5cm} \times\,\frac{\partial}{\partial \hat{\xi}} \left(\frac{P}{\hat{\xi}}\right)^l 
	\frac{\left(\ve{k} \times \ve{\hat{\xi}}\right)\cdot \ve{e}_3}{\hat{\xi}} \,\left|\ve{k} \times \ve{e}_3\right|^{l}\,U_{l-1}\left(x\right),
        \label{light_deflection_angle_term_6}
\end{eqnarray}

\noindent
where the argument $x$ is defined by Eq.~(\ref{Argument_x}) and $T_l$ and $U_l$ are Chebyshev polynomials of first and second kind, given by 
Eqs.~(\ref{Chebyshev_Polynomials_0}) - (\ref{Chebyshev_Polynomials_1}) and (\ref{Chebyshev_Polynomials_2}), respectively.  
Eqs.~(\ref{light_deflection_angle_term_1}) - (\ref{light_deflection_angle_term_3}) and (\ref{light_deflection_angle_term_4}) - (\ref{light_deflection_angle_term_6})
have to be inserted into (\ref{light_deflection_Final_M_L}) and (\ref{light_deflection_Final_S_L}) and yield the angle of light deflection at the observers position  
in terms of the global coordinate system. 
\begin{table*}[t]
\centering
\caption{The upper limit of light deflection at the Sun and the giant planets of the solar system caused by their mass-multipole structure according to
        Eq.~(\ref{upper_limit_M_L}). For the numerical value we assume a grazing light ray, $d_k = P$, that means the impact parameter equals the radius of
        the body. All values are given in micro-arcsecond (\muas). A blank entry indicates that the light deflection is less than $1$ nano-arcsecond (nas).}
\begin{tabular}{| c | c | c | c | c | c|}
\hline
&&&&&\\[-12pt]
Light deflection
&\hbox to 20mm{\hfill Sun \hfill}
&\hbox to 20mm{\hfill Jupiter \hfill}
&\hbox to 20mm{\hfill Saturn \hfill}
&\hbox to 20mm{\hfill Uranus \hfill}
&\hbox to 20mm{\hfill Neptune \hfill}\\[3pt]
\hline
&&&&&\\[-12pt]
        $|\delta(\ve{k}, \Delta\widehat{\ve{n}}_{M_0})|$ & $1.75 \times 10^{6}$ & $16.3 \times 10^{3}$ & $5.8 \times 10^{3}$ & $ 2.1 \times 10^{3} $  & $ 2.5 \times 10^{3} $ \\[3pt]
        $|\delta(\ve{k}, \Delta\widehat{\ve{n}}_{M_2})|$ & $ 0.455 $ & $ 239 $ & $ 94$ & $ 6.9 $  & $ 8.6$ \\[3pt]
        $|\delta(\ve{k}, \Delta\widehat{\ve{n}}_{M_4})|$ & $ 0.008 $ & $ 9.6 $ & $ 5.41 $ & $ 0.06 $  & $ 0.08 $ \\[3pt]
        $|\delta(\ve{k}, \Delta\widehat{\ve{n}}_{M_6})|$ & $ - $ & $ 0.55 $ & $ 0.50 $ & $ 0.001 $  & $ 0.001 $ \\[3pt]
        $|\delta(\ve{k}, \Delta\widehat{\ve{n}}_{M_8})|$ & $ - $ & $0.04 $ & $ 0.06 $ & $ - $  & $ - $ \\[3pt]
        $|\delta(\ve{k}, \Delta\widehat{\ve{n}}_{M_{10}})|$ & $ - $ & $0.003 $ & $ 0.01 $ & $ - $  & $ - $ \\[3pt]
\hline
\end{tabular}
\label{Table2}
\end{table*}
%
\begin{table*}[ht]
\centering 
\caption{The upper limit of light deflection at the Sun and the giant planets of the solar system caused by their spin-multipole structure to
Eqs.~(\ref{upper_limit_S_1}) - (\ref{upper_limit_S_L}). For the numerical value we assume a grazing light ray, $d_k = P$, that means the
impact parameter equals the radius of the body.
All values are given in micro-arcsecond (\muas). A blank entry indicates that the light deflection is less than $1$ nano-arcsecond (nas).}
\begin{tabular}{| c | c | c | c | c | c|}
\hline
&&&&&\\[-12pt]
Light deflection
&\hbox to 20mm{\hfill Sun \hfill}
&\hbox to 20mm{\hfill Jupiter \hfill}
&\hbox to 20mm{\hfill Saturn \hfill}
&\hbox to 20mm{\hfill Uranus \hfill}
&\hbox to 20mm{\hfill Neptune \hfill}\\[3pt]
\hline
&&&&&\\[-12pt]
        $|\delta(\ve{k}, \Delta\widehat{\ve{n}}_{S_1})|$ & $ 0.7 $ & $ 0.17 $ & $ 0.04 $ & $ 0.004 $  & $ 0.005 $ \\[3pt]
        $|\delta(\ve{k}, \Delta\widehat{\ve{n}}_{S_3})|$ & $ - $ & $ 0.026 $ & $ 0.008 $ & $ - $  & $ - $ \\[3pt]
        $|\delta(\ve{k}, \Delta\widehat{\ve{n}}_{S_5})|$ & $ - $ & $ 0.001 $ & $ - $ & $ - $  & $ - $ \\[3pt]
\hline
\end{tabular}
\label{Table3}
\end{table*}

\noindent 
According to Eqs.~(\ref{light_deflection_Final_M_L}) and (\ref{light_deflection_Final_S_L}) one has to build the sum of these expressions in 
Eqs.~(\ref{light_deflection_angle_term_1}) - (\ref{light_deflection_angle_term_3}) as well as of Eqs.~(\ref{light_deflection_angle_term_4}) - (\ref{light_deflection_angle_term_6}) 
in order to get the angle of light deflection of Eq.~(\ref{light_deflection_25}). By performing the derivative one gets 
\begin{eqnarray}
	\delta\left(\ve{k},\Delta\widehat{\ve{n}}_{M_L}\right) &=& - \frac{2 G M}{c^2 d_k}\,J_l\,F\left(\ve{x}_0,\ve{x}_1\right) 
	\nonumber\\ 
	&& \hspace{-2.0cm} \times \left(\frac{P}{d_k}\right)^l \left|\ve{k} \times \ve{e}_3\right|^{l}\;T_l\left(x\right), 
	\label{angle_M_L}
	\end{eqnarray}

\noindent
for the mass-multipoles and 
\begin{eqnarray}
	\delta\left(\ve{k},\Delta\widehat{\ve{n}}_{S_L}\right) &=& - \frac{4 G M}{c^3}\,\Omega\,J_{l-1} \,\frac{l}{l+4}\,F\left(\ve{x}_0,\ve{x}_1\right) 
	\left(\frac{P}{d_k}\right)^{l+1} 
        \nonumber\\
	&& \hspace{0.0cm} \times \frac{\left(\ve{k} \times \ve{d}_k\right)\cdot \ve{e}_3}{d_k} \,\left|\ve{k} \times \ve{e}_3\right|^{l}\,U_{l-1}\left(x\right),
	\label{angle_S_L}
\end{eqnarray}

\noindent 
for the spin-multipoles, where the parameter $\ve{\hat{\xi}}$ has been replaced by the impact vector $\ve{d}_k$. 
The dimensionless scalar function reads  
\begin{eqnarray}
	&& \hspace{-1.0cm} F\left(\ve{x}_0,\ve{x}_1\right) =  
	\frac{\ve{k} \cdot \ve{x}_1}{r_1} - \frac{1}{R} \left(\frac{\left(\ve{k} \cdot \ve{x}_1\right)^2}{r_1} - \frac{\left(\ve{k} \cdot \ve{x}_0\right)^2}{r_0}\right).
	\label{scalar_function_F_1}
\end{eqnarray}

\noindent 
The upper limit of this function is  
\begin{eqnarray}
        \left|F\left(\ve{x}_0,\ve{x}_1\right)\right| &\le& 2\,.  
	\label{scalar_function_F_2}
\end{eqnarray}

\noindent 
We note that the distance between the light source and the observer can be arbitrarily small and even be zero, i.e. extreme configurations with $R \rightarrow 0$ are entirely 
possible. By taking account of the upper limits (\ref{upper_limit_T_l}) and (\ref{upper_limit_U_l}) of Chebyshev polynomials of first and second kind, respectively, we obtain 
for the upper limits of Eqs.~(\ref{angle_M_L}) and (\ref{angle_S_L}),  
\begin{eqnarray}
        \left|\delta\left(\ve{k},\Delta\widehat{\ve{n}}_{M_L}\right)\right| &\le& \frac{4 G M}{c^2\,d_k}\,\left|J_l\right| \left(\frac{P}{d_k}\right)^l\,,
        \label{upper_limit_M_L}
        \\
        \left|\delta\left(\ve{k},\Delta\widehat{\ve{n}}_{S_1}\right)\right| &\le&
        \frac{4 G M}{c^3}\,\Omega\,\kappa^2 \left(\frac{P}{d_k}\right)^{2}\,, 
        \label{upper_limit_S_1}
	\\
        \left|\delta\left(\ve{k},\Delta\widehat{\ve{n}}_{S_L}\right)\right| &\le& \frac{8 G M}{c^3} \Omega \frac{l^2}{l+4} \left|J_{l-1}\right| \left(\frac{P}{d_k}\right)^{l+1},
        \label{upper_limit_S_L}
\end{eqnarray}

\noindent
where (\ref{upper_limit_M_L}) is valid for $l \ge 0$ and (\ref{upper_limit_S_L}) for $l > 1$, 
while (\ref{upper_limit_S_1}) is the upper limit of spin-dipole $l=1$ which has been obtained by inserting (\ref{delta_n_Spin_Dipole}) into (\ref{light_deflection_S_L}). 

The relations (\ref{upper_limit_M_L}) - (\ref{upper_limit_S_L}) represent strict upper limits of light deflection for the case of 
finite distances of source and observer from the massive solar system bodies, where the observer is assumed to be located somewhere near the Earth (e.g. at 
Lagrange point $L_2$), while the sources can be located arbitrarily. 
Numerical values of these upper limits of light deflection for the Sun and the giant planets of the solar system are presented by Table~\ref{Table2} and \ref{Table3}.
These numerical results show that astrometric measurements on the sub-micro-arcsecond level of accuracy need to account mass-multipoles of the orders $0 \le l \le 8$ and
spin-multipoles of the orders $1 \le l \le 3$ in the unit tangent vector (\ref{unit_tangent_vector_simplified_axisymmetric}).

\subsection{Light deflection at infinite spatial distances}

In this Subsection we will consider how these upper limits for the effect of light deflection of Eqs.~(\ref{upper_limit_M_L}) - (\ref{upper_limit_S_L}), 
where the source and the observer are located at finite spatial distances, are related to previously obtained results for the case of infinite spatial distances of source and 
observer \cite{Zschocke_Total_Light_Deflection_15PN}. In the limit of infinite spatial distances of source and observer the light deflection at finite distances of 
Eq.~(\ref{light_deflection_10}) becomes
\begin{eqnarray}
        \delta\left(\ve{\sigma},\ve{\nu}\right) &=& \arcsin \left|\ve{\sigma} \times \ve{\nu}\right|,  
        \label{total_light_deflection}
\end{eqnarray} 

\noindent 
where $\ve{\sigma}$ and $\ve{\nu}$ are the unit tangent vectors of the light ray at minus and plus infinity. The term (\ref{total_light_deflection}) is 
called {\it total light deflection} \cite{Kopeikin1997,Zschocke_Total_Light_Deflection_15PN}. The total light deflection (\ref{total_light_deflection}) 
has also been defined in the text above by Eq.~(44) in \cite{Kopeikin1997} and also by Eq.~(58) in \cite{Zschocke_Total_Light_Deflection_15PN}.

In fact, in the limit of infinite spatial distances of source and observer, $R \rightarrow \infty$, we get for the function (\ref{scalar_function_F_1})
\begin{eqnarray}
	\lim_{r_0 \rightarrow + \infty \atop r_1 \rightarrow + \infty}\,F\left(\ve{x}_0,\ve{x}_1\right) = 1 - \cos \delta\left(\ve{x}_0,\ve{x}_1\right) \rightarrow + 2\,, 
        \label{Limit_scalar_function_F_1}
\end{eqnarray}
 
\noindent
where in the last step of (\ref{Limit_scalar_function_F_1}) we have used that in the limit of spatial infinity $\cos \delta(\ve{x}_0,\ve{x}_1) \rightarrow -1$ if 
source and observer are located in opposite direction as seen from the body. 
Hence, by inserting (\ref{Limit_scalar_function_F_1}) into Eqs.~(\ref{angle_M_L}) and (\ref{angle_S_L}) one finds that these equations pass over to Eqs.~(114) and (121) 
in \cite{Zschocke_Total_Light_Deflection_15PN} in this limit. The remarkable fact, that these equations (114) and (121) in \cite{Zschocke_Total_Light_Deflection_15PN} 
are related to Chebyshev polynomials has, at the very first time, allowed for determining the numerical magnitude of light deflection in the gravitational field of an 
axi-symmetric body at rest up to any multipole order. They were given by Eqs.~(117) and (124) - (125) in our previous work \cite{Zschocke_Total_Light_Deflection_15PN} and read
\begin{eqnarray}
        \left|\delta\left(\ve{\sigma},\Delta\ve{\nu}_{M_L}\right)\right| &\le& \frac{4 G M}{c^2\,d_k}\,\left|J_l\right| \left(\frac{P}{d_k}\right)^l\,,
        \label{upper_limit_M_L_infinite}
        \\
	        \left|\delta\left(\ve{\sigma},\Delta\ve{\nu}_{S_1}\right)\right| &\le&
        \frac{4 G M}{c^3}\,\Omega\,\kappa^2 \left(\frac{P}{d_k}\right)^{2}\,,
        \label{upper_limit_S_1_infinite}
        \\
        \left|\delta\left(\ve{\sigma},\Delta\ve{\nu}_{S_L}\right)\right| &\le& \frac{8 G M}{c^3} \Omega \frac{l^2}{l+4} \left|J_{l-1}\right| \left(\frac{P}{d_k}\right)^{l+1}, 
        \label{upper_limit_S_L_infinite}
\end{eqnarray}
 
\noindent
where (\ref{upper_limit_M_L_infinite}) is valid for $l \ge 0$ and (\ref{upper_limit_S_L_infinite}) for $l > 1$, while (\ref{upper_limit_S_1_infinite}) is the upper limit of 
the spin-dipole $l=1$. These upper limits are valid for astrometric configurations with source and observer at infinite spatial distance from the gravitating solar system body. 
They coincide with the upper limits of light deflection in case of finite spatial distances, which are represented by Eqs.~(\ref{upper_limit_M_L}) - (\ref{upper_limit_S_L}).  
Thus, we have demonstrated that the upper limits of the total light deflection do also represent an upper limit for finite distances between source 
and observer from the solar system body. This fact confirms a corresponding statement made in the introductory section of our previous work \cite{Zschocke_Total_Light_Deflection_15PN}. 
The only restriction is that the observer has to be located somewhere nearby the Earth, for instance at Lagrange point $L_2$, as in the case of the {\it Gaia} mission \cite{Gaia1,Gaia2} 
or in case of the missions {\it GaiaNIR} \cite{Gaia_NIR} and {\it Theia} \cite{Theia} which have been proposed to ESA.

\section{Impact of 2PN and 3PN terms}\label{Section_2PN_3PN}

On the sub-micro-arcsecond level of precision also post-post-Newtonian (2PN) terms become relevant and it might be worthwhile to consider here briefly the order of magnitude 
of the dominant 2PN contributions. The expansion of the metric in (\ref{PN_Expansion_1PN_15PN}) in the 2PN approximation becomes   
\begin{eqnarray}
        && \hspace{-0.5cm} g_{\alpha\beta} = \eta_{\alpha\beta} + h_{\alpha\beta}^{\left(2\right)} + h_{\alpha\beta}^{\left(3\right)} + h_{\alpha\beta}^{\left(4\right)} + {\cal O}(c^{-5}),
        \label{PN_Expansion_1PN_15PN_2PN}
\end{eqnarray}

\noindent 
where $h_{\alpha\beta}^{\left(4\right)}$ are the 2PN perturbations of the metric tensor. This expansion implies a corresponding expansion of the light trajectory and 
its unit tangent vector (\ref{light_deflection}),
\begin{eqnarray}
	&& \hspace{-0.5cm} \ve{n} = \ve{k} + \Delta\ve{n}_{\rm 1PN} + \Delta\ve{n}_{\rm 1.5PN} + \Delta\ve{n}_{\rm 2PN} + {\cal O}(c^{-5}).
        \label{light_deflection_2PN}
\end{eqnarray}

\noindent
The dominant contributions of the 2PN terms are the mass-monopole and mass-quadrupole terms. The 2PN mass-monopole term, $\Delta\ve{n}_{\rm 2PN}^{M \times M}$, is 
given by Eq.~(87) in our work \cite{Article_Zschocke1} and the 2PN mass-quadrupole term, $\Delta\ve{n}_{\rm 2PN}^{M \times Q}$, has been given by Eq.~(30) in our 
investigation \cite{Zschocke_Basic_Transformation}. The effect of light deflection is defined by Eq.~(\ref{light_deflection_10}). 
The upper limits of these 2PN terms have been determined in our investigations \cite{Article_Zschocke1,Zschocke_Basic_Transformation}. For grazing rays they are given by 
\begin{eqnarray}
	\delta \left(\ve{k},\Delta\ve{n}_{\rm 2PN}^{{\rm M} \times {\rm M}}\right) &\le& 16\,\frac{G^2 M^2}{c^4}\,\frac{1}{(d_k)^2}\,\frac{r_1}{d_k}\,,
	\label{angle_2PN_Monopole}
	\\
	\delta\left(\ve{k},\Delta\ve{n}_{\rm 2PN}^{{\rm M} \times {\rm Q}}\right) &\le& 64\,\frac{G^2 M^2}{c^4}\,\frac{1}{(d_k)^2}\,|J_2|\,\frac{r_1}{d_k}\,.
	\label{angle_2PN_Quadrupole}
\end{eqnarray}

\noindent
The large factors $r_1/d_k \gg 1$ are the so-called {\it enhancing factors} which have been found in case of 2PN monopole by several independent 
investigations \cite{Article_Zschocke1,Enhanced_Term_1,Enhanced_Term_2}. Later these {\it enhancing factors} have also been found in case of 
2PN quadrupole in our work \cite{Zschocke_Basic_Transformation}. 
\begin{table}[t]
	\caption{The magnitude of the 2PN terms of Eqs.~(\ref{angle_2PN_Monopole}) and (\ref{angle_2PN_Quadrupole}) and of 3PN terms of Eq.~(\ref{light_delfection_3PN}) 
	to the angle of light deflection in case of a grazing ray at massive solar system bodies. All values are given in micro-arcseconds (\muas). A blank entry means 
	less than $1\,{\rm nas}$. In case of a Sun shield with an aspect angle of $45$ degree (like in the i{\it Gaia} mission) the 2PN terms (and the 3PN terms) of the 
	light rays in the gravitational field of the Sun contribute less than $1\,{\rm nas}$ \cite{Article_Zschocke1}.} 
\begin{tabular}{| c | c | c | c |}
\hline
	&&&\\[-12pt]
Body &\hbox to 21.5mm{\hfill $|\delta(\ve{k},\Delta\ve{n}_{\rm 2PN}^{{\rm M} \times {\rm M}})|$ \hfill} &\hbox to 21mm{\hfill $|\delta(\ve{k},\Delta\ve{n}_{\rm 2PN}^{{\rm M} \times {\rm Q}})|$ \hfill} 
	&\hbox to 25.5mm{\hfill $|\delta(\ve{k},\Delta\ve{n}_{\rm 3PN}^{{\rm M} \times {\rm M} \times {\rm M}})|$ \hfill} \\[3pt]
\hline
	&&&\\[-12pt]
	Sun & $3136.1$ & $-$ & $11.2$ \\[3pt]
	Jupiter & $16.11$  & $0.95$  & $0.014$ \\[3pt]
	Saturn & $4.42$  & $0.29$  & $ 0.004 $ \\[3pt]
	Uranus & $2.58$  & $0.04$  & $ 0.004 $ \\[3pt]
	Neptune & $5.83$  & $0.08$  & $ 0.023 $ \\[3pt]
\hline
\end{tabular}
\label{TableA}
\end{table}

\noindent 
The numerical magnitude of these terms is represented here again by Table~\ref{TableA}. 
These values in Table~\ref{TableA} show that the mass-monopole and mass-quadrupole terms need to be enclosed in the theory of light propagation for 
astrometry on the sub-micro-arcsecond level. For the moment it remains an open question of whether the mass-octupole term in the 2PN approximation 
needs also to be taken into account.  

A final word should also be done about the 3PN approximation. Higher multipoles in the 3PN approximation are certainly negligible and have, thus far, never been calculated. 
So we consider the gravitational field of a non-rotating spherically symmetric body at rest (i.e. there are no 1.5PN, 2.5PN, 3.5PN terms in the metric), where the expansion 
of the metric in (\ref{PN_Expansion_1PN_15PN}) in the 3PN approximation becomes 
\begin{eqnarray}
        && \hspace{-0.5cm} g_{\alpha\beta} = \eta_{\alpha\beta} + h_{\alpha\beta}^{\left(2\right)} + h_{\alpha\beta}^{\left(4\right)} + h_{\alpha\beta}^{\left(6\right)} + {\cal O}(c^{-8}),
        \label{PN_Expansion_1PN_15PN_2PN_3PN} 
\end{eqnarray}

\noindent
where $h_{\alpha\beta}^{\left(6\right)}$ are the 3PN perturbations of the metric tensor. 
This expansion implies a corresponding expansion of the light trajectory and its unit tangent vector (\ref{light_deflection}),
\begin{eqnarray} 
	\ve{n} &=& \ve{k} + \Delta\ve{n}_{\rm 1PN}^{\rm M} + \Delta\ve{n}_{\rm 2PN}^{{\rm M} \times {\rm M}} + \Delta\ve{n}_{\rm 3PN}^{{\rm M} \times {\rm M} \times {\rm M}} 
	+ {\cal O}(c^{-8}).
	\nonumber\\ 
        \label{light_deflection_3PN}
\end{eqnarray}

\noindent 
The magnitude of the angle of light deflection caused by 3PN mass-monopole terms of the unit tangent vector has been determined, at the first time, by means of a lens equation 
which is valid for finite distances \cite{Zschocke_Lense_Equation}, where the following formula (cf. Eq.~(27) ibid.) has been derived, 
\begin{eqnarray}
	&& \hspace{-0.5cm} \delta\left(\ve{k},\Delta\ve{n}_{\rm 3PN}^{{\rm M} \times {\rm M} \times {\rm M}}\right) 
	= 128\,\frac{G^3 M^3}{c^6}\,\frac{1}{(d_k)^3}\,\left(\frac{r_1}{d_k}\right)^2\,.
	\label{light_delfection_3PN}
	\end{eqnarray}

\noindent
Like in case of 2PN monopole and 2PN quadrupole, the {\it enhancing factor} $r_1/d_k \gg 1$ appears also in case of 3PN monopole in (\ref{light_delfection_3PN}).  
The same result (\ref{light_delfection_3PN}) has also been obtained by the approach of Time-Transfer-Function \cite{Linet_Teyssandier} (cf. Eqs.~(21) and (22) ibid.  
by taking the GR value $\gamma=1$ of this PPN parameter). Numerical values of these 3PN terms in (\ref{light_delfection_3PN}) in case of grazing rays are presented in Table~\ref{TableA}.

\section{Summary and Outlook}\label{Summary}

Astrometric measurements have recently achieved a level of a few micro-arcseconds in angular measurements of celestial objects by the ESA astrometry 
mission {\it Gaia} \cite{Gaia1,Gaia2}. One of the major goals in astrometry is to improve the accuracy of angular resolution, because it is directly related 
to the precision in determining the spatial positions of celestial objects. In fact, future space-astrometry missions, 
like {\it GaiaNIR} \cite{Gaia_NIR} and {\it Theia} \cite{Theia}, have been proposed to ESA which are aiming at the sub-micro-arcsecond level of precision. 
Other astrometry concepts are {\it TOLIMAN} \cite{Toliman}, an Australian-American project, and {\it SHERA} \cite{Shera}, an American project, which are
also aiming at the sub-\muas{} level of accuracy. As written already in the introductory Section, the term {\it sub-micro-arcsecond level} refers to an astrometric precision 
of angular measurements of $0.1$ micro-arcseconds. Such an accuracy necessitates a relativistic model of light propagation which is about $10$ times better 
than the end-of-mission accuracy. Accordingly, the given threshold of our model is $0.01$ micro-arcsecond ($10$ nano-arcsecond). 
Such an advancement in astrometric measurements needs a corresponding progress in the theory of light propagation in the curved space-time of the solar system. 

The determination of the spatial positions of celestial light sources is based on the observed direction of light signals, which are emitted by these sources. 
In other words, the determination of the spatial positions of celestial objects is based on measurements of the unit tangent vector of these light trajectories when they 
receive the observer. As outlined in the introductory Section, numerical integrations of the geodesic equation of light trajectories are not workable with regard to 
the huge amount of astrometric observations in astrometry missions like {\it Gaia} \cite{Gaia1,Gaia2} or {\it GaiaNIR} \cite{Gaia_NIR}. 
Instead of that, analytical solutions of the light trajectories are required for the treatment of such a huge amount of astrometric measurements. 
However, the analytical expressions of the unit tangent vector are rather cumbersome. In view of that involved structure of the tangent vector, 
it becomes clear how important it is to simplify the analytical solution of the tangent vector by neglecting all those terms that contribute less than $0.01\,\muas$ 
to the angle of light deflection. Such a considerably simplified expression of the unit tangent vector has been obtained in this investigation. 
The results of this investigation are summarized as follows:
\begin{enumerate}
	\item The unit tangent vector of the light ray in the field of an arbitrary body has been derived in the 1.5PN approximation, given by Eq.~(\ref{light_deflection_new}) 
	with the mass-multipole terms (\ref{delta_n_M}) - (\ref{delta_k3_M}) and the spin-multipole terms (\ref{delta_n_S}) - (\ref{delta_k6_S}). These expressions become 
	awful as soon as one performs the partial derivatives. 
	\item A simplified unit tangent vector of the light ray in the field of an arbitrary body has been obtained by Eq.~(\ref{unit_tangent_vector_simplified}) 
	with the mass-multipole terms of Eqs.~(\ref{hat_M_L}) - (\ref{Simplified_Tangent_Vector_3}) and spin-multipole terms 
	of Eqs.~(\ref{hat_S_L}) - (\ref{Simplified_Tangent_Vector_6}), where all partial derivatives have been performed already. 
        In the simplified tangent vector (\ref{unit_tangent_vector_simplified}) all those terms are neglected, which in total contribute less than a few nano-arcseconds
        to the angle of light deflection in all astrometric configurations. In particular:  
	\begin{enumerate} 
	\item The total sum of all neglected terms is less than $10$ nano-arcseconds for all solar system bodies and all configurations, except for grazing rays at Jupiter and Saturn.  
        \item In case of grazing rays at Jupiter, the total sum of all neglected terms is at most $36.2$ nano-arcseconds, as given by Eq.~(\ref{Total_Sum_1_Jupiter}). 
	\item In case of grazing rays at Saturn, the total sum of all neglected terms is at most $14.9$ nano-arcseconds, as given by Eq.~(\ref{Total_Sum_1_Saturn}). 
	\item If one implements the mass-quadrupole term, which is exact in the 1PN approximation, then one arrives at a simplified unit tangent vector where only terms are neglected which 
	contribute in total less than $10\,{\rm nas}$ for all possible astrometric configurations, including grazing rays at the Sun and at giant planets; see text below Eq.~(\ref{Total_Sum_2}). 
	\end{enumerate}
	The implementation of the simplified unit tangent vector of light rays into relativistic models for possible future astrometry missions would proceed in the same way,
        as it has been described in Section III of our recent work \cite{Zschocke_Basic_Transformation}.
	\item The simplified unit tangent vector of the light ray has been applied for the gravitational field of an axi-symmetric body. It has been demonstrated that in case 
	of an axi-symmetric body the unit tangent vector can be expressed in terms of Chebyshev polynomials, given by Eq.~(\ref{unit_tangent_vector_simplified_axisymmetric}) 
	with the mass-multipole terms of Eqs.~(\ref{hat_M_L_axisymmetric}) - (\ref{Simplified_n3}) and spin-multipole terms of Eqs.~(\ref{hat_S_L_axisymmetric}) - (\ref{Simplified_n6}). 
	The mass-monopole and spin-dipole terms are given by Eqs.~(\ref{delta_n_Mass_Monopole}) and (\ref{delta_n_Spin_Dipole}). 
	\item Upper limits of the effect of light deflection have been determined for an axi-symmetric body for the full set of mass-multipoles and 
	spin-multipoles, given by Eqs.~(\ref{upper_limit_M_L}) - (\ref{upper_limit_S_L}). 
	\item These these upper limits for finite distances of source and observer, given by Eqs.~(\ref{upper_limit_M_L}) - (\ref{upper_limit_S_L}), 
	coincide with the upper limits for infinite distances of source and observer, given by Eqs.~(\ref{upper_limit_M_L_infinite}) - (\ref{upper_limit_S_L_infinite}).
	Thus it has been demonstrated that the total light deflection, where light source and observer are located at infinite spatial
        distances from the body, represents an upper limit for bending of light. 
        \item It has been shown by numerical values presented in Tables~\ref{Table2} and \ref{Table3} that astrometric measurements on the sub-micro-arcsecond level of accuracy 
	necessitates to account for mass-multipoles of the order $0 \le l \le 8$ and for spin-multipoles of the order $0 \le l \le 3$ in the simplified unit tangent vector. 
	\item In Section~\ref{Section_2PN_3PN} the impact of those 2PN and 3PN terms is discussed, which are relevant on the sub-micro-arcsecond level. 
\end{enumerate}

\noindent 
The fifth point was already asserted in the introductory section in our recent investigation \cite{Zschocke_Total_Light_Deflection_15PN}.  
This conclusion has been shown here for the case of an observer which is located somewhere nearby the Earth, that means 
sufficiently far away from the massive solar system bodies, while the spatial position of the celestial light source is arbitrary.  

The sub-micro-arcsecond scale of accuracy will certainly be achieved in near future. For instance, the astrometry missions {\it GaiaNIR} \cite{Gaia_NIR} and {\it Theia} \cite{Theia}, 
recently proposed to ESA, are the most promising candidates to get realized within the next few decades \cite{Eric_Hoeg}. 
Both these missions are aiming at ultra highly precise angular measurements on the sub-micro-arcsecond level. In particular, the mission {\it GaiaNIR} 
aims for three-dimensional mapping of about $50$ billion ($50 \times 10^9$) stars in the near-infrared band. Such a huge amount of stellar objects implies the need for 
highly effective methods and algorithms of observational data reduction. The simplified expression for the unit tangent vector of light rays, which has been obtained in our investigation, 
aims at such a highly effective performance of data reduction of future space astrometry missions like {\it GaiaNIR} \cite{Gaia_NIR} or {\it Theia} \cite{Theia}.

\section{Acknowledgment}

This work was funded by the German Research Foundation (Deutsche Forschungsgemeinschaft DFG) under Grant No. 447922800. Sincere gratitude is expressed to Sergei A. Klioner, 
Michael H. Soffel, Ralf Sch\"utzhold, William G. Unruh, Jos H.J. de Bruijne, Anke Theuser, Lutz Graefe, G\"unter Plunien, Alexey Butkevich, J\"urgen Schreiber, Burkhard K\"ampfer, and 
Laszlo P. Csernai for kind support and inspiring discussions about astrometry and general theory of relativity.

\appendix

\section{Notation}\label{Appendix0}

The following notation is in use:
\begin{itemize}
\item Newtonian constant of gravitation: $G$.
\item vacuum speed of light in flat space-time: $c$.
\item Newtonian mass of the body: $M$.
\item Equatorial radius of the body: $P$.
\item Angular velocity of the body: $\Omega$.
\item Zonal harmonic coefficients of the body: $J_l$.
\item $\eta_{\alpha\beta}$ is the Minkowski metric. 
\item $g^{\alpha\beta}$ and $g_{\alpha\beta}$ are the contravariant and covariant components of the metric tensor with signature $\left(-,+,+,+\right)$.
\item $\displaystyle 1\,{\rm mas}\; ({\rm milli-arcsecond}) \simeq 4.85 \times 10^{-9}\,{\rm rad}$.
\item $\displaystyle 1\,\muas\; ({\rm micro-arcsecond}) \simeq 4.85 \times 10^{-12}\,{\rm rad}$.
\item $\displaystyle 1\,{\rm nas}\; ({\rm nano-arcsecond}) \simeq 4.85 \times 10^{-15}\,{\rm rad}$.
\item $\displaystyle 1\,{\rm pas}\; ({\rm pico-arcsecond}) \simeq 4.85 \times 10^{-18}\,{\rm rad}$.
\item $n! = n \left(n-1\right)\left(n-2\right)\cdot\cdot\cdot 2 \cdot 1$ is the factorial; by definition: $0! = 1$.
\item $n!! = n \left(n-2\right) \left(n-4\right)\cdot\cdot\cdot \left(2\;{\rm or}\;1\right)$ is the double factorial;
        by definition: $0 !! = 1$ and $(-1)!! = 1$.
\item $(2n)!! = 2^n n!$
\item $\displaystyle (2n-1)!! = \frac{(2n)!}{2^n n!}$
\item $\displaystyle (2n+1)!! = \frac{(2n+1)!}{2^n n!}$
\item Lower case Greek indices take values 0,1,2,3.
\item The contravariant components of four-vectors: $a^{\mu} = \left(a^0,a^1,a^2,a^3\right)$.
\item Lower case Latin indices take values 1,2,3.
\item The three-dimensional coordinate quantities (three-vectors) referred to
the spatial axes of the reference system are in boldface: $\ve{a}$.
\item The contravariant components of three-vectors: $a^{i} = \left(a^1,a^2,a^3\right)$.
\item The absolute value of a three-vector:
$a = |\ve{a}| = \sqrt{a^1\,a^1+a^2\,a^2+a^3\,a^3}$.
\item The scalar product of two three-vectors:
$\ve{a}\,\cdot\,\ve{b}=\delta_{ij}\,a^i\,b^j=a^i\,b^i$ with Kronecker delta $\delta_{ij}$.
\item The vector product of two three-vectors reads
$\left(\ve{a}\times\ve{b}\right)^i=\varepsilon_{ijk}\,a^j\,b^k$
with Levi-Civita symbol $\varepsilon_{ijk}$. 
\item The angle between two three-vectors $\ve{a}$ and $\ve{b}$ is designated as $\delta\left(\ve{a},\ve{b}\right)$. 
\item These angles can uniquely be computed by
\begin{eqnarray}
        \delta\left(\ve{a},\ve{b}\right) &=& \arccos \frac{\ve{a} \cdot \ve{b}}{|\ve{a}|\,|\ve{b}|}\,.
        \label{angular_relation}
\end{eqnarray}
\end{itemize}

\section{STF tensors}\label{Appendix_STF}

Here we will present some standard notations about symmetric trace-free (STF) tensors,
which are necessary for our considerations, while further STF relations can be found
in \cite{Poisson_Will,Thorne,Blanchet_Damour1,Multipole_Damour_2,Hartmann_Soffel}.
\begin{itemize}
\item $L=i_1 i_2 \dotsi_l$ is a Cartesian multi-index of a given tensor $T$, that means
$T_L \equiv T_{i_1 i_2 \,.\,.\,.\,i_l}$.
\item two identical multi-indices imply summation:
\begin{eqnarray}
        A_L\,B_L \equiv \sum\limits_{i_1\,.\,.\,.\,i_l}\,A_{i_1\,.\,.\,.\,i_l}\,B_{i_1\,.\,.\,.\,i_l}\;.
        \end{eqnarray}

\noindent
\item The symmetric part of a Cartesian tensor $T_L$ is (cf. Eq.~(2.1) in \cite{Thorne}):
\begin{eqnarray}
T_{\left(L\right)} &=& T_{\left(i_1 \dots i_l \right)} = \frac{1}{l!} \sum\limits_{\sigma}
T_{i_{\sigma\left(1\right)} \dots i_{\sigma\left(l\right)}}
\end{eqnarray}

\noindent
where $\sigma$ is running over {\it all permutations} of $\left(1,2,\dots,l\right)$.

For instance: let $T_{i_1 i_2 i_3}$ be a Cartesian tensor which is already symmetric in all of its Cartesian indices. Then
the symmetric part reads:
\begin{eqnarray}
        && \hspace{0.75cm} T_{\left(i_1 i_2 i_3\right)} = T_{i_1 i_2 i_3}\;.
\end{eqnarray}

\noindent
For instance: let $T_{i_1 i_2 i_3}$ be a Cartesian tensor which is not symmetric in any of its Cartesian indices. Then
the symmetric part reads:
\begin{eqnarray}
        && \hspace{0.75cm} T_{\left(i_1 i_2 i_3\right)} = \frac{1}{3!}
        \nonumber\\
        && \hspace{0.75cm} \times \left(T_{i_1 i_2 i_3}\!+\!T_{i_1 i_3 i_2}\!+\!T_{i_2 i_1 i_3}\!+\!T_{i_2 i_3 i_1}\!+\!T_{i_3 i_1 i_2}\!+\!T_{i_3 i_2 i_1}\right).
        \nonumber\\
\end{eqnarray}

\noindent
\item The un-normalized symmetric part of a Cartesian tensor $T_L$ is (cf. text above Eq.~(A19) in \cite{Blanchet_Damour1}): 
\begin{eqnarray}
        T_{\{L\}} &=& T_{\{i_1 \dots i_l \}} = \sum\limits_{\sigma \in S}
T_{i_{\sigma\left(1\right)} \dots i_{\sigma\left(l\right)}}
\end{eqnarray}

\noindent
where $S$ is the {\it smallest set of permutations} of $\left(1,2,\dots,l\right)$ which makes $T_{i_1} \dots i_{l}$
fully symmetric in its indices. Let us give some examples: 
\begin{eqnarray}
\delta_{\{i_1 i_2\}} &=& \delta_{i_1 i_2}\;,
        \\
\delta_{\{i_1 i_2}\,n_{i_3\}} &=& \delta_{i_1 i_2}\,n_{i_3} + \delta_{i_1 i_3}\,n_{i_2} + \delta_{i_2 i_3}\,n_{i_1}\;,
        \\
\delta_{\{i_1 i_2} \delta_{i_3 i_4\}} &=& \delta_{i_1 i_2} \delta_{i_3 i_4} + \delta_{i_1 i_3} \delta_{i_2 i_4} + \delta_{i_1 i_4} \delta_{i_2 i_3}\;.
\end{eqnarray}

\noindent
In case $T_{i_1 i_2 i_3}$ is a Cartesian tensor which is not symmetric in any of its indices, then the un-normalized symmetric part reads:
\begin{eqnarray}
        T_{\{i_1 i_2 i_3\}} &=& 3! \, T_{(i_1 i_2 i_3)}\,. 
\end{eqnarray}

\noindent
\item The symmetric trace-free part of a Cartesian tensor $T_L$ (notation: $\hat{T}_L \equiv {\rm STF}_L\,T_L = T_{<i_1 \dots i_l>}$) is (cf. Eq.~(2.2) in \cite{Thorne}):
\begin{eqnarray}
        && \hspace{0.75cm} \hat{T}_L = \sum_{k=0}^{\left[l/2\right]} a_{l k}\,\delta_{(i_1 i_2} \dots \delta_{i_{2k-1} i_{2k}}\,
        S_{i_{2k+1 \dots i_l)} \,a_1 a_1 \dots a_k a_k}
\nonumber\\
\label{anti_symmetric_1}
\end{eqnarray}

\noindent
and $S_L \equiv T_{\left(L\right)}$ abbreviates the symmetric part of tensor $T_L$. The coefficient in (\ref{anti_symmetric_1}) is given by
\begin{eqnarray}
a_{l k} &=& \left(-1\right)^k \frac{l!}{\left(l - 2 k\right)!}\,
\frac{\left(2 l - 2 k - 1\right)!!}{\left(2 l - 1\right)!! \left(2k\right)!!}\,.
\label{coefficient_anti_symmetric}
\end{eqnarray}
\end{itemize}

\noindent
For instance:
\begin{eqnarray}
	\hat{T}_{i_1 i_2 i_3} &\equiv& T_{< i_1 i_2 i_3 >}
        \nonumber\\
	&=& T_{\left(i_1 i_2 i_3\right)} 
        \nonumber\\ 
	&& - \frac{1}{5}
        \left(\delta_{i_1 i_2}\,T_{\left(i_3 kk\right)} + \delta_{i_2 i_3}\,T_{\left(i_1 kk\right)} + \delta_{i_3 i_1}\,T_{\left(i_2 kk\right)} \right)\,.
	\nonumber\\ 
\end{eqnarray}

\noindent
Three comments are in order about STF. First of all, the Kronecker delta has no symmetric trace-free part,
\begin{eqnarray}
        {\rm STF}_{ab} \,\delta^{ab} &=& 0\;.
        \label{STF_comment_1}
\end{eqnarray}

\noindent
Second, the symmetric trace-free part of any tensor which contains Kronecker delta is zero, if the Kronecker delta
has not any summation (dummy) index, for instance,
\begin{eqnarray}
        {\rm STF}_{abc} \,\delta^{ab}\,d_{k}^c &=& 0\;,
        \label{STF_comment_2a}
        \\
        {\rm STF}_{abc} \,\delta^{ab}\,k^c &=& 0\;.
        \label{STF_comment_2b}
\end{eqnarray}

\noindent
And third, the following relation is very useful \cite{Thorne,Poisson_Will,Blanchet_Damour1,Multipole_Damour_2} 
\begin{eqnarray}
        \underset{i_1 \dots i_l}{\rm STF}\,A_{i_1 \dots i_l}\;\underset{i_1 \dots i_l}{\rm STF}\,B_{i_1 \dots i_l}
        &=& A_{i_1 \dots i_l}\;\underset{i_1 \dots i_l}{\rm STF}\,B_{i_1 \dots i_l}\,,
	\label{STF_comment_3} 
\end{eqnarray}

\noindent
which allows for simplifying expressions on several occasion.
In particular, we need the following Cartesian STF tensor,
\begin{eqnarray}
        \hat{n}_L = \frac{x_{<\,i_1}}{r}\,\dots\,\frac{x_{i_l\,>}}{r} \;,
\label{Appendix_Cartesian_Tensor}
\end{eqnarray}

\noindent
where $x_i$ are the spatial coordinates of some arbitrary field point and $r = \left|\ve{x}\right|$; we note that $x_i = x^i$ and $\hat{n}_L = \hat{n}^L$. A very useful
relation for later purposes is the expansion of the Cartesian STF tensor $\hat{n}_L$ in terms of Cartesian tensor $n_L$ (cf. Eq.~(A20a) in \cite{Blanchet_Damour1}),
\begin{eqnarray}
        \hat{n}_L &=& \sum\limits_{k=0}^{[l/2]} \left(-1\right)^k \frac{\left(2 l - 2 k - 1\right)!!}{\left(2 l - 1\right)!!}
        \delta_{\{ i_1 i_2} \dots \delta_{i_{2 k - 1} i _{2 k}}  n_{i_{2 k + 1} \dots i_l \}} \,.
        \nonumber\\
\label{STF_n_L}
\end{eqnarray}


\section{Proof of relations (\ref{scale_delta_k4_S}) - (\ref{scale_delta_k5_k6_S})}\label{Appendix_Spin}

In this Appendix we will show the validity of relations (\ref{scale_delta_k4_S}) - (\ref{scale_delta_k5_k6_S}) in case of spin-dipole ($l=1$), which is the most dominant term, 
and we will determine the coefficients $C_1^{S_1}$ and $B_1^{S_1}$, which were given in the text below Eq.~(\ref{scale_delta_k5_k6_S}). For that we consider 
Eqs.~(\ref{delta_k4_S}) - (\ref{delta_k6_S}), which in case of spin-dipole read  
\begin{eqnarray}
        \Delta n^{4\,i}_{S_1} &=& + \frac{2 G \hat{S}_{b}}{c^3}\,\widehat{\epsilon}_{iab}\,
        \widehat{\partial}^{\tau_1}_{a}\,\frac{1}{r_1}\,,
        \label{Appendix_delta_k4_S}
        \\
        \Delta n^{5\,i}_{S_1} &=& + \frac{2 G \hat{S}_{b}}{c^3}\,\widehat{\epsilon}_{iab}\,\frac{1}{R}\,
        \widehat{\partial}^{\tau_1}_{a} \ln \left(r_1 - c \tau_1\right),
        \label{Appendix_delta_k5_S}
        \\
        \Delta n^{6\,i}_{S_1} &=& - \frac{2 G \hat{S}_{b}}{c^3}\,\widehat{\epsilon}_{iab}\,\frac{1}{R}\,
        \widehat{\partial}^{\tau_0}_{a} \ln \left(r_0 - c \tau_0\right),
        \label{Appendix_delta_k6_S}
\end{eqnarray}

\noindent 
where $\widehat{\epsilon}_{iab} = \epsilon_{iab} - k_i\,\epsilon_{jab}\,k^j$. The differential operators (\ref{Differential_Operator_k_0}) and (\ref{Differential_Operator_k_1}) 
for $l=1$ read 
\begin{eqnarray}
	\widehat{\partial}^{\tau_0}_{a} &=& P^{j_1}_{a}\,\frac{\partial}{\partial \xi^{j_1}} + k_a\,\frac{\partial}{\partial c\tau_0}\,,
	\label{differential_0}
	\\
        \widehat{\partial}^{\tau_1}_{a} &=& P^{j_1}_{a}\,\frac{\partial}{\partial \xi^{j_1}} + k_a\,\frac{\partial}{\partial c\tau_1}\,,
        \label{differential_1}
\end{eqnarray}

\noindent
and the spin-dipole (\ref{S}) is a three-vector,  
\begin{eqnarray}
	\ve{S} &=& - M\,\Omega \left(P\right)^2 J_0\,\kappa^2\,\ve{e}_{3}\,,
\end{eqnarray}

\noindent
where $\ve{e}_{3}$ is the principal axis of the body, which is the rotational axis of the axi-symmetric body; we recall that $J_0 = - 1$ (cf. Eq.~(\ref{zonal_harmonic_coefficients})). 

In order to determine the contribution of Eqs.~(\ref{Appendix_delta_k4_S}) - (\ref{Appendix_delta_k6_S}) to the angle of light deflection we make use of 
relation (\ref{light_deflection_S_L}). Then one finds that the second term of $\widehat{\epsilon}_{iab}$ and the first term on the right-hand side of 
(\ref{differential_0}) and (\ref{differential_1}) do not contribute to the angle of light deflection. Thus, by performing the derivatives, we get for the 
individual contributions to the angle of light deflection 
\begin{eqnarray}
	\delta\left(\ve{k},\Delta\ve{n}^{4}_{S_1}\right) &=& - \frac{2 G M}{c^3}\,\Omega\,\left(P\right)^2 \kappa^2 J_0 
	\left(\ve{k} \times \ve{e}_3\right) \cdot \frac{\ve{\hat{\xi}}}{\hat{\xi}} \frac{c\tau_1}{(r_1)^3}\,,
	\nonumber\\ 
        \label{Appendix_Spin_1}
        \\
        \delta\left(\ve{k},\Delta\ve{n}^{5}_{S_1}\right) &=& - \frac{2 G M}{c^3}\,\Omega\,\left(P\right)^2 \kappa^2 J_0 
	\left(\ve{k} \times \ve{e}_3\right) \cdot \frac{\ve{\hat{\xi}}}{\hat{\xi}} \frac{1}{R} \frac{1}{r_1},
	\nonumber\\ 
        \label{Appendix_Spin_2}
        \\
	\delta\left(\ve{k},\Delta\ve{n}^{6}_{S_1}\right) &=& + \frac{2 G M}{c^3}\,\Omega\,\left(P\right)^2 \kappa^2 J_0 
	\left(\ve{k} \times \ve{e}_3\right) \cdot \frac{\ve{\hat{\xi}}}{\hat{\xi}} \frac{1}{R} \frac{1}{r_0}. 
	\nonumber\\ 
        \label{Appendix_Spin_3}
\end{eqnarray}

\noindent
In order to determine the absolute value of these equations we use 
\begin{eqnarray}
	\left| \left(\ve{k} \times \ve{e}_3\right) \cdot \frac{\ve{\hat{\xi}}}{\hat{\xi}} \right| &\le& 1\,,
	\label{Absolute_Value_1}
	\\
	\left| \frac{c\tau_1}{r_1}\right| &\le& 1\,, 
	\label{Absolute_Value_2}
	\\
	\frac{1}{R} \left(\frac{1}{r_1} - \frac{1}{r_0}\right) &\le& \frac{1}{r_1}\,\frac{1}{\hat{\xi}}\,,
	\label{Absolute_Value_3}
\end{eqnarray}

\noindent 
and obtain 
\begin{eqnarray}
	&& \hspace{-1.0cm} \left|\delta\left(\ve{k},\Delta\ve{n}^{4}_{S_1}\right)\right| \le 2\,\kappa^2 \frac{G M}{c^3}\,\Omega\,|J_0| 
	 \left(\frac{P}{r_1}\right)^2,
        \label{Appendix_Spin_4}
        \\
	&& \hspace{-1.0cm} 
	\left|\delta\left(\ve{k},\Delta\ve{n}^{5+6}_{S_1}\right)\right| \le 2\,\kappa^2 \frac{G M}{c^3}\,\Omega\,|J_0| \left(\frac{P}{r_1} \right)\left(\frac{P}{d_k}\right),  
        \label{Appendix_Spin_5_6}
\end{eqnarray}

\noindent
where (\ref{Appendix_Spin_5_6}) is the absolute value of the sum of (\ref{Appendix_Spin_2}) and (\ref{Appendix_Spin_3}), and $\hat{\xi}$ has been replaced by $d_k$. These relations 
show the validity of Eqs.~(\ref{scale_delta_k4_S}) and (\ref{scale_delta_k5_k6_S}) in case of spin-dipole. The coefficients $C_1^{S_1} = 2\,\kappa^2$ and $B_1^{S_1} = 2\,\kappa^2$, 
can be read from (\ref{Appendix_Spin_4}) and (\ref{Appendix_Spin_5_6}) and they agree with the coefficients asserted in the text below Eq.~(\ref{scale_delta_k5_k6_S}). 
We note that the distance $R$ between the source and the observer is arbitrary and can also be zero. That is why one has to consider the sum of the terms 
(\ref{Appendix_Spin_2}) and (\ref{Appendix_Spin_3}), because each individual term would be infinite if $R$ tends to zero.


\section{Chebyshev polynomials}\label{Appendix_Chebyshev_Polynomials}

In this Section we will briefly review the Chebyshev polynomials \cite{Arfken_Weber,Gradstein_Ryshik,Abramowitz_Stegun}.
There are Chebyshev polynomials of first and second kind, $T_l\left(x\right)$ and $U_l\left(x\right)$, which form a sequence of orthogonal polynomials.
The power representation of Chebyshev polynomials of first kind reads (cf. Eqs.~(13.67) and (13.88a) in \cite{Arfken_Weber})
\begin{eqnarray}
        T_0 \left(x\right)  &=& 1 \;,
        \label{Chebyshev_Polynomials_0}
        \\
        T_l \left(x\right) &=& \frac{l}{2} \sum \limits_{n=0}^{[l/2]} \frac{\left(-1\right)^n}{n!} \,\frac{\left(l - n - 1\right)!}{\left(l - 2 n\right)!}
        \,\left(2 x\right)^{l - 2 n}\,,
        \label{Chebyshev_Polynomials_1}
\end{eqnarray}

\noindent
where $l \ge 1$.
The power representation of Chebyshev polynomials of second kind reads (cf. Eq.~(13.88b) in \cite{Arfken_Weber})
\begin{eqnarray}
        U_l \left(x\right) &=& \sum \limits_{n=0}^{[l/2]} \frac{\left(-1\right)^n}{n!} \,\frac{\left(l - n\right)!}{\left(l - 2 n\right)!}
        \,\left(2 x\right)^{l - 2 n}\;, 
        \label{Chebyshev_Polynomials_2}
\end{eqnarray}

\noindent
where $l \ge 0$. The argument of the Chebyshev polynomials is a real number of the closed interval $x \in \left[-1,+1\right]$. 
The Chebyshev polynomials of first and second kind are related by
\begin{eqnarray}
        T_l \left(x\right) &=& U_l\left(x\right) - x\,U_{l-1}\left(x\right),
        \label{Relation_Chebyshev_Polynomials_1}
        \\
        U_l \left(x\right) &=& \frac{x \,T_{l+1}\left(x\right) - T_{l+2}\left(x\right)}{1 - x^2} \;.
        \label{Relation_Chebyshev_Polynomials_2}
\end{eqnarray}

\noindent
The first derivative of Chebyshev polynomials of first kind (\ref{Chebyshev_Polynomials_1}) is related to
the Chebyshev polynomials of second kind as follows (see also p. $794$ in \cite{Arfken_Weber})
\begin{eqnarray}
        \frac{d\,T_l\left(x\right)}{d x} &=& l\,U_{l-1}\left(x\right). 
        \label{Chebyshev_polynomials_T_derivative}
\end{eqnarray}

\noindent
The first derivative of Chebyshev polynomials of second kind (\ref{Chebyshev_Polynomials_2}) is related to
the Chebyshev polynomials of first kind as follows
\begin{eqnarray}
        \frac{d\,U_l\left(x\right)}{d x} &=& \frac{x\,U_{l}\left(x\right) - \left(l + 1\right) T_{l+1}}{1-x^2}\;. 
        \label{Chebyshev_polynomials_U_derivative}
\end{eqnarray}

\noindent
The Chebyshev polynomials of first kind (\ref{Chebyshev_Polynomials_1}) can be written in terms of trigonometric functions (cf. Eq.~(13.83a) in \cite{Arfken_Weber})
\begin{eqnarray} 
        T_l \left(x\right) &=& \cos \left(l\,\arccos \,x\right) 
        \label{upper_limit_M_L_20}
\end{eqnarray}

\noindent
for $l \ge 0$. From (\ref{upper_limit_M_L_20}) follows 
\begin{eqnarray}
        T_l \left(x\right) &\le& 1\,.
        \label{upper_limit_T_l}
\end{eqnarray}

\noindent
The Chebyshev polynomials of second kind (\ref{Chebyshev_Polynomials_2}) can be written in terms of trigonometric functions (cf. Eqs.~(13.83b) and (13.85a) in \cite{Arfken_Weber})
\begin{eqnarray} 
        U_{l-1} \left(x\right) &=& \frac{1}{\sqrt{1-x^2}}\,\sin \left(l\,\arccos \,x\right).
        \label{upper_limit_S_L_20}
\end{eqnarray}

\noindent 
From (\ref{upper_limit_M_L_20}) follows
\begin{eqnarray}
	U_{l-1} \left(x\right) &\le& l\,.
        \label{upper_limit_U_l}
\end{eqnarray}


\section{Gegenbauer polynomials}\label{Appendix_Gegenbauer_Polynomials} 

The Gegenbauer polynomials, also known as ultraspherical polynomials, can be represented by hypergeometric functions \cite{Arfken_Weber,Gradstein_Ryshik,Abramowitz_Stegun}.
According to these representations, they can be written in the form
\begin{eqnarray}
	&& \hspace{-0.5cm} C_l^{\left(0\right)}\left(x\right) = \sum\limits_{n=0}^{[l/2]} \frac{\left(-1\right)^n}{n!}
        \frac{\Gamma\left(l-n\right)}{\left(l-2n\right)!}\, \left(2x\right)^{l-2n},
        \label{Gegenbauer0}
        \\
	&& \hspace{-0.5cm} C_l^{\left(\alpha\right)}\left(x\right) = \sum\limits_{n=0}^{[l/2]} \frac{\left(-1\right)^n}{n!}
        \frac{\Gamma\left(l+\alpha-n\right)}{\Gamma\left(\alpha\right)\,\left(l-2n\right)!}  
        \, \left(2x\right)^{l-2n},
        \label{Gegenbauer1}
\end{eqnarray}

\noindent
where $\alpha$ is the Gegenbauer index of the polynomial, $l$ is the degree of the polynomial and $x$ is a real number. 
In (\ref{Gegenbauer0}) the conditions are $\alpha = 0$ and $l \ge 1$, while in (\ref{Gegenbauer1}) the conditions are $\alpha \neq 0$ and $l \ge 0$. 
The Gamma function $\Gamma(w)$ is given by 
\begin{eqnarray}
        \Gamma\left(w\right) &=& \int\limits_{0}^{\infty} t^{w-1} e^{-t}\,dt\,, 
        \label{Gamma_Function}
\end{eqnarray}

\noindent
where $w \ge 0$ is a positive real number. Let $n$ be an integer. Then, the Gamma function is related to the factorials,
\begin{eqnarray}
        \Gamma\left(n\right) &=& \left(n-1\right)! \quad {\rm i.e.} \quad \Gamma\left(n+1\right) = n!\;.
        \label{Gamma_Functions_Integers}
\end{eqnarray}

\noindent
Let $n + 1/2$ be an half-integer. Then, the Gamma function is related to the factorials by 
\begin{eqnarray}
        \Gamma\left(n + \frac{1}{2}\right) &=& \sqrt{\pi}\;\frac{\left(2n\right)!}{2^{2n}\,n!} \;.
        \label{Gamma_Function_Half_Integers}
\end{eqnarray}

\noindent
By using this relation one may deduce from (\ref{Gegenbauer1}) the following form of Gegenbauer polynomials,
\begin{eqnarray}
	C_l^{\left(q + \frac{1}{2}\right)}\left(x\right) &=& \frac{q!}{\left(2q\right)!}\,\frac{1}{2^l} \sum\limits_{n=0}^{[l/2]} \frac{\left(-1\right)^n}{n!}
        \frac{\left(2l+2q-2n\right)!}{\left(l+q-n\right)!}
        \frac{\left(x\right)^{l-2n}}{\left(l-2n\right)!}
        \nonumber\\
        \label{Gegenbauer2}
\end{eqnarray}

\noindent
where $q \ge 0$ is a positive integer. These Gegenbauer polynomials satisfy the arithmetic operation 
\begin{eqnarray}
	C_l^{\left(q + \frac{1}{2}\right)}\left(-x\right) &=& \left(-1\right)^l C_l^{\left(q + \frac{1}{2}\right)}\left(x\right).
        \label{Gegenbauer2a}
\end{eqnarray}

\noindent 
We also note the value of Gegenbauer polynomials when the argument $x=1$ \cite{Arfken_Weber,Gradstein_Ryshik,Abramowitz_Stegun},
which is frequently be used as normalization constant,
\begin{eqnarray}
        C_l^{\left(q + \frac{1}{2}\right)}\left(1\right) &=& {l + 2q \choose l}\,.
        \label{Gegenbauer3}
\end{eqnarray}

\noindent 
In case of $q=0$ the Gegenbauer polynomials (\ref{Gegenbauer2}) become Legendre polynomials \cite{Arfken_Weber,Gradstein_Ryshik,Abramowitz_Stegun}
\begin{eqnarray}
	P_l \left(x\right) &=& C_l^{(1/2)}\left(x\right) = \frac{1}{2^l} \sum\limits_{n=0}^{\left[l/2\right]} \frac{\left(-1\right)^n}{n!}
        \frac{\left(2l-2n\right)!}{\left(l-n\right)!}\,\frac{x^{l-2n}}{\left(l-2n\right)!}\,.
	\nonumber\\ 
        \label{Legendre_Polynomials_2}
\end{eqnarray}

\noindent
By comparing (\ref{Gegenbauer0}) with (\ref{Chebyshev_Polynomials_1}) and  (\ref{Gegenbauer1}) with (\ref{Chebyshev_Polynomials_2})  
one finds that the Chebyshev polynomials are special Gegenbauer polynomials,
\begin{eqnarray}
	T_l \left(x\right) &=& \frac{l}{2}\,C_l^{(0)}\left(x\right) \quad\, {\rm for} \quad l \ge 1\,,
        \label{Gegenbauer_Chebyshev_1}
        \\
        U_l \left(x\right) &=& C_l^{(1)}\left(x\right) \quad\quad {\rm for} \quad l \ge 0\,.
        \label{Gegenbauer_Chebyshev_2}
\end{eqnarray}

\noindent


\section{Some relations}\label{Appendix_Relations}

First of all, we notice that 
\begin{eqnarray}
	\frac{\ve{k} \cdot \ve{x}_0}{r_0} &=& \frac{r_1^2 - r_0^2 - R^2}{2 R r_0}\,,
	\label{relation_angle_k_x0}
	\nonumber\\
	\frac{\ve{k} \cdot \ve{x}_1}{r_1} &=& \frac{r_1^2 - r_0^2 + R^2}{2 R r_1}\,. 
	\label{relation_angle_k_x1}
\end{eqnarray}

\noindent
By introducing the variable 
\begin{eqnarray}
 z &=& \frac{r_0}{r_1}\,, 
	\label{variable_z}
\end{eqnarray}

\noindent
with $z \in [0,\infty]$ we may rewrite these relations into the form 
\begin{eqnarray}
	\frac{\ve{k} \cdot \ve{x}_0}{r_0} &=& \frac{1 - z^2  - \widehat{R}^2}{2 z \widehat{R}}\,,
        \label{relation_angle_k_x0_z}
        \\
	\frac{\ve{k} \cdot \ve{x}_1}{r_1} &=& \frac{1 - z^2 + \widehat{R}^2}{2 \widehat{R}}\,,
        \label{relation_angle_k_x1_z}
\end{eqnarray}

\noindent
where 
\begin{eqnarray}
	\widehat{R} &=&  \sqrt{1 + z^2 - 2 z \cos \delta(\ve{x}_1, \ve{x}_0)} 
        \label{relation_widehat_R}
\end{eqnarray}

\noindent 
is a dimensionless term which we call {\it reduced distance} between source and observer, in order to distinguish it from the spatial distance $R$ between source and observer 
as defined by Eq.~(\ref{Spatial_Distance_1}). They are related to each other by $R = r_1 \widehat{R}$. The angle is in the interval $\delta(\ve{x}_1, \ve{x}_0) \in [0,2\pi]$. 

As usual, for the differential operators we are using the following notations, 
\begin{eqnarray}
        \partial_L &=& \frac{\partial}{\partial \xi^{i_1}}\,\dots\,\frac{\partial}{\partial \xi^{i_l}}\;,
        \\
	\widetilde{\partial}_L &=& P_{(i_1}^{j_1} \, \dots \, P_{i_l)}^{j_l}
        \frac{\partial}{\partial \xi^{j_1}}\,\dots\,\frac{\partial}{\partial \xi^{j_l}}\;,
        \\
        \widehat{\partial}_L &=& P_{<i_1}^{j_1} \, \dots \, P_{i_l>}^{j_l}
        \frac{\partial}{\partial \xi^{j_1}}\,\dots\,\frac{\partial}{\partial \xi^{j_l}}\;,
\end{eqnarray}

\noindent 
and the abbreviations
\begin{eqnarray}
        \xi_L &=& \xi_{i_1}\,\dots\,\xi_{i_l}\,,
\\
        \hat{\xi}_L &=& P^{j_1}_{i_1}\,\dots\,P^{j_l}_{i_l}\,\xi_{j_1}\,\dots\,\xi_{j_l}\,.
\end{eqnarray}

\noindent
We recall here that three-vector $\ve{\xi}$ has three independent spatial components, while three-vector $\ve{\hat{\xi}}$ is perpendicular to three-vector $\ve{k}$ 
and because of this restriction has only two independent spatial components. Therefore, three-vector $\ve{\hat{\xi}}$ equals impact vector $\ve{d}_k$. 

First, we consider the spatial derivatives acting on several functions, where the differential operator has been defined by Eq.~(\ref{Differential_Operator_p_0}).  
In particular, we need the following relations:  
\begin{eqnarray} 
	&& \partial_L\,\frac{1}{\xi} = \left(-1\right)^l \left(2l-1\right)!! \; \underset{i_1 \dots i_l}{\rm STF}\; \frac{\xi_L}{\left(\xi\right)^{2l+1}}\,,
	\label{Relation_1a}
\end{eqnarray}
\begin{eqnarray} 	
	&& \widehat{\partial}_L\,\frac{1}{\hat{\xi}} = \left(-1\right)^l \left(2l-1\right)!! \; \underset{i_1 \dots i_l}{\rm STF}\;
	\frac{\hat{\xi}_L}{(\hat{\xi})^{2l+1}}\,,
        \label{Relation_1b}
\end{eqnarray}
\begin{eqnarray} 
	&& \widehat{\partial}_L\,\frac{\hat{\xi}^{i}}{(\hat{\xi})^2} =  
	P^{ij}\,\frac{\partial}{\partial \xi^j} \left(-1\right)^{l+1} \underset{i_1 \dots i_l}{\rm STF} \sum\limits_{n=0}^{[l/2]} G^l_n
        \nonumber\\
	&& \hspace{1.5cm} \times \, \frac{P_{i_1 i_2} \, \dots \, P_{i_{2n-1} i_{2n}}\,
        \hat{\xi}_{i_{2n+1}} \,\dots\,\hat{\xi}_{i_l}}{(\hat{\xi})^{2l-2n}},
        \label{Relation_3}
\end{eqnarray}
\begin{eqnarray} 
	&& \widetilde{\partial}_{L}\,\frac{1}{r} = \left(-1\right)^{l} \sum\limits_{n=0}^{[l/2]} K^l_n 
	\nonumber\\
        && \hspace{1.5cm} \times\,\frac{P_{(i_1 i_2}\,\dots\,P_{i_{2n-1} i_{2n}}\,\hat{\xi}_{i_{2n+1}} \dots \hat{\xi}_{i_l)}}{\left(r\right)^{2l-2n+1}},
        \label{Relation_2}
\end{eqnarray}
\begin{eqnarray} 
	&& \widehat{\partial}_L\,\frac{1}{r} = \left(-1\right)^{l} \underset{i_1 \dots i_l}{\rm STF} \sum\limits_{n=0}^{[l/2]} K^l_n  
        \nonumber\\  
	&& \hspace{1.5cm} \times\,\frac{P_{i_1 i_2}\,\dots\,P_{i_{2n-1} i_{2n}}\,\hat{\xi}_{i_{2n+1}} \dots \hat{\xi}_{i_l}}{\left(r\right)^{2l-2n+1}}\,, 
        \label{Relation_2a}
	\end{eqnarray}
\begin{eqnarray} 
	&& \widehat{\partial}_L\,r = \left(-1\right)^{l+1} \underset{i_1 \dots i_l}{\rm STF} \sum\limits_{n=0}^{[l/2]} G^l_n\,W^l_n  
        \nonumber\\ 
	&& \hspace{1.5cm} \times\,\frac{P_{i_1 i_2}\,\dots\,P_{i_{2n-1} i_{2n}}\,\hat{\xi}_{i_{2n+1}} \dots \hat{\xi}_{i_l}}{\left(r\right)^{2l-2n-1}}\,,
	\label{Relation_2b}
\end{eqnarray}
\begin{eqnarray} 
	&& \widehat{\partial}_L\,\frac{\hat{\xi}^{i}}{(\hat{\xi})^2}\,\frac{c\tau}{r} 
	= P^{ij}\,\frac{\partial}{\partial \xi^j} \left(-1\right)^{l+1} \underset{i_1 \dots i_l}{\rm STF} \sum\limits_{n=0}^{[l/2]} G^l_n\,F^l_n(c\tau,r)  
	\nonumber\\ 
	&& \hspace{1.5cm} \times \,\frac{P_{i_1 i_2} \, \dots \, P_{i_{2n-1} i_{2n}}\,
	\hat{\xi}_{i_{2n+1}} \,\dots\,\hat{\xi}_{i_l}}{(\hat{\xi})^{2l-2n}}\,,
        \label{Relation_4}
\end{eqnarray}
 
\noindent
where 
\begin{eqnarray} 
	\left[l/2\right] &=& l/2 \quad\quad\quad\;\;{\rm for}\;\,{\rm even}\;\,{\rm values}\;\,{\rm of}\;\,l\,, 
        \label{l1_even}
	\\
	\left[l/2\right] &=& (l-1)/2 \quad {\rm for}\;\,{\rm odd}\;\,{\rm values}\;\,{\rm of}\;\,l\,,
        \label{l1_odd}
\end{eqnarray}

\noindent
is the largest natural number less than or equal to $l/2$. These relations are valid for $n \ge 1$. 
The coefficients $G^l_n$ and $W^l_n$ are given by Eqs.~(\ref{Coefficients_G_l_n}) and (\ref{Coefficients_W_l_n}), 
while the coefficient $K^l_n$ reads 
\begin{eqnarray} 
	K^l_n &=& \frac{l!}{2^l}\,\frac{\left(-1\right)^{n}}{n!}\,\frac{\left(2l-2n\right)!}{(l-n)!}\,\frac{1}{(l-2n)!} \,.
        \label{K_l_n}
\end{eqnarray}

\noindent 
These coefficients are identical with the coefficients of the Legendre polynomials (\ref{Legendre_Polynomials_2}) times the faculty of the multipole index $l$. 
The scalar function in (\ref{Relation_4}) is defined by 
\begin{eqnarray}
        F^l_n\left(c\tau,r\right) &=& \frac{c\tau}{r} \sum\limits_{k=0}^{l-n-1} \frac{1}{2^{2k}} {2k \choose k} \left(\frac{\hat{\xi}}{r}\right)^{2k}, 
        \label{Function_F1}
\end{eqnarray}

\noindent 
where we note that $\hat{\xi} = \sqrt{(r)^2 - (c\tau)^2}$. There are congruences of these polynomials \cite{Polynomial1,Polynomial2} and one may express them  
into ordinary hypergeometric functions \cite{Polynomial3,Polynomial4}. The coefficients $\displaystyle {2k \choose k}$ are the central binomial coefficients. By means of the 
generating function of these coefficients,
\begin{eqnarray}
        \frac{1}{\sqrt{1-x^2}} &=& \sum\limits_{k=0}^{\infty} \frac{1}{2^{2k}} {2k \choose k} \left(x\right)^{2k}\,,
        \label{generating_function_F_l_n}
\end{eqnarray}

\noindent
one may show that $|F^l_n(c\tau,r)| \le 1$. These polynomials in Eq.~(\ref{Function_F1}) can also be written in the form 
\begin{eqnarray}
        F^l_n\left(c\tau,r\right) &=&  \frac{c\tau}{r} + \accentset{\ast}{F}^l_n\left(c\tau,r\right), 
        \label{Function_F2}
\end{eqnarray}

\noindent
where the scalar function $\accentset{\ast}{F}^l_n(c\tau,r)$ is defined by 
\begin{eqnarray}
        \accentset{\ast}{F}^l_n\left(c\tau,r\right) &=& \frac{c\tau}{r} \sum\limits_{k=1}^{l-n-1} \frac{1}{2^{2k}} {2k \choose k} \left(\frac{\hat{\xi}}{r}\right)^{2k}\,.
        \label{Function_F_Star}
\end{eqnarray}

\noindent
Relations like (\ref{Relation_1a}) - (\ref{Relation_2b}) were given in several articles, 
e.g. \cite{Blanchet_Damour1,Kopeikin1997,Klioner_Soffel_2000,Zschocke_Total_Light_Deflection_15PN}. 
A proof of relation (\ref{Relation_3}) has been shown in \cite{Zschocke_Total_Light_Deflection_15PN}, while the proof of relation (\ref{Relation_4}) is very similar to 
the approach in \cite{Zschocke_Total_Light_Deflection_15PN}. 

Second, we consider scalar derivatives. Especially, we need the following relations   
\begin{eqnarray}
        && \hspace{-0.5cm} \left(\frac{\partial}{\partial c\tau}\right)^{p-1} \frac{1}{\left(r\right)^m} = 
        \sum\limits_{k=1}^{\{p/2\}} L^{p,m}_k \frac{\left(c \tau\right)^{p-2k+1}}{\left(r\right)^{m+2p-2k}},
        \label{Proof_Mass_5a}
	\\
	&& \hspace{-0.5cm} \left(\frac{\partial}{\partial c\tau}\right)^{p-2} \frac{1}{\left(r\right)^m} = 
	\sum\limits_{k=1}^{[p/2]} Q^{p,m}_k \frac{\left(c \tau\right)^{p-2k}}{\left(r\right)^{m+2p-2k-2}},
        \label{Proof_Mass_5b}
\end{eqnarray}

\noindent
where (\ref{Proof_Mass_5a}) and (\ref{Proof_Mass_5b}) are valid for $p \ge 1$ and $p \ge 2$, respectively. The upper limit of the sum in (\ref{Proof_Mass_5a}) means 
\begin{eqnarray}
        \{p/2\} &=& p/2 \quad\quad\quad\;\;{\rm for}\;\,{\rm even}\;\,{\rm values}\;\,{\rm of}\;\,p\,,
        \label{l2_even}
        \\
        \{p/2\} &=& (p+1)/2 \quad {\rm for}\;\,{\rm odd}\;\,{\rm values}\;\,{\rm of}\;\,p\,,
        \label{l2_odd}
\end{eqnarray}

\noindent
which is the smallest natural number larger than or equal to $p/2$, while the upper limit of the sum in (\ref{Proof_Mass_5b}) 
has been explained by Eqs.~(\ref{l1_even}) and (\ref{l1_odd}). The coefficients in (\ref{Proof_Mass_5a}) are given by
\begin{eqnarray}
        L^{p,m}_k &=& \left(-1\right)^{p+k} {p-1 \choose p-2k+1} \left(2k-3\right)!!
        \nonumber\\
        && \times \frac{\left(m+2p-2k-2\right)!!}{\left(m-2\right)!!}\,, 
        \label{Coefficients_L_p_m_k}
\end{eqnarray}

\noindent 
and the coefficients in (\ref{Proof_Mass_5b}) are given by 
\begin{eqnarray}
        Q^{p,m}_k &=& \left(-1\right)^{p+k+1} {p-2 \choose p-2k} \left(2k-3\right)!!
        \nonumber\\
        && \times \frac{\left(m+2p-2k-4\right)!!}{\left(m-2\right)!!}\,.
        \label{Coefficients_Q_p_m_k}
\end{eqnarray}


\section{Proof of relations (\ref{Relation_n_1}) and (\ref{Relation_n_2})}\label{Appendix_Relation_1_and_2}

In this Appendix we will show the validity of relations (\ref{Relation_n_1}) and (\ref{Relation_n_2}). By inserting the relations
\begin{eqnarray}
        \frac{\ve{\hat{\xi}}}{r - c \tau} \frac{1}{r} &=& \frac{\partial}{\partial \ve{\hat{\xi}}}\,\ln\left(r - c\tau\right), 
        \label{Appendix_Relation_A1}
        \\
        \frac{\ve{\hat{\xi}}}{r - c \tau} &=& \frac{\partial}{\partial \ve{\hat{\xi}}}\left[\left(c \tau\right) \ln\left(r - c\tau\right) + r\right], 
        \label{Appendix_Relation_B1}
\end{eqnarray}

\noindent
into the left-hand side of Eqs.~(\ref{Relation_n_1}) and (\ref{Relation_n_2}), we get the following two functions
\begin{eqnarray}
	&& \hspace{-0.5cm} \ve{U}_L = \frac{\partial}{\partial \ve{\hat{\xi}}}\, \widehat{\partial}_L \,\ln\left(r - c\tau\right),
        \label{Appendix_Relation_A3}
        \\
	&& \hspace{-0.5cm} \ve{V}_L = \left(c\tau\right) \ve{U}_L + \frac{\partial}{\partial \ve{\hat{\xi}}}\,\widehat{\partial}_{L}\,r\,,
        \label{Appendix_Relation_B3}
\end{eqnarray}

\noindent
where we have inserted the differential operator of Eq.~(\ref{Differential_Operator_p_0}).
Using
\begin{eqnarray}
        \frac{\partial}{\partial \xi^{j_1}}\,\ln\left(r - c\tau\right) &=& \frac{\xi^{j_1}}{r - c\tau}\,\frac{1}{r} \,,
        \label{Appendix_Relation_A4}
\end{eqnarray}

\noindent
which is just relation (\ref{Appendix_Relation_A1}) without the projector, one obtains for expression (\ref{Appendix_Relation_A3})
\begin{eqnarray}
	\ve{U}_L = \frac{\partial}{\partial \ve{\hat{\xi}}}\;\underset{i_1 \dots i_l}{\rm STF}\;P_{i_{2}}^{j_{2}}\, \dots \,P_{i_l}^{j_l} 
        \frac{\partial}{\partial \xi^{j_2}}\,\dots\,\frac{\partial}{\partial \xi^{j_{l}}}\,\frac{\hat{\xi}^{j_1}}{(\hat{\xi})^2}\left(1 + \frac{c\tau}{r}\right),
        \nonumber\\ 
        \label{Appendix_Relation_A5}
\end{eqnarray}

\noindent
where $(\hat{\xi})^2 = r^2 - (c\tau)^2$ has been inserted. By inserting relations (\ref{Relation_3}) and (\ref{Relation_4}) into (\ref{Appendix_Relation_A5}) one recovers
Eq.~(\ref{Relation_n_1}). Similarly, by inserting (\ref{Relation_3}) and (\ref{Relation_4}) and relation (\ref{Relation_2b}) into (\ref{Appendix_Relation_B3}) one
recovers Eq.~(\ref{Relation_n_2}).


\section{The scalar function $S^{l,p}_n$}\label{Scalar_Function}

If one determines the upper limits of neglected terms, one encounters in the subsequent appendices the following dimensionless scalar function, 
\begin{eqnarray}
        && \hspace{-0.5cm} S^l_{p,n} = k_{i_1}\,\dots\,k_{i_{p+2n}}\,\frac{\hat{\xi}_{i_{p+2n+1}} \dots \hat{\xi}_{i_l}}{(\hat{\xi})^{l-p-2n}}\, 
        \delta^{\;3}_{<{i_1}}\,\dots\,\delta^3_{{i_l}>}\,.
        \label{Function_S}
\end{eqnarray}

\noindent
The tensor $k_{i_1}\,\dots\,k_{i_{p+2n}}\,\hat{\xi}_{i_{p+2n+1}} \dots \hat{\xi}_{i_l}$ originates from the action of the differential operators (\ref{Differential_Operator_tau_0_p}) 
and (\ref{Differential_Operator_tau_1_p}) on some scalar functions, while the STF tensor $\delta^{\;3}_{<{i_1}}\,\dots\,\delta^3_{{i_l}>}$ originates from the mass-multipoles 
(\ref{M_L}) and spin-multipoles (\ref{S_L}) of an axi-symmetric body. The scalar function (\ref{Function_S}) carries the natural number $l$ which is the multipole-order. 
In addition, the scalar function (\ref{Function_S}) carries the natural numbers $p$ and $n$, where $p$ originates from these differential operators, while index $n$ originates  
from relations like (\ref{Relation_3}) - (\ref{Relation_4}). 

In this Appendix we will determine the upper limit of the absolute value of (\ref{Function_S}),
\begin{eqnarray}
        && \hspace{-0.5cm} \left|S^l_{p,n}\right| = \left| k_{i_1}\,\dots\,k_{i_{p+2n}}\,\frac{\hat{\xi}_{i_{p+2n+1}} \dots \hat{\xi}_{i_l}}{(\hat{\xi})^{l-p-2n}}\, 
        \delta^{\;3}_{<{i_1}}\,\dots\,\delta^3_{{i_l}>}\right|.
        \label{Absolute_Function_S}
\end{eqnarray}

\noindent
Clearly, an upper limit of (\ref{Absolute_Function_S}) is given by
\begin{eqnarray}
        \left|S^l_{p,n} \right| \le 1 
        \label{Upper_Limit_Function_S}
\end{eqnarray}

\noindent
for all possible values of $l,p,n$. As we will see below, such an upper limit is by far too inaccurate for a correct determination of the upper limits of the neglected terms of the 
unit tangent vector. That is why we have to consider this expression (\ref{Function_S}) more carefully. By implementing the STF tensor (\ref{STF_Expansion}) into (\ref{Function_S}) 
one obtains
\begin{eqnarray}
        S^{l}_{p,n} &=& \sum\limits_{m=0}^{\left[\frac{l-p-2n}{2}\right]} M^{l,p,n}_{m}  
        \left(\frac{\ve{\hat{\xi}} \cdot \ve{e}_3}{\hat{\xi}}\right)^{l-p-2n-2m}
        \nonumber\\ 
        && \hspace{-0.5cm} \times \sum\limits_{s=m}^{\left[\frac{p+2n+2m}{2}\right]} H^l_s\,
        N^{p,n}_{m,s} \left(\ve{k} \cdot \ve{e}_3\right)^{p+2n+2m-2s}, 
        \label{Function_S_4}
\end{eqnarray}

\noindent
with the coefficients
\begin{eqnarray}
        M^{l,p,n}_{m} &=& \frac{\left(l-p-2n\right)!}{\left(l-p-2n-2m\right)!\;\left(2m\right)!!}\,,
        \label{Coefficient_M_l_p_n_m}
        \\
        N^{p,n}_{m,s} &=& \frac{\left(p+2n\right)!}{\left(p+2n+2m-2s\right)!\;\left(2s-2m\right)!!}\,.
        \label{Coefficient_N_p_n_m_s}
\end{eqnarray}

\noindent
As an explicit example we consider the case of mass-quadrupole $l=2$. From (\ref{Function_S_4}) we get
\begin{eqnarray}
        S^{l=2}_{p=0,n=0} &=& \left(\frac{\ve{\hat{\xi}} \cdot \ve{e}_3}{\hat{\xi}}\right)^2 - \frac{1}{3}\,,
        \label{S_l_p_Quadrupole_A}
        \\
        S^{l=2}_{p=0,n=1} &=& \left(\ve{k} \cdot \ve{e}_3\right)^2 - \frac{1}{3}\,,
        \label{S_l_p_Quadrupole_B}
        \\
        S^{l=2}_{p=1,n=0} &=& \left(\frac{\ve{\hat{\xi}} \cdot \ve{e}_3}{\hat{\xi}}\right) \, \left(\ve{k} \cdot \ve{e}_3\right).
        \label{S_l_p_Quadrupole_C}
\end{eqnarray}

\noindent 
Furthermore, in two specific cases the function $S^{l}_{p,n}$ in (\ref{Function_S_4}) becomes a Legendre polynomial (\ref{Legendre_Polynomials_2}),
\begin{eqnarray}
        S^l_{p=l,n=0} &=& \frac{l!}{(2l-1)!!} \; P_l\left(\ve{e}_3 \cdot \ve{k}\right),
        \label{P_l_1}
        \\
	S^l_{p=0,n=0} &=& \frac{l!}{(2l-1)!!} \; P_l\left(\frac{\ve{e}_3 \cdot \ve{\hat{\xi}}}{\hat{\xi}}\right).
        \label{P_l_2}
\end{eqnarray}

\noindent
The sum over variable $s$ in (\ref{Function_S_4}) leads to Gegenbauer polynomials, 
\begin{eqnarray}
        && \sum\limits_{s=m}^{\left[\frac{p+2n+2m}{2}\right]} H^l_s\,N^{p,n}_{m,s} \left(\ve{k} \cdot \ve{e}_3\right)^{p+2n+2m-2s} 
        \nonumber\\ 
        && = \left(-1\right)^m 2^{p+2n+m-1} \frac{\left(l-1\right)!}{\left(2l-1\right)!} \frac{\left(2l-p-2n-2m\right)!}{\left(l-p-2n-m\right)!} 
        \nonumber\\ 
        && \quad \times \frac{C^{l-p-2n-m+1/2}_{p+2n}\left(\ve{k} \cdot \ve{e}_3\right)}{C^{l-p-2n-m+1/2}_{p+2n}\left(1\right)}, 
        \label{Inner_Summation_S1} 
\end{eqnarray}

\noindent
where the Gegenbauer polynomials are given by Eq.~(\ref{Gegenbauer2}). The arguments of these polynomials are within the closed interval $\ve{k}\cdot\ve{e}_3 \in \left[-1,+1\right]$. 
Within this interval, the maximal and minimal values of the Gegenbauer polynomials are reached at the boundary values $\ve{k}\cdot\ve{e}_3 = \pm 1$. Thus, using the maximal and 
minimal values of Gegenbauer polynomials as given by Eq.~(\ref{Gegenbauer3}), one may show that the upper limit of (\ref{Function_S_4}) is given by 
\begin{eqnarray}
        \left|S^{l}_{p,n}\right| &\le& \left|\sum\limits_{m=0}^{\left[\frac{l-p-2n}{2}\right]} M^{l,p,n}_{m}\; 
        \sum\limits_{s=m}^{\left[\frac{p+2n+2m}{2}\right]} H^l_s\,N^{p,n}_{m,s}\right|. 
        \label{Absolute_Value_S}
\end{eqnarray}

\noindent
A Fortran 90 code has been employed \cite{Fortran} as independent check that (\ref{Absolute_Value_S}) represents the upper limit of (\ref{Function_S_4}).
In addition, the relations (\ref{P_l_1}) and (\ref{P_l_2}) have been used to check this numerical Fortran 90 code. In view of its importance,
the computer algebra systems {\it Mathematica} \cite{Mathematica} and {\it Maple} \cite{Maple} have also been used to confirm relation (\ref{Absolute_Value_S}).
By using (\ref{Inner_Summation_S1}) one finds that the sum over variable $s$ in (\ref{Absolute_Value_S}) reads
\begin{eqnarray}
        && \hspace{-1.5cm} \sum\limits_{s=m}^{\left[\frac{p+2n+2m}{2}\right]} H^l_s\,N^{p,n}_{m,s} = \left(-1\right)^m \frac{\left(l-1\right)!}{\left(2l-1\right)!}  
        \nonumber\\ 
        && \times\, 2^{p+2n+m-1} \frac{\left(2l-p-2n-2m\right)!}{\left(l-p-2n-m\right)!}. 
        \label{inner_summation}
\end{eqnarray}

\noindent
Thus, by inserting (\ref{inner_summation}) into (\ref{Absolute_Value_S}) we obtain the upper limit
\begin{eqnarray}
        && \left|S^{l}_{p,n}\right| \le 2^{p+2n-1} \frac{\left(l-1\right)!}{\left(2l-1\right)!} \left(l-p-2n\right)! 
        \nonumber\\
        && \times \sum\limits_{m=0}^{\left[\frac{l-p-2n}{2}\right]} \frac{\left(-1\right)^m}{\left(l-p-2n-2m\right)!}\,
        \frac{\left(2l-p-2n-2m\right)!}{\left(l-p-2n-m\right)!\;m!}  
        \nonumber\\ 
        \label{Absolute_Value_S_5}
\end{eqnarray}

\noindent
where $(2m)!! = 2^m m!$ has been used. This upper limit depends on the chosen values of $l,p,n$ and is much more fine-tuned than the estimate as given above by 
Eq.~(\ref{Upper_Limit_Function_S}) which is too inaccurate. As explicit example we consider the absolute values of (\ref{S_l_p_Quadrupole_A}) - (\ref{S_l_p_Quadrupole_C}), 
which read  
\begin{eqnarray}
	\left|S^{l=2}_{p=0,n=0}\right| &\le& \frac{2}{3}\,,
        \label{Absolute_Value_S_l_p_Quadrupole_A}
	\\
	\left|S^{l=2}_{p=0,n=1}\right| &\le& \frac{2}{3}\,,
        \label{Absolute_Value_S_l_p_Quadrupole_B}
        \\
	\left|S^{l=2}_{p=1,n=0}\right| &\le& 1\,.
        \label{Absolute_Value_S_l_p_Quadrupole_C}
\end{eqnarray}

\noindent 
The application of relation (\ref{Absolute_Value_S_5}) yields the same upper limits (\ref{Absolute_Value_S_l_p_Quadrupole_A}) - (\ref{Absolute_Value_S_l_p_Quadrupole_C}). 
However, an additional comment needs to be done about the absolute value in (\ref{Absolute_Value_S_l_p_Quadrupole_C}). The impact vector $\ve{\hat{\xi}}$ and the unit-vector of 
the unperturbed light ray $\ve{k}$ are perpendicular to each other: $\ve{\hat{\xi}} \cdot \ve{k}=0$. Therefore, the possible values of their components are not independent of 
each other, but they are restricted by this constraint. This fine-tuning fact has no impact of the upper limits of (\ref{Absolute_Value_S_l_p_Quadrupole_A}) and 
(\ref{Absolute_Value_S_l_p_Quadrupole_B}), but it has an impact on the upper limit of (\ref{Absolute_Value_S_l_p_Quadrupole_C}). Namely, if one takes into account that 
$\ve{\hat{\xi}}$ and $\ve{k}$ are perpendicular to each other, then it actually follows from (\ref{S_l_p_Quadrupole_C}) that 
\begin{eqnarray}
	\left|S^{l=2}_{p=1,n=0}\right| &\le& \frac{1}{2}\,.
	\label{Absolute_Value_S_l_p_Quadrupole_D}
\end{eqnarray}

\noindent
That relation is needed below and has, for instance, also been used in \cite{Zschocke7} (cf. Eq.~(64) ibid.). The upper limit in (\ref{Absolute_Value_S_l_p_Quadrupole_D}) is by a 
factor $2$ smaller than the upper limit in (\ref{Absolute_Value_S_l_p_Quadrupole_C}). This example shows that relation (\ref{Absolute_Value_S_5}) is a correct upper limit. But if 
one takes account of the fact that $\ve{\hat{\xi}}$ and $\ve{k}$ are perpendicular to each other, then the values for $|S^{l}_{p,n}|$ are even smaller that (\ref{Absolute_Value_S_5}). 
However, this fine-tuning fact, that $\ve{\hat{\xi}}$ and $\ve{k}$ are perpendicular to each other, turns out to be of no relevance for the determination of the upper limit of the 
neglected terms of higher multipole order $l>2$, because these terms are below the given threshold of $10\,{\rm nas}$ anyway. But for the quadrupole $l=2$ we will take the upper 
limit of (\ref{Absolute_Value_S_l_p_Quadrupole_D}) instead of (\ref{Absolute_Value_S_l_p_Quadrupole_C}), because the neglected terms of quadrupole are very near this threshold.

\section{Proof of relation (\ref{scaling_delta_k1_M_p_neq_0})}\label{Proof1}

The perturbation of the unit tangent vector in the left-hand side of relation (\ref{scaling_delta_k1_M_p_neq_0}) is given by Eq.~(\ref{delta_k1_M_p_neq_0}) and reads 
\begin{eqnarray} 
        \Delta \ve{n}^{1, p\neq 0}_{M_L} &=& - \frac{2 G \hat{M}_{L}}{c^2} \frac{\left(-1\right)^l}{l!}\,  
	\widehat{\partial}^{\tau_1,p \neq 0}_{L}\,\frac{\ve{\hat{\xi}}}{r_1 - c \tau_1} \frac{1}{r_1},
        \label{Proof_Mass_1} 
\end{eqnarray}

\noindent 
with $l\ge2$ and where the differential operator is given by Eq.~(\ref{Differential_Operator_tau_1_p}). We will determine the impact of (\ref{Proof_Mass_1}) 
on the angle of light deflection. Using relation (\ref{Relation_A}) as well as the relations 
\begin{eqnarray}
	\frac{\partial}{ \partial c \tau_1} \, \frac{\ve{\hat{\xi}}}{r_1 - c \tau_1} \frac{1}{r_1} &=& + \frac{\ve{\hat{\xi}}}{\left(r_1\right)^3}\,,
	\label{Relation_Proof1_A}
	\\
	\frac{\partial}{\partial \ve{\hat{\xi}}}\,\frac{1}{r_1} &=& - \frac{\ve{\hat{\xi}}}{\left(r_1\right)^3}\,,
	\label{Relation_Proof1_B}
\end{eqnarray}

\noindent 
one obtains for the angle between $\ve{k}$ and $\Delta \ve{n}^{1, p\neq 0}_{M_L}$ the expression   
\begin{eqnarray}
	\delta\left(\ve{k},\Delta \ve{n}^{1, p\neq 0}_{M_L}\right) &=& - \frac{2 G \hat{M}_{L}}{c^2} \frac{\left(-1\right)^l}{l!} \frac{\partial}{\partial \hat{\xi}} 
        \sum\limits_{p=1}^{l} {l \choose p} \left(\frac{\partial}{\partial c\tau_1}\right)^{p-1} 
        \nonumber\\
        && \hspace{-2.5cm} \times\,\underset{i_1 \dots i_l}{\rm STF}\; 
        k_{i_1}\,\dots\,k_{i_p}\,P_{i_{p+1}}^{j_{p+1}}\, \dots \,P_{i_l}^{j_l} \frac{\partial}{\partial \xi^{j_{p+1}}}\, \dots \,
       \frac{\partial}{\partial \xi^{j_{l}}}\,\frac{1}{r_1}\,.
        \label{Proof_Mass_3}
\end{eqnarray}

\noindent
By means of relation (\ref{Relation_2}) one obtains   
\begin{eqnarray}
        \delta\left(\ve{k},\Delta \ve{n}^{1, p\neq 0}_{M_L}\right) &=& - \frac{2 G \hat{M}_{L}}{c^2} \frac{1}{l!}\,\frac{\partial}{\partial \hat{\xi}}  
        \sum\limits_{p=1}^{l} {l \choose p} \left(\frac{\partial}{\partial c\tau_1}\right)^{p-1}   
        \nonumber\\
	&& \hspace{-2.5cm} \times \frac{\left(-1\right)^p}{\left(r_1\right)^{l-p+1}} \sum\limits_{n=0}^{[(l-p)/2]} \left(-1\right)^n K^{l-p}_n\,  
	\left(\frac{\hat{\xi}}{r_1}\right)^{l-p-2n} \,S^{i_1 \dots i_l}_{p,n}
	\nonumber\\ 
        \label{Proof_Mass_4}
\end{eqnarray}

\noindent
with the dimensionless tensorial function 
\begin{eqnarray}
	S^{i_1 \dots i_l}_{p,n} &=& \underset{i_1 \dots i_l}{\rm STF}\,k_{i_1}\,\dots\,k_{i_{p+2n}}\,\frac{\hat{\xi}_{i_{p+2n+1}} \dots \hat{\xi}_{i_l}}{(\hat{\xi})^{l-p-2n}}\,. 
	\label{Function_S_1}
\end{eqnarray}

\noindent 
Here it has been taken into account that a contraction of the mass-multipoles with a Kronecker symbol vanishes, hence the projectors can be simplified by 
means of relation (\ref{Replacement}). By implementing the mass-multipoles (\ref{M_L}) of an axi-symmetric body one gets
\begin{eqnarray}
	\delta\left(\ve{k},\Delta \ve{n}^{1, p\neq 0}_{M_L}\right) &=& + \frac{2 G M}{c^2} \frac{(P)^l}{l!} J_l \frac{\partial}{\partial \hat{\xi}} 
        \sum\limits_{p=1}^{l} {l \choose p} \left(\frac{\partial}{\partial c\tau_1}\right)^{p-1}   
        \nonumber\\
	&& \hspace{-2.25cm} \times \frac{\left(-1\right)^p}{\left(r_1\right)^{l-p+1}} \sum\limits_{n=0}^{[(l-p)/2]} \left(-1\right)^n K^{l-p}_n\,  
        \left(\frac{\hat{\xi}}{r_1}\right)^{l-p-2n} S^{l}_{p,n} 
        \nonumber\\ 
        \label{Proof_Mass_6}
\end{eqnarray}

\noindent 
with the dimensionless scalar function  
\begin{eqnarray}
	&& \hspace{-0.5cm} S^l_{p,n} = k_{i_1}\,\dots\,k_{i_{p+2n}}\,\frac{\hat{\xi}_{i_{p+2n+1}} \dots \hat{\xi}_{i_l}}{(\hat{\xi})^{l-p-2n}}\, 
	\delta^{\;3}_{<{i_1}}\,\dots\,\delta^3_{{i_l}>}\,.
	\label{Function_S_2}
\end{eqnarray}

\noindent 
In the step from Eq.~(\ref{Function_S_1}) to Eq.~(\ref{Function_S_2}) the STF operation has been omitted in view of relation (\ref{STF_comment_3}). 
The function of Eq.~(\ref{Function_S_2}) is considered in Appendix~\ref{Scalar_Function}. Here it is only noticed that $S^{l}_{p,n}$ does not depend 
on variable $c\tau_1$, hence the derivatives with respect to variable $c \tau_1$ of Eq.~(\ref{Proof_Mass_6}) do not act on $S^l_{p,n}$. Furthermore, 
$S^{l}_{p,n}$ depends on the unit direction of three-vector $\ve{\hat{\xi}}$, but not on its absolute value $\hat{\xi}$. In other words, according to 
relation (\ref{Relation_C}), the derivative of $S^l_{p,n}$ with respect to $\hat{\xi}$ of Eq.~(\ref{Proof_Mass_6}) vanishes too. 
Thus, by performing the partial derivative with respect to $\hat{\xi}$ and afterwards the partial derivatives with respect to $c \tau_1$ by means of 
relation (\ref{Proof_Mass_5a}), one obtains 
\begin{eqnarray}
        \delta\left(\ve{k},\Delta \ve{n}^{1, p\neq 0}_{M_L}\right) &=& \delta_1 + \delta_2\,,
	\label{Proof_Mass_7}
\end{eqnarray}

\noindent
with the dimensionless terms  
\begin{eqnarray}
	\delta_1 &=& + \frac{2 G M}{c^2 r_1} \left(\frac{P}{r_1}\right)^l \frac{J_l}{l!} 
        \sum\limits_{p=1}^{l} \left(-1\right)^p {l \choose p} 
	\nonumber\\ 
	&& \times \sum\limits_{n=0}^{[(l-p)/2]} \left(-1\right)^n K^{l-p}_n \; S^{l}_{p,n} \,\left(l-p-2n\right) 
	\nonumber\\ 
	&& \times \sum\limits_{k=1}^{\{p/2\}} L^{p,m_1}_{k} 
	\left(\frac{\hat{\xi}}{r_1}\right)^{l-p-2n-1} \left(\frac{c \tau_1}{r_1}\right)^{p-2k+1}  
        \nonumber\\
        \label{delta_1}
\end{eqnarray}

\noindent 
and
\begin{eqnarray}
        \delta_2 &=& - \frac{2 G M}{c^2 r_1} \left(\frac{P}{r_1}\right)^l \frac{J_l}{l!}
        \sum\limits_{p=1}^{l} \left(-1\right)^p {l \choose p}
        \nonumber\\
        && \times \sum\limits_{n=0}^{[(l-p)/2]} \left(-1\right)^n K^{l-p}_n \; S^{l}_{p,n} \, \left(2l-2p-2n+1\right)
        \nonumber\\
        && \times \sum\limits_{k=1}^{\{p/2\}} L^{p,m_2}_{k}
        \left(\frac{\hat{\xi}}{r_1}\right)^{l-p-2n+1} \left(\frac{c \tau_1}{r_1}\right)^{p-2k+1}. 
        \nonumber\\
        \label{delta_2}
\end{eqnarray}

\noindent 
The coefficients $L^{p,m}_{k}$ are given by Eq.~(\ref{Coefficients_L_p_m_k}), where $m$ is replaced by $m_1=2l-2p-2n+1$ and $m_2=2l-2p-2n+3$, respectively. 
The summation over $k$ of Eqs.~(\ref{delta_1}) and (\ref{delta_2}) leads to Gegenbauer polynomials (\ref{Gegenbauer1}), that means 
\begin{eqnarray}
	\sum\limits_{k=1}^{\{p/2\}} L^{p,m_1}_{k} \left(\frac{c \tau_1}{r_1}\right)^{p-2k+1} &=& (-1)^p \,(p-1)! \; C^{m_1/2}_{p-1}\left(\frac{c\tau_1}{r_1}\right),
	\nonumber\\ 
        \label{Summation_delta_1_k}
        \\
	\sum\limits_{k=1}^{\{p/2\}} L^{p,m_2}_{k} \left(\frac{c\tau_1}{r_1}\right)^{p-2k+1} &=& (-1)^p \,(p-1)! \; C^{m_2/2}_{p-1}\left(\frac{c \tau_1}{r_1}\right). 
	\nonumber\\ 
        \label{Summation_delta_2_k}
\end{eqnarray}

\noindent
Inserting (\ref{Summation_delta_1_k}) and (\ref{Summation_delta_2_k}) into (\ref{delta_1}) and (\ref{delta_2}) yields for the absolute value of (\ref{Proof_Mass_7})
\begin{eqnarray}
	\left|\delta\left(\ve{k},\Delta \ve{n}^{1, p\neq 0}_{M_L}\right)\right| &=& C_1^{M_L} \frac{G M}{c^2 r_1} \left|J_l\right| \left(\frac{P}{r_1}\right)^l \,,
        \label{Proof_Mass_8}
\end{eqnarray}

\noindent 
which shows the validity of relation (\ref{scaling_delta_k1_M_p_neq_0}). The coefficients are given by 
\begin{eqnarray}
	C_1^{M_L} &=& \frac{2}{l!}\,\sum\limits_{p=1}^{l} {l \choose p} \sum\limits_{n=0}^{[(l-p)/2]} \left(-1\right)^n K^{l-p}_n \, \left|S^{l}_{p,n}\right| (p-1)! 
        \nonumber\\ 
	&& \hspace{-1.00cm} \times \Bigg|\left(l-p-2n\right) \left(\frac{\hat{\xi}}{r_1}\right)^{l-p-2n-1} C^{l-p-n+1/2}_{p-1}\left(\frac{c \tau_1}{r_1}\right)
	\nonumber\\ 
	&& \hspace{-1.0cm} - \left(2l-2p-2n+1\right) \left(\frac{\hat{\xi}}{r_1}\right)^{l-p-2n+1} C^{l-p-n+3/2}_{p-1}\left(\frac{c\tau_1}{r_1}\right) \Bigg|. 
	\nonumber\\ 
        \label{Proof_Mass_9}
\end{eqnarray}

\noindent 
The coefficients $K^{l-p}_n$ are given by Eq.~(\ref{K_l_n}) where $l$ is replaced by $l-p$. It is noticed that this coefficient contains a factor $(-1)^n$ 
so that (\ref{Proof_Mass_9}) represents a double sum over positive terms. The insertion of (\ref{Absolute_Value_S_5}) into (\ref{Proof_Mass_9}) results into 
three sums over the variables $p,n,m$. In order to determine the term of Eq.~(\ref{Proof_Mass_9}) a Fortran 90 code has been employed \cite{Fortran} and the relations 
\begin{eqnarray}
	\frac{\hat{\xi}}{r_1} &=& \sin \delta(\ve{k},\ve{x}_1),
	\label{angle_k_r1}
	\\
	\frac{c \tau_1}{r_1} &=& \cos \delta(\ve{k},\ve{x}_1),
\end{eqnarray}

\noindent
have been inserted with the angle $\delta(\ve{k},\ve{x}_1) \in [0,\pi]$. The maximal values of these coefficients (\ref{Proof_Mass_9}) are given in Table~\ref{Table_C_1_M}. 
\begin{table}[t]
	\caption{The maximal values of the coefficients $C_1^{M_L}$ of Eq.~(\ref{Proof_Mass_9}). The values are simplified to one digit after the decimal point. 
	These coefficients are the prefactors of relation (\ref{scaling_delta_k1_M_p_neq_0}).} 
\begin{tabular}{| c | c |}
\hline
&\\[-12pt]
$l$ &\hbox to 20mm{\hfill $C_1^{M_L}$ \hfill}\\[3pt]
\hline
&\\[-12pt]
$2$ & $2.0$ \\[3pt]
$4$ & $13.1$ \\[3pt]
$6$ & $77.3$ \\[3pt]
$8$ & $431.5$ \\[3pt]
$10$ & $2366.8$ \\[3pt]
\hline
\end{tabular}
\label{Table_C_1_M}
\end{table}

\noindent 
By inserting these numerical results for the coefficients into (\ref{Proof_Mass_8}) one finds 
\begin{eqnarray}
	\left|\delta\left(\ve{k},\Delta \ve{n}^{1, p\neq 0}_{M_L}\right)\right| &\ll& 1\,{\rm nas}
        \label{Proof_Mass_10}
\end{eqnarray}

\noindent
to any multipole order and any solar system body and can safely be neglected. For instance, for Jupiter we get $2 \times 10^{-7}\,{\rm nas}$ for $l=2$.


\section{Proof of relation (\ref{scaling_delta_k2_k3_M_p_neq_0})}\label{Proof2} 

The term on the left-hand side of relation (\ref{scaling_delta_k2_k3_M_p_neq_0}) is given by the sum of Eqs.~(\ref{delta_k2_M_p_neq_0}) and (\ref{delta_k3_M_p_neq_0})  
\begin{eqnarray}
	\Delta \ve{n}^{2+3, p\neq 0}_{M_L} &=& \Delta \ve{n}^{2, p\neq 0}_{M_L} + \Delta \ve{n}^{3, p\neq 0}_{M_L},
	\label{Proof_Mass_2} 
\end{eqnarray}

\noindent 
and reads 
\begin{eqnarray}
	\Delta \ve{n}^{2+3, p\neq 0}_{M_L} &=& + \frac{2 G \hat{M}_{L}}{c^2} \frac{\left(-1\right)^l}{l!}\,\frac{1}{R} 
	\nonumber\\ 
	&& \hspace{-1.5cm} 
	\times \left(\widehat{\partial}^{\tau_1,p \neq 0}_{L}\,\frac{\ve{\hat{\xi}}}{r_1 - c \tau_1} - \widehat{\partial}^{\tau_0,p \neq 0}_{L}\,
	\frac{\ve{\hat{\xi}}}{r_0 - c \tau_0}\right),  
        \label{Proof_Mass_20}
\end{eqnarray}

\noindent
with $l\ge2$ and where the differential operators are given by Eqs.~(\ref{Differential_Operator_tau_0_p}) and (\ref{Differential_Operator_tau_1_p}). In this  
Appendix, we will determine the impact of (\ref{Proof_Mass_2}) on the angle of light deflection. By means of the relation 
\begin{eqnarray}
	\frac{\partial}{ \partial c \tau} \, \frac{\ve{\hat{\xi}}}{r - c \tau} &=& \frac{\ve{\hat{\xi}}}{r - c \tau}\,\frac{1}{r}\,,
\end{eqnarray}

\noindent
and afterwards making use of (\ref{Appendix_Relation_A1}) and (\ref{Relation_A}), one obtains for the angle between $\ve{k}$ and $\Delta \ve{n}^{2+3, p\neq 0}_{M_L}$ the 
following expression  
\begin{eqnarray}
	\delta\left(\ve{k},\Delta \ve{n}^{2+3, p\neq 0}_{M_L}\right) &=& - \frac{2 G \hat{M}_{L}}{c^2}\,\frac{1}{R}\,\frac{\left(-1\right)^l}{l!} \frac{\partial}{\partial \hat{\xi}} 
        \sum\limits_{p=1}^{l} {l \choose p} 
        \nonumber\\
        && \hspace{-3.0cm} \times\,\underset{i_1 \dots i_l}{\rm STF}\; 
        k_{i_1}\,\dots\,k_{i_p}\,P_{i_{p+1}}^{j_{p+1}}\, \dots \,P_{i_l}^{j_l} \frac{\partial}{\partial \xi^{j_{p+1}}}\, \dots \,
       \frac{\partial}{\partial \xi^{j_{l}}} 
	\nonumber\\
	&& \hspace{-3.0cm} \times 
	\left[\left(\frac{\partial}{\partial c\tau_1}\right)^{p-1} \ln\left(r_1-c\tau_1\right)-\left(\frac{\partial}{\partial c\tau_0}\right)^{p-1} \ln\left(r_0-c\tau_0\right)\right].
	\nonumber\\ 
        \label{Proof_Mass_25}
\end{eqnarray}

\noindent
We separate (\ref{Proof_Mass_25}) into two terms, 
\begin{eqnarray}
        \delta(\ve{k},\Delta \ve{n}^{2+3, p\neq 0}_{M_L}) &=& 
	\delta(\ve{k},\Delta \ve{n}^{2+3, p=1}_{M_L}) + \delta(\ve{k},\Delta \ve{n}^{2+3, p \ge 2}_{M_L}),
	\nonumber\\ 
        \label{Proof_Mass_30}
\end{eqnarray}

\noindent
where the term with $p = 1$ reads   
\begin{eqnarray}
	&& \hspace{-0.5cm} 
	\delta\left(\ve{k},\Delta \ve{n}^{2+3, p=1}_{M_L}\right) = - \frac{2 G \hat{M}_{L}}{c^2}\,\frac{1}{R}\,\frac{\left(-1\right)^l}{(l-1)!} \frac{\partial}{\partial \hat{\xi}}
        \nonumber\\
        &&  \hspace{-0.25cm} \times\,\underset{i_1 \dots i_l}{\rm STF}\;k_{i_1}\,P_{i_{2}}^{j_{2}}\, \dots \,P_{i_l}^{j_l} 
	\frac{\partial}{\partial \xi^{j_{2}}}\,\dots\,\frac{\partial}{\partial \xi^{j_{l}}}\;\ln \frac{r_1 - c\tau_1}{r_0 - c\tau_0},  
        \label{Proof_Mass_30_A}
\end{eqnarray}

\noindent 
while the term with $p \ge 2$ reads  
\begin{eqnarray}
        \delta\left(\ve{k},\Delta \ve{n}^{2+3, p \ge 2}_{M_L}\right) &=& + \frac{2 G \hat{M}_{L}}{c^2}\,\frac{1}{R}\,\frac{\left(-1\right)^l}{l!} \frac{\partial}{\partial \hat{\xi}}
        \sum\limits_{p=2}^{l} {l \choose p}
        \nonumber\\
        && \hspace{-3.0cm} \times\,\underset{i_1 \dots i_l}{\rm STF}\;
        k_{i_1}\,\dots\,k_{i_p}\,P_{i_{p+1}}^{j_{p+1}}\, \dots \,P_{i_l}^{j_l} \frac{\partial}{\partial \xi^{j_{p+1}}}\, \dots \,
       \frac{\partial}{\partial \xi^{j_{l}}}
        \nonumber\\
        && \hspace{-3.0cm} \times
	\left[\left(\frac{\partial}{\partial c\tau_1}\right)^{p-2} \frac{1}{r_1} -\left(\frac{\partial}{\partial c\tau_0}\right)^{p-2} \frac{1}{r_0}\right]. 
        \label{Proof_Mass_30_B}
\end{eqnarray}

\noindent
In (\ref{Proof_Mass_30_B}) the relation 
\begin{eqnarray}
	\frac{\partial}{\partial c \tau}\, \ln \left(r - c\tau\right) &=& - \frac{1}{r} 
	\label{Relation_Proof2_A} 
\end{eqnarray}

\noindent 
has been used. 

First of all, we consider (\ref{Proof_Mass_30_A}) which we denote as $\delta_3$. Using relations (\ref{Relation_3}) and (\ref{Relation_4}) we get  
\begin{eqnarray}
	&& \hspace{-0.5cm} P^{(i_2}_{j_2} \dots P^{i_l)}_{j_l}\, 
	\frac{\partial}{\partial \xi^{j_{2}}}\,\dots\,\frac{\partial}{\partial \xi^{j_{l}}}\;\ln \frac{r_1 - c\tau_1}{r_0 - c\tau_0} 
        \nonumber\\ 
	&=& \left(-1\right)^l \sum\limits_{n=0}^{\left[\frac{l-1}{2}\right]} G^{l-1}_n\,\frac{P_{(i_2 i_3} \, \dots \, P_{i_{2n} i_{2n+1}}\,
	\hat{\xi}_{i_{2n+2}} \,\dots\,\hat{\xi}_{i_l)}}{(\hat{\xi})^{2l-2n-2}}
	\nonumber\\
	&& \times \left( F^{l-1}_n(c\tau_1,r_1) - F^{l-1}_n(c\tau_0,r_0) \right). 
        \label{Relation_delta_3}
\end{eqnarray}

\noindent 
By inserting (\ref{Relation_delta_3}) into (\ref{Proof_Mass_30_A}) follows  
\begin{eqnarray}
	\delta_3 &=& - \frac{2 G \hat{M}_{L}}{c^2}\,\frac{1}{R}\,\frac{1}{(l-1)!} \frac{\partial}{\partial \hat{\xi}}\,\frac{1}{(\hat{\xi})^{l-1}} 
	\sum\limits_{n=0}^{\left[\frac{l-1}{2}\right]} \left(-1\right)^{n} G^{l-1}_n 
	\nonumber\\
	&& \times\, S^{i_1 \dots i_l}_{p=1,n} \left( F^{l-1}_n(c\tau_1,r_1) - F^{l-1}_n(c\tau_0,r_0) \right),
        \label{delta_3_A}
\end{eqnarray}

\noindent
where the dimensionless scalar function $F^{l-1}_n\left(c\tau,r\right)$ is given by Eq.~(\ref{Function_F1}) and the dimensionless tensorial function reads 
\begin{eqnarray}
        S^{i_1 \dots i_l}_{p=1,n} &=& \underset{i_1 \dots i_l}{\rm STF}\,k_{i_1}\,\dots\,k_{i_{2n+1}}\,\frac{\hat{\xi}_{i_{2n+2}} \dots \hat{\xi}_{i_l}}{(\hat{\xi})^{l-2n-1}}\,,
        \label{Function_S_3}
\end{eqnarray}

\noindent
which is the special case $p=1$ of (\ref{Function_S_1}). Because a contraction of the STF mass multipoles $M_L$ with a Kronecker symbol vanishes, we have simplified the projectors 
in (\ref{Relation_delta_3}) by means of (\ref{Replacement}) with $p=1$ in order to get (\ref{delta_3_A}) with (\ref{Function_S_3}). By implementing these mass-multipoles (\ref{M_L}) 
of an axi-symmetric body one gets 
\begin{eqnarray}
	\delta_3 &=& + \frac{2 G M}{c^2}\,\frac{(P)^l}{R}\,J_l\,\frac{1}{(l-1)!} \frac{\partial}{\partial \hat{\xi}}\,\frac{1}{(\hat{\xi})^{l-1}} 
        \nonumber\\
	&& \hspace{-0.75cm} \times \sum\limits_{n=0}^{\left[\frac{l-1}{2}\right]} \left(-1\right)^{n} G^{l-1}_n S^{l}_{p=1,n} 
        \left( F^{l-1}_n(c\tau_1,r_1) - F^{l-1}_n(c\tau_0,r_0) \right),
	\nonumber\\ 
        \label{delta_3_B}
\end{eqnarray}

\noindent
with the dimensionless scalar function 
\begin{eqnarray}
        && \hspace{-0.5cm} S^l_{p=1,n} = k_{i_1}\,\dots\,k_{i_{2n+1}}\,\frac{\hat{\xi}_{i_{2n+2}} \dots \hat{\xi}_{i_l}}{(\hat{\xi})^{l-2n-1}}\, 
        \delta^{\;3}_{<{i_1}}\,\dots\,\delta^3_{{i_l}>}\,, 
        \label{Function_S_5}
\end{eqnarray}

\noindent
which is the special case $p=1$ of (\ref{Function_S}). In the step from Eq.~(\ref{Function_S_3}) to Eq.~(\ref{Function_S_5}) the STF operation has been omitted 
due to relation (\ref{STF_comment_3}). The function of Eq.~(\ref{Function_S_5}) is considered in Appendix~\ref{Scalar_Function}. As it was mentioned 
in the text below Eq.~(\ref{Function_S_2}), the derivative of (\ref{Function_S_5}) with respect $\hat{\xi}$ vanishes. 
Thus, the partial derivative in (\ref{delta_3_B}) with respect to $\hat{\xi}$ yields 
\begin{eqnarray}
	\delta_3 &=& - \frac{2 G M}{c^2 r_1} \left(\frac{P}{\hat{\xi}}\right)^l J_l \sum\limits_{n=0}^{\left[\frac{l-1}{2}\right]} \left(-1\right)^n G^{l-1}_n S^{l}_{p=1,n}\, 
	 T\left(z,\cos\psi\right)
	\nonumber\\ 
        \label{delta_3_C}
\end{eqnarray}
 
\noindent
where $\psi = \delta(\ve{x}_1,\ve{x}_0)$ is the angle between $\ve{x}_0$ and $\ve{x}_1$. The function $T$ in (\ref{delta_3_C}) reads 
\begin{eqnarray}
	T\left(z,\cos\psi\right) &=& \frac{1}{\widehat{R}} \bigg[\frac{1}{(l-2)!} \left( F^{l-1}_n(c\tau_1,r_1) - F^{l-1}_n(c\tau_0,r_0) \right) 
	\nonumber\\
	&& \hspace{-1.00cm} - \frac{1}{(l-1)!}\left(E^{l-1}_n(c\tau_1,r_1) - E^{l-1}_n(c\tau_0,r_0) \right)\bigg], 
        \label{delta_3_D}
\end{eqnarray}
 
\noindent
where $F^{l-1}_n(c\tau,r)$ is given by Eq.~(\ref{Function_F1}), while $E^{l-1}_n(c\tau,r)$ is $\hat{\xi}$ times the derivative of $F^{l-1}_n (c\tau,r)$ with 
respect to $\hat{\xi}$, 
\begin{eqnarray}
	&& \hspace{-1.0cm} E^{l-1}_n\left(c\tau,r\right) = - F^{l-1}_n\left(c\tau,r\right) 
	\nonumber\\ 
	&& \hspace{0.5cm} + \left(\frac{c\tau}{r}\right)^3 \; \sum\limits_{k=0}^{l-n-2} \frac{2k+1}{2^{2k}} {2k \choose k} \left(\frac{\hat{\xi}}{r}\right)^{2k},
        \label{Function_E}
\end{eqnarray}

\noindent 
and $\widehat{R}$ is the reduced distance (\ref{relation_widehat_R}). The function (\ref{delta_3_D}) can be written in the more explicit form 
\begin{eqnarray}
	T\left(z,\cos\psi\right) &=& \frac{1}{(l-1)!} \left(T_1 - T_2\right), 
        \label{delta_3_E}
\end{eqnarray}

\noindent 
with 
\begin{eqnarray}
	T_1 &=& \sum\limits_{k=0}^{l-n-2} \frac{l}{2^{2k}} {2k \choose k} 
	\left(\frac{\hat{\xi}}{r_0}\right)^{2k} \frac{1}{\widehat{R}} \left[\frac{\ve{k} \cdot \ve{x}_1}{r_1}\,z^{2k} - \frac{\ve{k} \cdot \ve{x}_0}{r_0} \right],  
	\nonumber\\ 
        \label{function_T1}
	\\
	T_2 &=& \sum\limits_{k=0}^{l-n-2} \frac{2k+1}{2^{2k}} 
        {2k \choose k} \left(\frac{\hat{\xi}}{r_0}\right)^{2k}
        \nonumber\\
	&& \times\,\frac{1}{\widehat{R}} \left[\left(\frac{\ve{k} \cdot \ve{x}_1}{r_1}\right)^3 \,z^{2k} - \left(\frac{\ve{k} \cdot \ve{x}_0}{r_0}\right)^3 \right], 
        \label{function_T2}
\end{eqnarray}

\noindent 
where $c \tau_0$ and $c \tau_1$ have been replaced by $\ve{k} \cdot \ve{x}_0$ and $\ve{k} \cdot \ve{x}_1$, respectively. 
In view of $\hat{\xi}/r_0 = \sin\delta(\ve{k},\ve{x}_0)$ and by means of relations (\ref{relation_angle_k_x0_z}) and (\ref{relation_angle_k_x1_z}), it becomes apparent that 
the function (\ref{delta_3_E}) depends on two variables only: $z$ and $\psi$. For $l=2$ one obtains $|T(z,\cos\psi)| \le 2.1$, while for higher multipole orders $l>2$ we have used 
the computer algebra system {\it Maple} \cite{Maple} and have found that the absolute value of (\ref{delta_3_E}) is maximized when $z=1$,  
\begin{eqnarray}
	\left|T\left(z,\cos\psi\right)\right| &\le& \left|T\left(z=1,\cos\psi\right) \right| \quad {\rm for} \quad l > 2\,. 
        \label{delta_3_F}
\end{eqnarray}

\noindent 
By inserting the limits
\begin{eqnarray}
	&& \lim_{z \to 1} \frac{1}{\widehat{R}} \left[\frac{\ve{k} \cdot \ve{x}_1}{r_1} \,z^{2k} - \frac{\ve{k} \cdot \ve{x}_0}{r_0} \right] = 1\,, 
	\label{Relation_delta_3_A}
	\\
	&& \lim_{z \to 1} \frac{1}{\widehat{R}} \left[\left(\frac{\ve{k} \cdot \ve{x}_1}{r_1}\right)^3 \,z^{2k} - \left(\frac{\ve{k} \cdot \ve{x}_0}{r_0}\right)^3 \right] 
	= \frac{1}{2} \left(1 - \cos \psi\right), 
	\nonumber\\ 
	\label{Relation_delta_3_B}
\end{eqnarray}

\noindent 
into (\ref{function_T1}) and (\ref{function_T2}) one arrives at the function 
\begin{eqnarray}
        T\left(z=1,\cos\psi\right) &=& \frac{1}{(l-1)!} \left(L_1 - L_2\right),
        \label{delta_3_G}
\end{eqnarray}

\noindent 
which depends on one variable, $\psi$, and where  
\begin{eqnarray}
	&& \hspace{-0.5cm} L_1 = \sum\limits_{k=0}^{l-n-2} \frac{l}{2^{2k}} {2k \choose k} \left(\frac{\hat{\xi}}{r_0}\right)^{2k}\,,
        \label{function_L1}
        \\
	&& \hspace{-0.5cm} L_2 = \sum\limits_{k=0}^{l-n-2} \frac{2k+1}{2^{2k}} {2k \choose k} \left(\frac{\hat{\xi}}{r_0}\right)^{2k} \frac{1}{2} \left(1 - \cos \psi\right). 
        \label{function_L2}
\end{eqnarray}

\noindent 
Both these polynomials are positive, $L_1 > 0$ and $L_2 \ge 0$, hence 
\begin{equation} 
\begin{array}[c]{l}
\displaystyle
\left|L_1 - L_2\right| \le \left \{ \begin{array}[c]{l}
\hspace{0.3cm} L_1 
\quad \mbox{if} \quad \quad L_1 \ge L_2 \\
\\
\displaystyle
\hspace{0.3cm} L_2
	\quad \mbox{if} \quad \quad L_2 \ge L_1 
\end{array} \right \} \;.
\end{array}
\label{L_1_L_2}
\end{equation} 

\noindent 
The largest values of $L_1$ and $L_2$ are achieved for $\hat{\xi}/r_0 = 1$. Hence, we get 
\begin{eqnarray}
	L_1 &\le& \sum\limits_{k=0}^{l-n-2} \frac{l}{2^{2k}} {2k \choose k},
        \label{function_L1_limit}
        \\
	L_2 &\le& \sum\limits_{k=0}^{l-n-2} \frac{2k+1}{2^{2k}} {2k \choose k}. 
        \label{function_L2_limt}
\end{eqnarray}

\noindent 
These terms can be evaluated by using the relations \cite{Polynomial3} 
\begin{eqnarray}
        \sum\limits_{k=0}^{m} \frac{1}{2^{2k}} {2k \choose k} &=& \frac{2m + 1}{2^{2m}} {2m \choose m}, 
        \label{Function_Sum_F1}
	\\
	\sum\limits_{k=0}^{m} \frac{k}{2^{2k}} {2k \choose k} &=& \frac{m}{3}\,\frac{2m + 1}{2^{2m}} {2m \choose m},
        \label{Function_Sum_F2}
\end{eqnarray}

\noindent
from which one concludes that $L_1 > L_2$ because of the factor $l$ in (\ref{function_L1_limit}). Thus, in view of (\ref{L_1_L_2}), we have only to consider the term 
\begin{eqnarray}
        L_1 &\le& l\,\frac{2l-2n-3}{2^{2l-2n-4}} {2l-2n-4 \choose l-n-2} 
        \label{function_L1_limit_final}
\end{eqnarray}

\noindent 
and obtain the following upper limit 
\begin{eqnarray}
	\hspace{-0.75cm} \left|T\left(z,\cos\psi\right)\right| \le \frac{l}{(l-1)!} \frac{2l-2n-3}{2^{2l-2n-4}} {2l-2n-4 \choose l-n-2}. 
        \label{delta_3_H}
\end{eqnarray}

\noindent 
By inserting (\ref{delta_3_H}) into (\ref{delta_3_B}) we get for the absolute value of $\delta_3$ the following upper limit 
\begin{eqnarray}
	\left|\delta_3\right| &\le& B_{1,A}^{M_L}\,\frac{G M}{c^2 r_1} \left(\frac{P}{\hat{\xi}}\right)^l \left|J_l\right|, 
        \label{upper_limit_delta_3}
\end{eqnarray}

\noindent
where
\begin{eqnarray}
	B_{1,A}^{M_L} &=& \frac{2\,l}{(l-1)!} \,\sum\limits_{n=0}^{\left[\frac{l-1}{2}\right]} \left(-1\right)^n G^{l-1}_n \left|S^{l}_{p=1,n}\right|
	\nonumber\\ 
	&& \times \, \frac{2l-2n-3}{2^{2l-2n-4}} {2l-2n-4 \choose l-n-2}.
        \label{delta_3_L}
\end{eqnarray}

\noindent
The coefficients $G^{l-1}_n$ are given by Eq.~(\ref{Coefficients_G_l_n}) and for $\left|S^{l}_{p=1,n}\right|$ we take relation (\ref{Absolute_Value_S_5}), 
except for the case of $l=2$, where we use relation (\ref{Absolute_Value_S_l_p_Quadrupole_D}). 
In order to determine the sum over variable $n$ of Eq.~(\ref{delta_3_L}) a Fortran 90 code has been employed \cite{Fortran}.   
The values of these coefficients (\ref{delta_3_L}) are given in Table~\ref{Table_B_2_A_M}.
\begin{table}[t]
	\caption{The coefficients $B_{1,A}^{M_L}$ of Eq.~(\ref{delta_3_L}). The values are simplified to one digit after the decimal point. 
	The relation (\ref{Absolute_Value_S_l_p_Quadrupole_D}) needs to be used, in order to get the coefficient $B_{1,A}^{M_L}$ for $l=2$.}
\begin{tabular}{| c | c |}
\hline
&\\[-12pt]
	$l$ &\hbox to 20mm{\hfill $B_{1,A}^{M_L}$ \hfill}\\[3pt]
\hline
&\\[-12pt]
$2$ & $2.1$ \\[3pt]
$4$ & $18.3$ \\[3pt]
$6$ & $85.6$ \\[3pt]
$8$ & $387.4$ \\[3pt]
$10$ & $1711.5$ \\[3pt]
\hline
\end{tabular}
\label{Table_B_2_A_M}
\end{table}

\noindent 
Now we consider (\ref{Proof_Mass_30_B}). The mathematical structure of the individual terms with parameters $\left(c\tau_0,r_0\right)$ and $\left(c\tau_1,r_1\right)$ 
of Eq.~(\ref{Proof_Mass_30_B}) is almost identical to Eq.~(\ref{Proof_Mass_3}). In fact, we may use the very same steps as performed from Eq.~(\ref{Proof_Mass_3}) to 
Eq.~(\ref{Proof_Mass_7}) and use relation (\ref{Proof_Mass_5b}). Then we get 
\begin{eqnarray}
	\delta\left(\ve{k},\Delta \ve{n}^{2+3, p \ge 2}_{M_L}\right) &=& \delta_4 + \delta_5 
	\label{Proof_Mass_35}
\end{eqnarray}

\noindent 
with the dimensionless terms 
\begin{eqnarray}
	\delta_4 &=& - \frac{2 G M}{c^2 r_1} \left(\frac{P}{\hat{\xi}}\right)^l \frac{J_l}{l!}  
        \sum\limits_{p=2}^{l} \left(-1\right)^p {l \choose p}
        \nonumber\\
        && \times \sum\limits_{n=0}^{[(l-p)/2]} \left(-1\right)^n K^{l-p}_n \; S^{l}_{p,n} \,\left(l-p-2n\right)
        \nonumber\\
	&& \times \sum\limits_{k=1}^{[p/2]} Q^{p,m_1}_{k} \; T_4  
        \label{delta_4}
\end{eqnarray}

\noindent
and
\begin{eqnarray}
	\delta_5 &=& + \frac{2 G M}{c^2 r_1} \left(\frac{P}{\hat{\xi}}\right)^l \frac{J_l}{l!}
        \sum\limits_{p=2}^{l} \left(-1\right)^p {l \choose p}
        \nonumber\\
	&& \times \sum\limits_{n=0}^{[(l-p)/2]} \left(-1\right)^n K^{l-p}_n \; S^{l}_{p,n} \, \left(2l-2p-2n+1\right)
        \nonumber\\
	&& \times \sum\limits_{k=1}^{[p/2]} Q^{p,m_2}_{k} \; T_5 \,. 
        \label{delta_5}
\end{eqnarray}

\noindent
The coefficients $Q^{p,m}_{k}$ are given by Eq.~(\ref{Coefficients_Q_p_m_k}), where $m$ is replaced by
$m_1=2l-2p-2n+1$ and $m_2=2l-2p-2n+3$, respectively, and the dimensionless functions read 
\begin{eqnarray}
	T_4 &=& \frac{(\hat{\xi})^{2l-p-2n-1}}{\widehat{R}} \left[\frac{(\ve{k}\cdot\ve{x}_1)^{p-2k}}{(r_1)^{2l-2n-2k-1}} - \frac{(\ve{k}\cdot\ve{x}_0)^{p-2k}}{(r_0)^{2l-2n-2k-1}}\right],
	\nonumber\\
	\label{function_T3}
	\\
	T_5 &=& \frac{(\hat{\xi})^{2l-p-2n+1}}{\widehat{R}} \left[\frac{(\ve{k}\cdot\ve{x}_1)^{p-2k}}{(r_1)^{2l-2n-2k+1}} - \frac{(\ve{k}\cdot\ve{x}_0)^{p-2k}}{(r_0)^{2l-2n-2k+1}}\right],
	\nonumber\\
	\label{function_T4}
\end{eqnarray}

\noindent
where the terms $c \tau_0$ and $c \tau_1$ have been replaced by $\ve{k} \cdot \ve{x}_0$ and $\ve{k} \cdot \ve{x}_1$, respectively. In view of relations (\ref{relation_angle_k_x0_z}) 
and (\ref{relation_angle_k_x1_z}), it becomes apparent that these functions (\ref{function_T3}) and (\ref{function_T4}) depend on two variables only: $z$ and $\psi$. 
Using the computer algebra system {\it Maple} \cite{Maple} one finds that their maximal absolute values are given by 
$\left|T_4\right| \le 1.0$ and $\left|T_5\right| \le 1.1$ for $l=2$, while for higher multipole order we get 
\begin{eqnarray}
	 \left|T_4\right| &\le& 2.0 \quad {\rm for} \quad l > 2\,,
        \label{upper_limit_T3}
	\\ 
        \left|T_5\right| &\le& 2.0 \quad {\rm for} \quad l > 2\,.
        \label{upper_limit_T4}
\end{eqnarray}

\noindent 
These upper limits are valid for any set of parameters $p,k,n$ which is within the region of possible values. The region of possible values is 
given by the upper limits of the sum of these parameters. 

Thus we get for the upper limits of (\ref{delta_4}) and (\ref{delta_5})  
\begin{eqnarray}
	\left|\delta_4\right| &\le& B_{1,B}^{M_L}\,\frac{G M}{c^2 r_1} \left(\frac{P}{\hat{\xi}}\right)^l \left|J_l\right|,
        \label{upper_limit_delta_4}
        \\
	\left|\delta_5\right| &\le& B_{1,C}^{M_L}\,\frac{G M}{c^2 r_1} \left(\frac{P}{\hat{\xi}}\right)^l \left|J_l\right|,
	\label{upper_limit_delta_5} 
\end{eqnarray}

\noindent
with 
\begin{eqnarray}
	B_{1,B}^{M_L} &=& \frac{2}{l!} \left|T_4\right|\, \sum\limits_{p=2}^{l} \left(-1\right)^p {l \choose p}
        \sum\limits_{n=0}^{[(l-p)/2]} \left(-1\right)^n K^{l-p}_n \; \left|S^{l}_{p,n}\right|
        \nonumber\\
        && \times \left(l-p-2n\right) \sum\limits_{k=1}^{[p/2]} Q^{p,2l-2p-2n+1}_{k} \,,
	\label{S_4}
	\\ 
	B_{1,C}^{M_L} &=& \frac{2}{l!} \left|T_5\right|\, \sum\limits_{p=2}^{l} \left(-1\right)^p {l \choose p}
        \sum\limits_{n=0}^{[(l-p)/2]} \left(-1\right)^n K^{l-p}_n \; \left|S^{l}_{p,n}\right| 
        \nonumber\\
        && \times \left(2l-2p-2n+1\right) \sum\limits_{k=1}^{[p/2]} Q^{p,2l-2p-2n+3}_{k}\,.
	\label{S_5}
\end{eqnarray}

\noindent
These terms can be simplified by means of 
\begin{eqnarray}
	&& \hspace{-1.0cm} \sum\limits_{k=1}^{[p/2]} Q^{p,2l-2p-2n+1}_{k} = \left(-1\right)^p \frac{\left(2l-p-2n-1\right)!}{\left(2l-2p-2n+1\right)!}\,,
\nonumber\\
	&& \hspace{-1.0cm} \sum\limits_{k=1}^{[p/2]} Q^{p,2l-2p-2n+3}_{k} = \left(-1\right)^p \frac{\left(2l-p-2n\right)!}{\left(2l-2p-2n+2\right)!}\,.
\end{eqnarray}

\noindent
In order to perform the summation over variables $p,n$ of Eqs.~(\ref{S_4}) and (\ref{S_5}) a Fortran 90 code has been employed \cite{Fortran}. 
The values of these coefficients of Eqs.~(\ref{S_4}) and (\ref{S_5}) are given in Table~\ref{Table_B_2_B_B_M}.
\begin{table}[t]
        \caption{The coefficients $B_{1,B}^{M_L}$ and $B_{1,C}^{M_L}$ of Eqs.~(\ref{S_4}) and (\ref{S_5}). The values are simplified to one digit after the decimal point.}
\begin{tabular}{| c | c | c |} 
\hline
	&&\\[-12pt]
        $l$ &\hbox to 20mm{\hfill $B_{1,B}^{M_L}$ \hfill} &\hbox to 20mm{\hfill $B_{1,C}^{M_L}$ \hfill} \\[3pt]
\hline
	&&\\[-12pt]
$2$ & $0.0$ & $0.7$ \\[3pt]
$4$ & $1.5$ & $3.3$ \\[3pt]
$6$ & $3.9$ & $10.3$ \\[3pt]
$8$ & $15.4$ & $39.2$ \\[3pt]
$10$ & $54.5$ & $142.3$ \\[3pt]
\hline
\end{tabular}
\label{Table_B_2_B_B_M}
\end{table}

\noindent
According to Eq.~(\ref{Proof_Mass_30}) we may add the terms (\ref{upper_limit_delta_3}) as well as (\ref{upper_limit_delta_4}) and (\ref{upper_limit_delta_5}) together 
and obtain 
\begin{eqnarray}
        \left|\delta(\ve{k},\Delta \ve{n}^{2+3, p\neq 0}_{M_L})\right| &\le& B_{1}^{M_L}\,\frac{G M}{c^2 r_1} \left(\frac{P}{d_k}\right)^l \left|J_l\right|, 
	\label{Proof2_50}
\end{eqnarray}

\noindent
where $\hat{\xi}$ has been replaced by the impact parameter $d_k$ and the coefficients of Eq.~(\ref{scaling_delta_k2_k3_M_p_neq_0}) are given by  
\begin{eqnarray}
	B_{1}^{M_L} &=& B_{1,A}^{M_L} + B_{1,B}^{M_L} + B_{1,C}^{M_L}.
	\label{Sum_B_2_M_L}
\end{eqnarray}

\noindent 
These coefficients (\ref{Sum_B_2_M_L}) are summarized in Table~\ref{Table_B_2_M}. 
\begin{table}[t]
	\caption{The maximal values of the coefficients $B_{1}^{M_L}$ of Eq.~(\ref{Sum_B_2_M_L}). The values are simplified to one digit after the decimal point. 
	These coefficients are the prefactors of relation (\ref{scaling_delta_k2_k3_M_p_neq_0}).} 
\begin{tabular}{| c | c |}
\hline
        &\\[-12pt]
        $l$ &\hbox to 20mm{\hfill $B_{1}^{M_L}$ \hfill} \\[3pt]
\hline
       &\\[-12pt]
$2$ & $2.8$  \\[3pt]
$4$ & $23.1$  \\[3pt]
$6$ & $99.8$  \\[3pt]
$8$ & $442.0$ \\[3pt]
$10$ & $1908.3$ \\[3pt]
\hline
\end{tabular}
\label{Table_B_2_M}
\end{table}

\noindent 
Eq.~(\ref{Proof2_50}) agrees with Eq.~(\ref{scaling_delta_k2_k3_M_p_neq_0}). By inserting the coefficients of Table~\ref{Table_B_2_M} into (\ref{Proof2_50}) 
one obtains the contribution of this term to the angle of light deflection. Numerical values for grazing light ray at Jupiter and Saturn are presented in Table~\ref{Table_delta_3_4_5}. 


\section{Proof of relation (\ref{Proof_T2_35})}\label{Appendix_Neglection_1} 

In this Appendix we will show that the maximal contribution of the term (\ref{Proof_T2_1}) to the angle of light deflection is given by Eq.~(\ref{Proof_T2_35}).
In view of relation (\ref{light_deflection_M_L}) and by means of (\ref{Relation_A}) we find that the contribution of (\ref{Proof_T2_1}) to the angle of light
deflection is given by 
\begin{eqnarray}
	\delta(\ve{k},\ve{U}_2^{M_L}) &=& + \frac{2 G M_L\left(t\right)}{c^2} \frac{1}{l!}\,\frac{\partial}{\partial \hat{\xi}} \sum\limits_{n=0}^{[l/2]} 
	\accentset{\ast}{F}^{l}_n\left(c\tau_1,r_1\right)  
	\nonumber\\ 
	&& \hspace{-1.0cm} \times \,G_n^l \,P_{i_1 i_2}\,\dots\,P_{i_{2n-1} i_{2n}}
        \,\frac{\hat{\xi}_{i_{2n+1}} \dots \hat{\xi}_{i_l}}{(\hat{\xi})^{2l-2n}}\,. 
	\label{Appendix_Proof_T2_5}
\end{eqnarray}

\noindent
Using the mass-multipoles of an axi-symmetric body as given by Eq.~(\ref{M_L}) and taking account of relation (\ref{Replacement}) we obtain 
\begin{eqnarray}
	\delta(\ve{k},\ve{U}_2^{M_L}) &=& - \frac{2 G M}{c^2} \frac{J_l}{l!} \sum\limits_{n=0}^{[l/2]} \left(-1\right)^n G_n^l S^l_{p=0,n}
        \nonumber\\
	&& \times\,\frac{\partial}{\partial \hat{\xi}}\,
        \left(\frac{P}{\hat{\xi}}\right)^l \accentset{\ast}{F}^{l}_n\left(c\tau_1,r_1\right)
        \label{Appendix_Proof_T2_10}
\end{eqnarray}

\noindent
where $S^l_{p=0,n}$ is the special case $p=0$ of the scalar function of Eq.~(\ref{Function_S_2}). In Eq.~(\ref{Appendix_Proof_T2_10}) it has been taken into account that the 
derivative of this scalar function (\ref{Function_S_2}) with respect to $\hat{\xi}$ vanishes. By performing the derivative in (\ref{Appendix_Proof_T2_10}) we get 
\begin{eqnarray}
	\delta(\ve{k},\ve{U}_2^{M_L}) &=& \frac{2 G M}{c^2 \hat{\xi}} \frac{J_l}{l!} \left(\frac{P}{\hat{\xi}}\right)^l \sum\limits_{n=0}^{[l/2]} \left(-1\right)^n G_n^l S^l_{p=0,n} 
        \nonumber\\
	&& \times \bigg[l\,\accentset{\ast}{F}^{l}_n\left(c\tau_1,r_1\right)  - \accentset{\ast}{E}^{l}_n\left(c\tau_1,r_1\right)\bigg],
        \label{Appendix_Proof_T2_15}
\end{eqnarray}

\noindent
where $\accentset{\ast}{F}^{l}_n(c\tau_1,r_1)$ is given by Eq.~(\ref{Function_F_Star}) and $\accentset{\ast}{E}^l_n(c\tau_1,r_1)$ reads 
\begin{eqnarray}
	\accentset{\ast}{E}^{l}_n\left(c\tau_1,r_1\right) &=& - \accentset{\ast}{F}^{l}_n\left(c\tau_1,r_1\right) 
	\nonumber\\
	&& \hspace{-1.0cm} + \left(\frac{c\tau_1}{r_1}\right)^3\;\sum\limits_{k=1}^{l-n-1} \frac{2k+1}{2^{2k}} {2k \choose k} \left(\frac{\hat{\xi}}{r_1}\right)^{2k}, 
        \label{Function_E_Star}
\end{eqnarray}

\noindent 
which resembles a similar structure encountered above by Eq.~(\ref{Function_E}). 
Using $0 \le \hat{\xi}/r_1 \le 1$ and $(-1)^n\,G_n^l > 0$ we find for the upper limit of the absolute value of (\ref{Appendix_Proof_T2_15})   
\begin{eqnarray}
	\left|\delta(\ve{k},\ve{U}_2^{M_L})\right| &\le& \frac{2 G M}{c^2 r_1} \left(\frac{P}{r_1}\right) \frac{|J_l|}{l!} \left(\frac{P}{\hat{\xi}}\right)^{l-1}
        \sum\limits_{n=0}^{[l/2]} \left(-1\right)^n G_n^l 
        \nonumber\\
	&& \hspace{-1.5cm} 
	\times \left|S^l_{p=0,n}\right| \bigg(l\; \left|\accentset{\ast}{F}^l_n\left(c\tau_1,r_1\right)\right| + \left|\accentset{\ast}{E}^l_n\left(c\tau_1,r_1\right)\right|\bigg).  
        \label{Appendix_Proof_T2_25}
\end{eqnarray}

\noindent
Furthermore, since $|c \tau_1/r_1| \le 1$ we conclude from (\ref{Function_F_Star}) and (\ref{Function_E_Star}) that 
\begin{eqnarray}
	&& \hspace{-0.75cm} \left|\accentset{\ast}{F}^l_n\left(c\tau_1,r_1\right)\right| \le \sum\limits_{k=1}^{l-n-1} \frac{1}{2^{2k}} {2k \choose k},
        \label{Appendix_Function_F_prime_limit}
        \\
	&& \hspace{-0.75cm} \left|\accentset{\ast}{E}^l_n\left(c\tau_1,r_1\right)\right| \le \left|\accentset{\ast}{F}^l_n\left(c\tau_1,r_1\right)\right| + 
	\sum\limits_{k=1}^{l-n-1} \frac{2k+1}{2^{2k}} {2k \choose k}.  
        \label{Appendix_Function_E_prime_limit}
\end{eqnarray}

\noindent 
Both these polynomials can be evaluated by means of relations (\ref{Function_Sum_F1}) and (\ref{Function_Sum_F2}). One finds that 
\begin{eqnarray}
	\left(l+1\right) \left|\accentset{\ast}{F}^l_n\left(c\tau_1,r_1\right)\right| \ge \sum\limits_{k=1}^{l-n-1} \frac{2k+1}{2^{2k}} {2k \choose k}. 
\end{eqnarray}

\noindent 
Therefore,  
\begin{eqnarray}
        \left|\delta(\ve{k},\ve{U}_2^{M_L})\right| &\le& \frac{4 G M}{c^2 r_1} \left(\frac{P}{r_1}\right) \frac{\left|J_l\right|}{l!} \left(\frac{P}{\hat{\xi}}\right)^{l-1}
        \sum\limits_{n=0}^{[l/2]} \left(-1\right)^n G_n^l 
        \nonumber\\
	&& \hspace{-0.5cm} \times \left|S^l_{p=0,n}\right| \left(l+1\right) \left|\accentset{\ast}{F}^l_n\left(c\tau_1,r_1\right)\right|. 
        \label{Appendix_Proof_T2_30}
\end{eqnarray}

\noindent 
For (\ref{Appendix_Function_F_prime_limit}) we get from (\ref{Function_Sum_F1})  
\begin{eqnarray}
	\sum\limits_{k=0}^{l-n-1} \frac{1}{2^{2k}} {2k \choose k} &=& \frac{2l-2n-1}{2^{2l-2n-2}} {2l-2n-2 \choose l-n-1} - 1\,. 
	\nonumber\\ 
        \label{Appendix_Function_F_prime_upper_limit_1}
\end{eqnarray}

\noindent
Hence 
\begin{eqnarray}
	\accentset{\ast}{F}^l_n\left(c\tau_1,r_1\right) &\le& \frac{2l-2n-1}{2^{2l-2n-2}} {2l-2n-2 \choose l-n-1} 
	\label{Appendix_Function_F_prime_upper_limit_2}
\end{eqnarray}

\noindent 
because the first term on the right-hand side of (\ref{Appendix_Function_F_prime_upper_limit_1}) is larger than $1$.  
Inserting (\ref{Appendix_Function_F_prime_upper_limit_2}) into (\ref{Appendix_Proof_T2_30}) yields   
\begin{eqnarray}
        \left|\delta(\ve{k},\ve{U}_2^{M_L})\right| &\le& C_2^{M_L}\,\frac{G M}{c^2 r_1} \left(\frac{P}{r_1}\right) \left|J_l\right| \left(\frac{P}{\hat{\xi}}\right)^{l-1}  
        \label{Appendix_Proof_T2_35}
\end{eqnarray}

\noindent 
with the coefficient 
\begin{eqnarray}
	C_2^{M_L} &=& 4\,\frac{l+1}{l!} \sum\limits_{n=0}^{[l/2]} \left(-1\right)^n G_n^l \left|S^l_{p=0,n}\right| 
	\nonumber\\ 
	&& \times \, \frac{2l-2n-1}{2^{2l-2n-2}} {2l-2n-2 \choose l-n-1}. 
        \label{Coefficient_A_l}
\end{eqnarray}

\noindent 
In order to perform the summation over variable $n$ of Eq.~(\ref{Coefficient_A_l}) a Fortran 90 code has been employed \cite{Fortran}. Numerical values of these coefficients 
are presented in Table~\ref{Table_A_l}. 
\begin{table}[t]
	\caption{The values of the coefficients $C_2^{M_L}$ of Eq.~(\ref{Coefficient_A_l}). The numerical values were rounded to integers.}
\begin{tabular}{| c | c |}
\hline
&\\[-12pt]
$l$ &\hbox to 20mm{\hfill $C_2^{M_L}$ \hfill}\\[3pt]
\hline
&\\[-12pt]
$2$ & $32$ \\[3pt]
$4$ & $154$ \\[3pt]
$6$ & $720$ \\[3pt]
$8$ & $3224$ \\[3pt]
$10$ & $14156$ \\[3pt]
\hline
\end{tabular}
\label{Table_A_l}
\end{table}

\noindent
By inserting these numerical results for the coefficients into (\ref{Appendix_Proof_T2_35}) one finds 
\begin{eqnarray}
        \left|\delta(\ve{k},\ve{U}_2^{M_L})\right| &\ll& 1\,{\rm nas}
        \label{Appendix_Proof_T2_40}
\end{eqnarray}

\noindent
to any multipole order and any solar system body and can safely be neglected. For instance, for Jupiter we get $3 \times 10^{-6}\,{\rm nas}$ for $l=2$.

\section{Proof of relation (\ref{Proof_S2_35})}\label{Appendix_Neglection_2} 

In this Appendix we will show that the maximal contribution of the term (\ref{Proof_S2_1}) to the angle of light deflection is given by Eq.~(\ref{Proof_S2_35}). 
In view of relation (\ref{light_deflection_M_L}) and by means of (\ref{Relation_A}) we find that the contribution of (\ref{Proof_S2_1}) to the angle of light 
deflection is given by 
\begin{eqnarray}
	\delta(\ve{k},\ve{V}^{M_L}_2) &=& - \frac{2 G \hat{M}_L\left(t\right)}{c^2}\,\frac{1}{l!} \frac{1}{R} \left(V^2_L\left(c\tau_1,r_1\right) - V^2_L\left(c\tau_0,r_0\right)\right)
	\nonumber\\ 
        \label{Appendix_Proof_S2_5} 
\end{eqnarray}

\noindent
with 
\begin{eqnarray}
	V^2_L\left(c\tau,r\right) &=& \frac{\partial}{\partial \hat{\xi}} \sum\limits_{n=0}^{[l/2]} G_n^l \,P_{i_1 i_2}\,\dots\,P_{i_{2n-1} i_{2n}} 
	\nonumber\\ 
	&& \hspace{-2.0cm} \times \left[c\tau\,\accentset{\ast}{F}^l_n\left(c\tau,r\right) + r W^l_n \left(\frac{\hat{\xi}}{r}\right)^{2l-2n}\right] 
        \frac{\hat{\xi}_{i_{2n+1}} \dots \hat{\xi}_{i_l}}{\left(\hat{\xi}\right)^{2l-2n}}\,. 
        \nonumber\\
        \label{Appendix_Proof_S2_10} 
\end{eqnarray}

\noindent
The function $\accentset{\ast}{F}^l_n(c\tau,r)$ is given by Eq.~(\ref{Function_F_Star}) and the coefficients $G_n^l$ and $W^l_n$ are given by Eqs.~(\ref{Coefficients_G_l_n}) 
and (\ref{Coefficients_W_l_n}). In order to determine the magnitude of (\ref{Appendix_Proof_S2_5}) we insert the mass-multipole (\ref{M_L}) of an axi-symmetric body and get 
\begin{eqnarray}
	\delta(\ve{k},\ve{V}^{M_L}_2) &=& + \frac{2 G M}{c^2} \frac{J_l}{l!} \frac{1}{R} \left(V^2_L\left(c\tau_1,r_1\right) - V^2_L\left(c\tau_0,r_0\right)\right)
	\nonumber\\ 
        \label{Appendix_Proof_S2_15} 
\end{eqnarray}

\noindent 
with
\begin{eqnarray}
	V^2_L\left(c\tau,r\right) &=& \sum\limits_{n=0}^{[l/2]} \left(-1\right)^n G_n^l \,S^l_{p=0,n}  
        \nonumber\\ 
	&& \hspace{-1.5cm} \times \frac{\partial}{\partial \hat{\xi}} \left(\frac{P}{\hat{\xi}}\right)^l 
	\left[c\tau\,\accentset{\ast}{F}^l_n\left(c\tau,r\right) + r W^l_n \left(\frac{\hat{\xi}}{r}\right)^{2l-2n}\right],  
        \nonumber\\
        \label{Appendix_Proof_S2_20} 
\end{eqnarray}

\noindent
where we have used relation (\ref{Replacement}) and the dimensionless scalar function $S^l_{p,n}$ is given by Eq.~(\ref{Function_S}). Furthermore, it has been taken into account 
that the derivative of this scalar function with respect to $\hat{\xi}$ vanishes. The difference between (\ref{Appendix_Proof_S2_10}) and (\ref{Appendix_Proof_S2_20}) is the STF 
tensor of Eq.~(\ref{STF_Expansion}) and a term $(P)^l$, which is the radius of the body to power $l$, but we shall avoid a new notation and keep the notation $V^2_L(c\tau,r)$ 
of Eq.~(\ref{Appendix_Proof_S2_20}) as is. By performing the derivative in (\ref{Appendix_Proof_S2_20}) and inserting into (\ref{Appendix_Proof_S2_15}) we get 
\begin{eqnarray}
        \delta(\ve{k},\ve{V}^{M_L}_2) &=& B_2^{M_L}\,\frac{G M}{c^2 r_1}\,J_l \left(\frac{P}{\hat{\xi}}\right)^{l}, 
        \label{Appendix_Proof_S2_35}
\end{eqnarray}

\noindent
which coincides with the asserted relation (\ref{Proof_S2_35}). The dimensionless coefficients are given by 
\begin{eqnarray}
	B_2^{M_L} &=& B_2^{M_L}\left(c\tau_1,r_1\right) - B_2^{M_L}\left(c\tau_0,r_0\right), 
     \label{Coefficient_B_l} 
\end{eqnarray}

\noindent
with 
\begin{eqnarray}
	B_2^{M_L}\left(c\tau,r\right) &=& - \frac{2}{l!}\,\frac{1}{\widehat{R}} \sum\limits_{n=0}^{[l/2]} \left(-1\right)^n G_n^l \,S^l_{p=0,n}
	\nonumber\\
	&& \hspace{-2.0cm} \times \bigg[l\,f_1(\hat{\xi},r) - \left(l-1\right)\,f_2(\hat{\xi},r) - f_3(\hat{\xi},r) - g_1(\hat{\xi},r)  
        \nonumber\\
	&& \hspace{-1.5cm} + 2\,g_2(\hat{\xi},r) - g_3(\hat{\xi},r) + l\,W^l_n\,\left(\frac{\hat{\xi}}{r}\right)^{2l-2n-1} 
	\nonumber\\
	&& \hspace{-1.5cm} - W^l_n\,\left(\frac{\hat{\xi}}{r}\right)^{2l-2n+1} - \left(2l - 2n\right) W^l_n\,\left(\frac{\hat{\xi}}{r}\right)^{2l-2n+1} 
        \nonumber\\
	&& \hspace{-1.5cm} + \left(2l - 2n\right) W^l_n\,\left(\frac{\hat{\xi}}{r}\right)^{2l-2n+3} \bigg], 
	\label{Coefficient_B_l_1}
\end{eqnarray}

\noindent
where $\widehat{R}$ is the reduced distance (\ref{relation_widehat_R}) between source and observer. The dimensionless function $S^l_{p=0,n}$ is a special case of Eq.~(\ref{Function_S_4}). The 
coefficients $G_n^l$ and $W^l_n$ are defined by Eqs.~(\ref{Coefficients_G_l_n}) and (\ref{Coefficients_W_l_n}). In order to obtain (\ref{Coefficient_B_l_1}) from (\ref{Appendix_Proof_S2_20}) we 
have used the relations 
\begin{eqnarray}
	\left(c \tau_1\right)^2 &=& \left(r_1\right)^2 - (\hat{\xi})^2,
	\label{Rel_1}
	\\
	\left(c \tau_0\right)^2 &=& \left(r_0\right)^2 - (\hat{\xi})^2.
	\label{Rel_2}
\end{eqnarray}

\noindent
The functions in (\ref{Coefficient_B_l_1}) are given by 
\begin{eqnarray}
	f_a(\hat{\xi},r) &=& \sum\limits_{k=1}^{l-n-1} \frac{1}{2^{2k}} {2k \choose k} \left(\frac{\hat{\xi}}{r}\right)^{2k+2a-3}\,,
	\label{f_n}
	\\
	g_a(\hat{\xi},r) &=& \sum\limits_{k=1}^{l-n-1} \frac{2k}{2^{2k}} {2k \choose k} \left(\frac{\hat{\xi}}{r}\right)^{2k+2a-3}\,.
        \label{g_n}
\end{eqnarray}

\noindent 
According to Eq.~(\ref{Coefficient_B_l_1}) the coefficients $B_2^{M_L}$ in (\ref{Coefficient_B_l}) contain only terms of the structure 
\begin{eqnarray}
	\frac{1}{\widehat{R}} \left[\left(\frac{\hat{\xi}}{r_1}\right)^b - \left(\frac{\hat{\xi}}{r_0}\right)^b\right] &=& 
	\frac{r_0 - r_1}{R}\,\sum\limits_{j=0}^{b-1} \left(\frac{\hat{\xi}}{r_1}\right)^{j} \left(\frac{\hat{\xi}}{r_0}\right)^{b-j} 
	\nonumber\\ 
	\label{Coefficient_B_l_2} 
\end{eqnarray}

\noindent 
where $b \ge 1$ is a natural number and on the right-hand side of (\ref{Coefficient_B_l_2}) we have used the third binomial theorem. The distance $R$ between source and 
observer of Eq.~(\ref{Spatial_Distance_1}) and the dimensionless reduced distance $\widehat{R}$ of Eq.~(\ref{relation_widehat_R}) are related to each other by $R = r_1\,\widehat{R}$. 
We may write the identity (\ref{Coefficient_B_l_2}) in the following form 
\begin{eqnarray}
        \frac{1}{\widehat{R}} \left[\left(\frac{\hat{\xi}}{r_1}\right)^b - \left(\frac{\hat{\xi}}{r_0}\right)^b\right] &=&
	\frac{r_0 - r_1}{R} \left(\frac{\hat{\xi}}{r_0}\right)^b + {\cal O}\left(\frac{\hat{\xi}}{r_1}\right). 
        \nonumber\\
        \label{Coefficient_B_l_4}
\end{eqnarray}

\noindent
The terms of the order ${\cal O}(\hat{\xi}/r_1)$ in (\ref{Coefficient_B_l_4}) turn out to be terms of the order ${\cal O}\left(\frac{GM}{c^2 r_1} \frac{J_l P}{r_1}\right)$ in the 
final expression (\ref{Appendix_Proof_S2_35}). These expressions are at least by a factor $10^{-4}$ smaller than the first term on the right-hand side of (\ref{Coefficient_B_l_4}) 
and contributes less than $10^{-3}\,{\rm nas}$ to the angle of light deflection and can safely be neglected. Then, by inserting (\ref{Coefficient_B_l_1}) into the 
coefficients (\ref{Coefficient_B_l}) and using relation (\ref{Coefficient_B_l_4}) one obtains for these coefficients 
\begin{eqnarray}
        B_2^{M_L} &=& + \frac{2}{l!}\,\frac{r_1-r_0}{R} \sum\limits_{n=0}^{[l/2]} \left(-1\right)^n G_n^l \,S^l_{p=0,n}\;D^l_n  
        \label{Coefficient_B_l_5}
\end{eqnarray}

\noindent 
with the function
\begin{eqnarray}
	D^l_n &=& l\,f_1(\hat{\xi},r_0) - \left(l-1\right)\,f_2(\hat{\xi},r_0) - f_3(\hat{\xi},r_0) - g_1(\hat{\xi},r_0)  
        \nonumber\\
        && + 2\,g_2(\hat{\xi},r_0) - g_3(\hat{\xi},r_0) + l\,W^l_n\,\left(\frac{\hat{\xi}}{r_0}\right)^{2l-2n-1} 
        \nonumber\\
        && - W^l_n\,\left(\frac{\hat{\xi}}{r_0}\right)^{2l-2n+1} - \left(2l - 2n\right) W^l_n\,\left(\frac{\hat{\xi}}{r_0}\right)^{2l-2n+1} 
        \nonumber\\
        && + \left(2l - 2n\right) W^l_n\,\left(\frac{\hat{\xi}}{r_0}\right)^{2l-2n+3}, 
        \label{Function_D}
\end{eqnarray}

\noindent
where  
\begin{eqnarray}
        f_a(\hat{\xi},r_0) &=& \sum\limits_{k=1}^{l-n-1} \frac{1}{2^{2k}} {2k \choose k} \left(\frac{\hat{\xi}}{r_0}\right)^{2k+2a-3}\,,
        \label{f_n_r0}
        \\
        g_a(\hat{\xi},r_0) &=& \sum\limits_{k=1}^{l-n-1} \frac{2k}{2^{2k}} {2k \choose k} \left(\frac{\hat{\xi}}{r_0}\right)^{2k+2a-3}\,.
        \label{g_n_r0}
\end{eqnarray}

\noindent
For the ratio of impact parameter, $\hat{\xi}$, and distance between body and source, $r_0$, we introduce the variable 
\begin{eqnarray}
        y = \frac{\hat{\xi}}{r_0}   
        \label{Ratio}
\end{eqnarray}

\noindent
and notice that this real-valued variable is in the closed interval $y \in \left[0,1\right]$. 
\begin{table}[t]
        \caption{The absolute value of the coefficients $B_2^{M_L}$ of Eq.~(\ref{Coefficient_B_l_6}). For the case $l=2$ the upper limit is given by Eq.~(\ref{Upper_Limit_Coefficient_B_l_equal_2}).
        The values are simplified to one digit after the decimal point.}
\begin{tabular}{| c | c |}
\hline 
&\\[-12pt]
$l$ &\hbox to 20mm{\hfill $|B_2^{M_L}|$ \hfill}\\[3pt]
\hline
&\\[-12pt]
$2$ & $0.7$ \\[3pt] 
$4$ & $4.7$ \\[3pt]
$6$ & $19.2$ \\[3pt]
$8$ & $77.9$ \\[3pt]
$10$ & $315.5$ \\[3pt]
\hline
\end{tabular}
\label{Table_B_l}
\end{table}

\noindent 
The case of mass-quadrupole $l=2$ needs special care. So, before we proceed further, we will consider the function $D^l_n$ in (\ref{Function_D}) 
explicitly for $l=2$. We get 
\begin{eqnarray}
        D^{l=2}_{n=0} &=& \frac{5}{2}\,y^3 - 4\,y^5 + 2\,y^7\,,
        \label{Function_D_l_2_n_0}
	\\
	D^{l=2}_{n=1} &=& 2\,y - 3\,y^3 + 2\,y^5. 
        \label{Function_D_l_2_n_1}
\end{eqnarray}

\noindent
Inserting (\ref{Function_D_l_2_n_0}) and (\ref{Function_D_l_2_n_1}) into (\ref{Coefficient_B_l_5}) we get for the coefficients $B_2^{M_L}$ 
\begin{eqnarray}
	B^{M_2}_{2} &=& \frac{r_1-r_0}{R} \left(\left(e_3^x\right)^2 - \frac{1}{3}\right) 
	\left(5\,y^3 - 8\,y^5 + 4\,y^7\right)
	\nonumber\\ 
	&& \hspace{-0.5cm} + \frac{r_1-r_0}{R} \left(\left(e_3^y\right)^2 - \frac{1}{3}\right) \left(2\,y - 3\,y^3 + 2\,y^5\right),
        \label{Coefficient_B_l_equal_2}
\end{eqnarray}

\noindent
where we have used (\ref{S_l_p_Quadrupole_A}) and (\ref{S_l_p_Quadrupole_B}) as well as the coefficients (\ref{Coefficients_G_l_n}) have been inserted, which is 
$G^{l=2}_{n=0} = 2$ and $G^{l=2}_{n=1} = -1$. In addition, the coordinate system has been rotated such that the $x^1$-axis is aligned with $\hat{\ve{\xi}}$ and 
the $x^2$-axis is aligned with $\ve{k}$, hence $\ve{\hat{\xi}} \cdot \ve{e}_3/\hat{\xi} = e_3^x$ and $\ve{k}\cdot\ve{e}_3 = e_3^y$. This 
orientation of the coordinate system is always possible, because the vectors $\ve{\hat{\xi}}$ and $\ve{k}$ are perpendicular to each other. The components 
of the unit-vector $\ve{e}_3$ are not independent of each other, but restricted by the condition: $(e_3^x)^2 + (e_3^y)^2 + (e_3^z)^2 = 1$. 
Taking into account this fact and using the inequality 
\begin{eqnarray}
        \left|\frac{r_1 - r_0}{R} \right| \le 1
        \label{Coefficient_B_l_7}
\end{eqnarray}

\noindent 
one finds that that the upper limit of $B_{l=2}$ is given by 
\begin{eqnarray}
	\left|B^{M_{2}}_{2}\right| &\le&  \frac{2}{3}\,.
	\label{Upper_Limit_Coefficient_B_l_equal_2}
\end{eqnarray}

\noindent 
A Fortran 90 code has been employed \cite{Fortran}, in order to show that the function (\ref{Function_D}) is larger or equal to zero, $D^l_n \ge 0$. 
The maximal value of function (\ref{Function_D}) is given by 
\begin{eqnarray}
        D^l_n &\le& c_l\,\left(l - 1\right) W^l_n\,. 
        \label{Function_D_Maximum}
\end{eqnarray}

\noindent
The coefficients $c_l$ have been determined numerically. The results are: $c_2 = 1.0$, $c_4 = 1.1$, $c_6=1.3$, $c_8=1.5$, $c_{10}=1.6$. 
These maximal values (\ref{Function_D_Maximum}) have been checked for all possible values of $l,n$ with $2 \le l \le 10$. 
Inserting (\ref{Function_D_Maximum}) into (\ref{Coefficient_B_l_5}) and taking the absolute value yields 
\begin{eqnarray}
	\left|B_2^{M_L}\right| &\le& \frac{2}{l!}\,c_l\,\left(l-1\right) \sum\limits_{n=0}^{[l/2]} \left(-1\right)^n G_n^l \left|S^l_{p=0,n}\right|\,W^l_n\,.
	\nonumber\\ 
        \label{Coefficient_B_l_6}
\end{eqnarray}

\noindent 
Furthermore, the inequality (\ref{Coefficient_B_l_7}) has been used. The numerical values of the upper limits of Eq.~(\ref{Coefficient_B_l_6}) are presented in Table~\ref{Table_B_l} 
and the contribution of the term (\ref{Appendix_Proof_S2_35}) to the angle of light deflection is presented in Table~\ref{Table_T2_35_S2_35}. It is mentioned that relation 
(\ref{Coefficient_B_l_6}) yields $B^{M_2}_{2} = 1.3$ instead of the more sophisticated estimation in (\ref{Upper_Limit_Coefficient_B_l_equal_2}). The reason is that 
relation (\ref{Coefficient_B_l_6}) we have not taken into account that $\hat{\ve{\xi}}$ and $\ve{k}$ are perpendicular to each other. Furthermore, the upper limit presented 
in (\ref{Function_D_Maximum}) slightly overestimates the upper limit of $D^l_n$. Both these effects together lead to the fact that relation (\ref{Coefficient_B_l_6}) overestimates 
the upper limits of $B^l_n$ by approximately a factor of $2$. Nevertheless, the tiny values presented in Table~\ref{Table_T2_35_S2_35} show that relation (\ref{Coefficient_B_l_6}) is 
sufficient to demonstrate that the contribution of $\ve{V}_L^2$ of Eq.~(\ref{S_2}) is negligible for astrometry on the sub-micro-arcsecond level.



\begin{thebibliography}{99}
\bibitem{Comment1}
The post-Newtonian expansion is an analytical expansion in inverse powers of the speed of light up to the order ${\cal O}\left(c^{-8}\right)$;
see Ref.~\cite{Blanchet_Damour1}.
\bibitem{Comment2}
Terms which contain the first time-derivative of the mass-multipoles, $\dot{M}_L$, can be neglected because they contribute much less than $1\,{\rm nas}$ to the angle of light deflection. Terms 
which contain the first time-derivative of the spin-multipoles, $\dot{S}_L$, are beyond the 1.5PN approximation. They contribute also much less than $1\,{\rm nas}$ to the angle of light deflection.
\bibitem{Comment4}
This value of $25.3\,{\rm nas}$ for the neglected mass-quadrupole terms slightly improves a previous estimation of $32.6\,{\rm nas}$ for the neglected mass-quadrupole terms 
for grazing rays at Jupiter in our previous investigations \cite{Zschocke6,Zschocke7}. In particular, by Eq.~(15) in \cite{Zschocke6} and by Eq.~(80) in \cite{Zschocke7} 
a factor $4.5$ in front of the dominant neglected quadrupole terms has been found. Here we get $B_1^{M_L} = 2.8$ for the coefficient of Eq.~(\ref{scaling_delta_k2_k3_M_p_neq_0}) 
and $B_2^{M_L} = 0.7$ for the coefficient of Eq.~(\ref{Proof_S2_35}) for mass-quadrupole, which in total gives a factor $3.5$ in front of the dominant neglected quadrupole terms, 
which is a bit smaller than the factor $4.5$ found in \cite{Zschocke6,Zschocke7}.
\end{thebibliography}
\end{document}